%% file: WHIM_draftFinal.tex
\documentclass[structabstract,longauth]{aa}

\input Planck.tex

\usepackage{graphicx}
\usepackage{txfonts}   
\usepackage{dsfont}
\usepackage{natbib} 
\usepackage{url}                                                          %

\begin{document}
\title{Planck intermediate results. VIII. Filaments between interacting clusters}
   \subtitle{Planck intermediate results}
   \titlerunning{SZ in merging clusters}
\input{PIP_14_authors_and_institutes.tex}

  \abstract
   {About half  of the baryons of the Universe are  expected to be in the form of filaments of hot and low-density intergalactic medium. Most of these baryons remain undetected even by the most advanced X-ray observatories, which are limited in sensitivity to the diffuse low-density medium. }
   {The \Planck\ satellite has provided hundreds of detections of the hot gas in clusters of galaxies via the thermal Sunyaev-Zel'dovich (tSZ) effect and is an ideal instrument for studying extended low-density media through the tSZ effect. In this paper we use the \Planck\ data to search for signatures of a fraction of these missing baryons between pairs of galaxy clusters.}
   {Cluster pairs are good candidates for searching for the hotter and denser phase of the intergalactic medium (which is more easily observed through the SZ effect). Using an X-ray catalogue of clusters and the Planck data, we selected  physical pairs of clusters as candidates. Using the Planck data, we constructed a local map of the tSZ effect centred on each pair of galaxy clusters. \ROSAT\ data were used to construct X-ray maps of these pairs. After hmodelling and subtracting the tSZ effect and X-ray emission for each cluster in the pair, we studied the residuals on both the SZ and X-ray maps.}
   {For the merging cluster pair A399-A401 we observe a significant tSZ effect signal in the intercluster region beyond the virial radii of the clusters. A joint X-ray  SZ analysis allows us to constrain the temperature and density of this intercluster medium. We obtain a temperature of  $kT = 7.1 \pm 0.9$ \, keV (consistent with previous estimates) and a baryon density of $(3.7 \pm 0.2) \times 10^{-4}$ \, cm$^{-3}$.}
  {The \Planck\  satellite mission has provided the first SZ detection of the hot and diffuse intercluster gas.}


   \maketitle
%

\section{Introduction}
\label{introduction}

A sizeable fraction of the baryons of the Universe are expected to be in the form of the WHIM (warm-hot intergalactic medium) and remain undetected 
at low redshifts. The WHIM is expected to exist mostly in filaments but also around and between massive clusters.
These missing baryons are supposed to be in a low-density, low-temperature phase (overdensities between 5 to 200 times the critical density and 
$T = 10^5-10^7$ \, K, \cite{Cen99}), making the amount of X-rays produced by the WHIM too small to be detected with current X-ray facilities.  
By contrast, their detection could be possible via the Sunyaev-Zel'dovich (SZ hereafter) effect \citep{SZ} produced by the inverse Compton scattering between the cosmic microwave background (CMB) photons  and the electrons of the WHIM. As the SZ effect is proportional to the electron pressure in the medium, low-density and low-temperature regions can be detected provided their integrated signal is strong enough. \Planck's relatively poor resolution becomes an advantage in this situation since it permits scanning of wide regions of the sky that can later be integrated to increase the signal-to-noise ratio of the diffuse (but intrinsically large-scale) SZ signal. 

The full-sky coverage and wide frequency range of the \Planck\ satellite mission makes it possible to produce reliable maps of the tSZ emission \citep{2011A&A...536A...8P,2011A&A...536A...9P,2011A&A...536A..10P}. In particular, Planck is better suited than ground experiments to detecting diffuse SZ signals, such as the WHIM, which can extend over relatively large angular scales. Ground experiments can be affected at large angular scales by atmospheric fluctuations that need to be removed. This removal process can distort the modes that include the large angular scale signals. \Planck\ data do not suffer from these limitations and can use their relatively poor angular resolution (when compared to some ground experiments) to its advantage. Indeed, diffuse low surface brightness objects can be resolved and detected by \Planck.
Finally, the wide frequency coverage and extremely high sensitivity of \Planck\ allows for detailed foreground (and CMB) removal that otherwise would overwhelm the weak signal of the WHIM. 

The gas around clusters is expected to be hotter and denser than the WHIM in filaments, making direct detection of the cluster gas more likely.
In addition, the increase of pressure caused by the merging process enhances the SZ signal, making it easier to detect the gas between pairs of interacting clusters. In the process of hierarchical formation clusters assemble via continuous accretion and 
merger events. Therefore, the bridge of intercluster matter between them is expected to be of higher density, 
temperature, and thus thermal pressure than the average WHIM matter found in cosmic filaments \citep{dolag2006}.  
\\

The \Planck\  satellite (\cite{planckmission}) has the potential to detect these filamentary structures directly via the SZ effect. 
Suitable targets for Planck are close objects that subtend large solid angles and therefore have high integrated SZ fluxes.
Alternatively, regions between mergers (filaments between pairs of clusters) or extremely deep gravitational wells (superclusters such as the Shapley 
or Corona Borealis, \cite{cacho2009}) 
will contain diffuse gas with increased pressure that could be detected by \Planck.  For this work, we concentrate on searching for diffuse filamentary-like structure between
 pairs of merging clusters. We used the MCXC (Meta-Catalog of X-Ray Detected Clusters) catalogue of clusters of galaxies \citep{Piffaretti2011} and the \Planck\ data to select a 
sample of pairs of merging clusters to study the properties of the gas in the intercluster region. \\

Indirect WHIM detections have been claimed through absorption lines in the X-ray (and UV) band  \citep{Richter08}. 
There is also evidence of filamentary structure in the intercluster region from X-ray observations of several well-known merging cluster pairs such as A222-A223 \citep{Werner2008}, A399-A401 \citep{Sakelliou2004}, A3391-A3395 \citep{Tittley2001},  and from the double cluster A1758 \citep{Durret}. 
The pairs of clusters A3391-A3395 (separated by about 50$\arcm$ on the sky and at redshifts z=0.051 and z=0.057, respectively \citep{Tittley2001}) 
and more specially, A399-A401 (separated by about 40$\arcm$ on the sky and at redshifts  z=0.0724  and z=0.0737, respectively) 
are of particular interest for the purpose of this paper, given their geometry and angular separation. This is sufficient to allow \Planck\ to resolve the individual cluster components.
 
For A399-A401, earlier observations show an excess of X-ray emission above the background level in the intercluster region. 
Using XMM data, \cite{Sakelliou2004} obtained  best-fitting models in the intercluster region that indicated such an excess. 
Both clusters are classified as non-cool-core clusters and show weak radio halos \citep{Murgia2010}. These two facts could be an indication of a past interaction between the two clusters. \cite{Fujita1996}  analysed  \ASCA\ data of the intercluster region and found a relatively high temperature in this region. They suggested a pre-merger scenario but did not rule out a past interaction. \cite{Fabian1997} used HRI \ROSAT\ data and found a prominent linear feature in A399 pointing towards A401. They suggested that this could be evidence of a past interaction.
Using \Suzaku\  observations, \cite{Fujita2008} found that the intercluster region has a relatively high metallicity of 0.2 solar. 
These works estimated that the filamentary bridge has an electron density of $n_e \sim 10^{-4}$ \, cm$^{-3}$ \citep{Fujita1996,Sakelliou2004,Fujita2008}.  \\

In this paper we concentrate on pairs of merging clusters including A399-A401 and we study the physical properties of the gas in the intercluster region
via a combined analysis of the  tSZ effect and the X-ray emission. 
The paper is organized as follows. Section \ref{P_Data} gives a brief description of the \Planck\ data used for this study.
In Section~\ref{selection} we describe the selection procedure used to identify the most suitable pairs of clusters for the analysis. We search for pairs of clusters
for which the contribution of the SZ effect to the signal is significant in the intercluster medium.
Section~\ref{xrays} describes the X-ray \ROSAT\ observations for the selected pairs of clusters.
In section~\ref{Models} we model the SZ and X-ray emission from the clusters assuming spherical symmetry and subtract them from the data.
In section~\ref{analysis} the SZ and X-ray residuals are fitted to a simplified filament model to characterize the physical properties of the intercluster region.
Section~\ref{discussions} discusses our main results focusing on the limitations imposed by the cluster spherical symmetry assumption and alternative
non-symmetric scenarios. Finally, we conclude in Section~\ref{conclusions}.
        
\section{Planck data}
\label{P_Data}

\begin{table*}[tmb]                 
\begingroup
\newdimen\tblskip \tblskip=5pt
\caption{Main physical parameters of the selected pairs of clusters. Values quoted with NA are unknown.}
\label{tbl_sel}    
\nointerlineskip
\vskip -3mm
\footnotesize
\setbox\tablebox=\vbox{
   \newdimen\digitwidth 
   \setbox0=\hbox{\rm 0} 
   \digitwidth=\wd0 
   \catcode`*=\active 
   \def*{\kern\digitwidth}
   \newdimen\signwidth 
   \setbox0=\hbox{+} 
   \signwidth=\wd0 
   \catcode`!=\active 
   \def!{\kern\signwidth}
%
{\tabskip=1em                                 
\halign{\hfil#\hfil&                       
        \hfil#\hfil&
        \hfil#\hfil&
        \hfil#\hfil&
        \hfil#\hfil&
        \hfil#\hfil&
        \hfil#\hfil&
        \hfil#\hfil&
        \hfil#\hfil&
        \hfil#\hfil&
        \hfil#\hfil \cr
\noalign{\vskip 3pt\hrule\vskip 1.5pt\hrule\vskip 5pt}
Cluster A & GLon  & GLat & z  & $\theta_{500}$  (')  & Cluster B & GLon   & GLat & z &   $\theta_{500}$  (')   &  $\theta_{12}$ (') \cr
\noalign{\vskip 3pt\hrule\vskip 5pt}
A0399& 164.315  & $-$39.458  & 0.0722  & 13.53  &A0401& 164.184  & $-$38.869  & 0.0739 & 14.73 & 35.8  \cr
A3391& 262.377  & $-$25.148  & 0.0514  & 14.91  &A3395& 263.243  & $-$25.188  & 0.0506  & 15.67 & 47.0  \cr 
A2029& 6.437  & 50.53  &  0.0766  & 15.32  &A2033& 7.308  & 50.795  & 0.0817  & 9.94 & 36.6  \cr
MKW3s& 11.393  & 49.458  & 0.0442  & 18.14  &A2063& 12.811  & 49.681  & 0.0355  & 21.29 & 56.8  \cr
A2147& 28.970  & 44.535  & 0.0353  & 22.19  &A2152& 29.925  & 43.979  & 0.0370  & 13.12 & 52.9  \cr
A2256& 111.014  & 31.759  & 0.0581  & 16.62  &A2271& 110.047  & 31.276  & 0.0584  & 10.55 & 57.3  \cr
A0209& 159.878  & $-$73.507  & 0.2060  & 5.60  &A222& 162.494  & $-$72.221  & 0.21430  & 4.35 & 89.9  \cr
\noalign{\vskip 3pt\hrule\vskip 5pt}
A0021& 114.819  & $-$33.711  & 0.0940  & 8.74  &IVZw015& 114.953  & $-$34.357  & 0.0948  & 8.00 & 39.3  \cr 
RXJ  & 271.597  & $-$12.509  & 0.0620  & 13.86  &  RXJ & 272.087  & $-$11.451  & 0.0610  & 11.67 & 69.6\cr
RXJ  & 303.215  & 31.603  & 0.0535  & 13.72  &  RXJ  & 304.324  & 31.534  & 0.0561 & 10.98 & 56.8  \cr 
A3558& 311.987  & 30.726  & 0.0480  & 19.50  &A3562& 313.329  & 30.358  & 0.0490  & 16.09 & 72.7  \cr
A2259& 50.385  & 31.163  & 0.1640  & 6.23  &RXJ17.33+26.62& 49.221  & 30.859  & 0.1644  & 7.18 & 62.5  \cr 
A3694& 8.793  & $-$35.204  & 0.0936  & 9.19  &A3695& 6.701  & $-$35.547  & 0.0894  & 10.62 & 104.4 \cr 
A3854& 8.456  & $-$56.330  & 0.1486  & 6.80  &A3866& 9.408  & $-$56.946  & 0.1544  & 7.23 & 48.5  \cr
MS2215& 58.636  & $-$46.675  & 0.0901  & 9.44  & RXJ  & 59.757  & $-$46.262  & 0.0902  & 8.00 & 52.5  \cr 
RXJ  & 59.757  & $-$46.262  & 0.0902  & 8.00  &A2440& 62.405  & $-$46.431  & 0.0906  & 9.58 & 110.1  \cr 
\noalign{\vskip 3pt\hrule\vskip 5pt}
A222& 162.494  & $-$72.221  & 0.2143  & 4.35  &A223& 162.435  & $-$72.006  & 0.2108  & 4.61 & 12.9 \cr
A3528N& 303.709  & 33.844  & 0.0542  & 12.67  &A3528S& 303.784  & 33.643  & 0.0544 & 13.94 & 12.6  \cr 
A3530& 303.990  & 32.532  & 0.0541  & 12.73  &A3532& 304.426  & 32.477  & 0.0554  & 14.24 & 22.2  \cr
RXJ  & 109.848  & 53.046  & 0.3320  & 3.04  &Zw1358.1+6245& 109.941  & 52.846  & 0.3259  & 3.71 & 12.4  \cr 
A2061 & 48.130  & 57.161  & 0.0777  & 11.06  &A2067& 48.548  & 56.776  & 0.0756  & 8.46 & 26.8  \cr
NPM1G+30.0  & 58.259  & 18.810  & 0.0717  & 9.42  &  RXJ & 58.309  & 18.547  & 0.0645 & 12.71 & 16.0  \cr 
A2384B& 33.321  & $-$48.441  & 0.0963  & 7.87  &A2384A& 33.540  & $-$48.431  & 0.0943  & 8.98 & 8.7  \cr
A2572a& 93.857  & $-$38.800  & 0.0422  & 15.34  &A2572& 94.232  & $-$38.933  & 0.0403  & 14.67 & 19.2  \cr
\noalign{\vskip 3pt\hrule\vskip 5pt}
 PLCK12.8+49.7 &   12.818  &    49.699   &  NA  & NA &  PLCK11.3+49.43  &   11.368 &     49.431 & NA  &  NA & 58.68   \cr
\noalign{\vskip 5pt\hrule\vskip 3pt}}}  
}                                   
\endPlancktablewide                 
\endgroup
\end{table*}                        

\Planck\ \citep{tauber2010a, Planck2011-1.1} is the third-generation
space mission to measure the anisotropy of the CMB.  
It observes the sky in nine frequency bands
covering 30--857\,GHz with high sensitivity and angular resolution
from 31\arcm\ to 5\arcm.  The Low Frequency Instrument (LFI;
\cite{Mandolesi2010, Bersanelli2010, Planck2011-1.4} covers the 30,
44, and 70\,GHz bands with amplifiers cooled to 20\,\hbox{K}.  The
High Frequency Instrument (HFI; \citealt{Lamarre2010,
Planck2011-1.5}) covers the 100, 143, 217, 353, 545, and 857\,GHz
bands with bolometers cooled to 0.1\,\hbox{K}.  Polarization is
measured in all but the highest two bands \citep{Leahy2010,
Rosset2010}.  A combination of radiative cooling and three mechanical
coolers produces the temperatures needed for the detectors and optics
\citep{Planck2011-1.3}.  Two data processing centres (DPCs) check and
calibrate the data and make maps of the sky \citep{Planck2011-1.7,
Planck2011-1.6}.  \Planck's sensitivity, angular resolution, and
frequency coverage make it a powerful instrument for galactic and
extragalactic astrophysics as well as cosmology.  Early astrophysics
results are given in Planck Collaboration VIII--XXVI 2011, based on data
taken between 13~August 2009 and 7~June 2010.  Intermediate
astrophysics results are now being presented in a series of papers
based on data taken between 13~August 2009 and 27~November 2010. 

This paper is based on  \Planck's first 15.5-month survey mission. 
The whole sky has been covered more than two times. We refer
to the \cite{2011A&A...536A...6P} and \cite{2011A&A...536A...5Z}
for the generic scheme of TOI processing and map-making, as
well as for the technical characteristics of the maps used. This
work is based on the nominal survey full-sky maps in the nine
\Planck\ frequency bands provided in HEALPIX \citep{healpix} 
with nside=2048 and full resolution. An error map is associated with
each frequency band and is obtained from the difference of the
first half and second half of the \Planck\ rings for a given position
of the satellite. The resulting maps are basically free from
astrophysical emission, but they are a good representation
of the statistical instrumental noise and systematic error. We
adopted circular Gaussian beam patterns with FWHM values
of 32.6, 27.0, 13.0, 9.88, 7.18, 4.87, 4.65, 4.72, and
4.39\,\arcm for channel frequencies of 30, 44, 70, 100, 143, 217, 353, 545, and 857 \, GHz, respectively. 
The uncertainties in flux measurements
due to beam corrections, map calibrations and uncertainties in
bandpasses are expected to be small, as discussed extensively in
\cite{2011A&A...536A...8P,2011A&A...536A...9P,2011A&A...536A..10P} \\

\section{Pairs of merging clusters in \Planck}
\label{selection}

  \begin{figure*}
   \centering
      \includegraphics[width=4cm,height=8cm,angle=90]{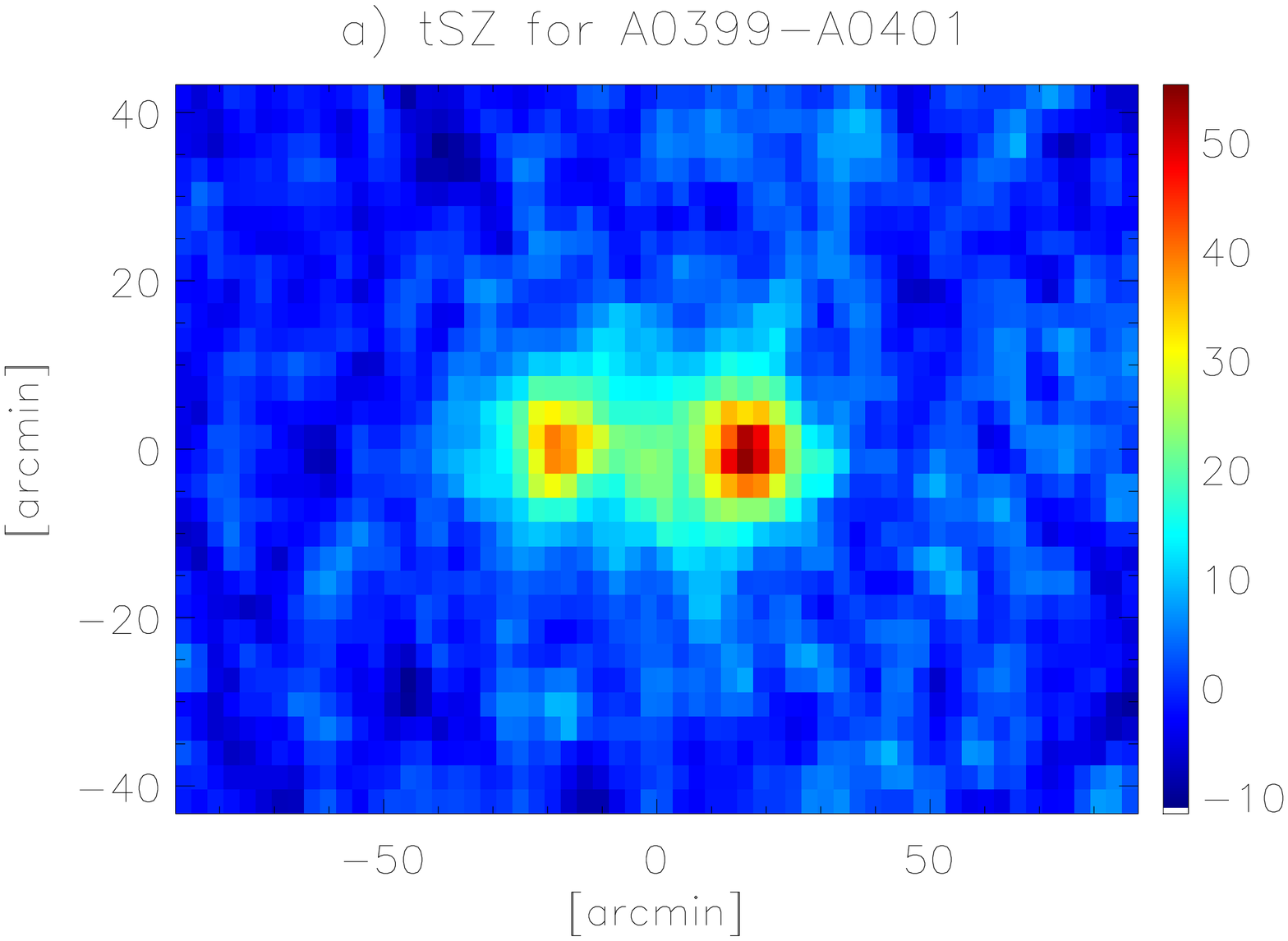}
      \includegraphics[width=4cm,height=8cm,angle=90]{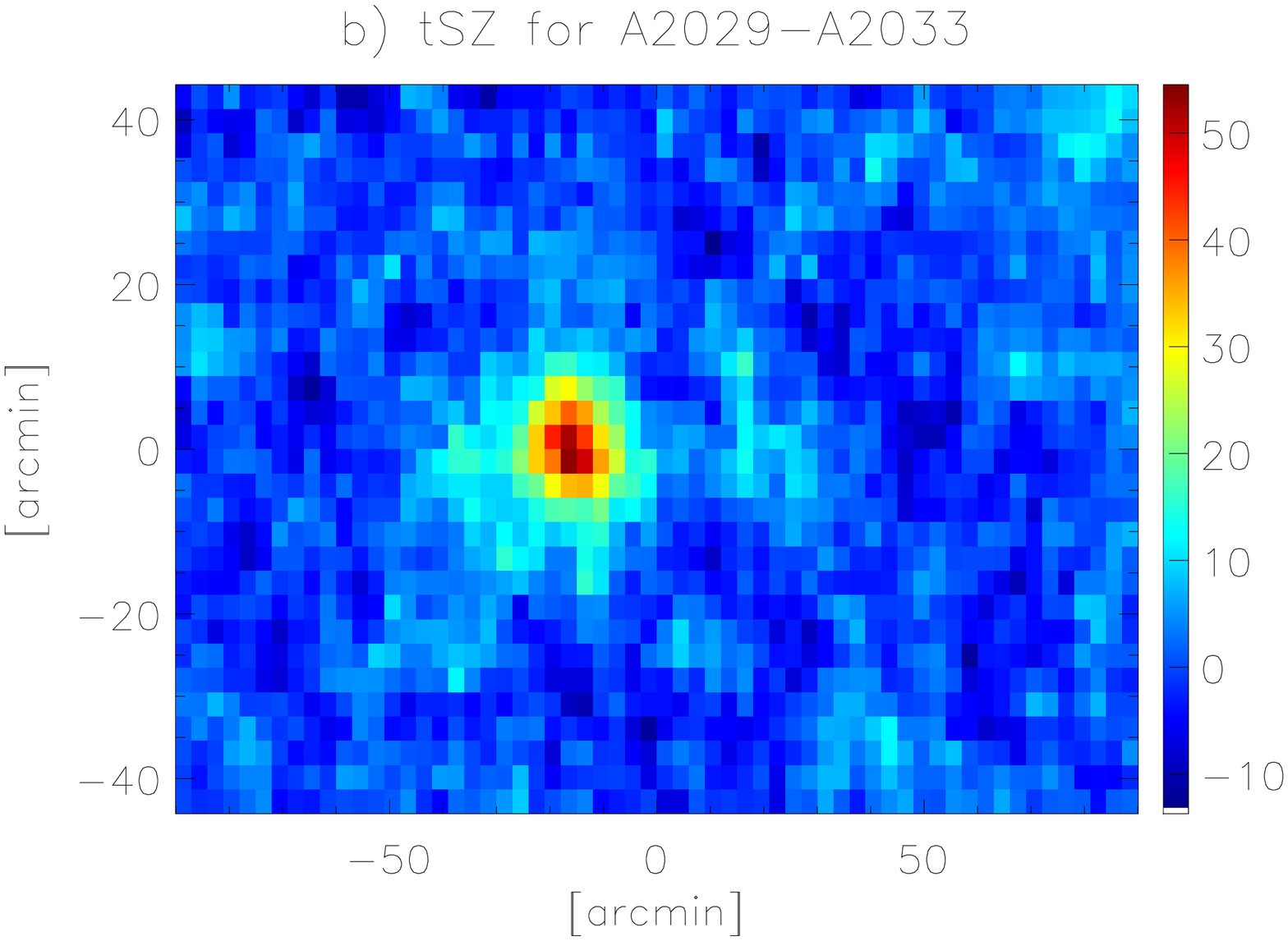}
      \includegraphics[width=4cm,height=8cm,angle=90]{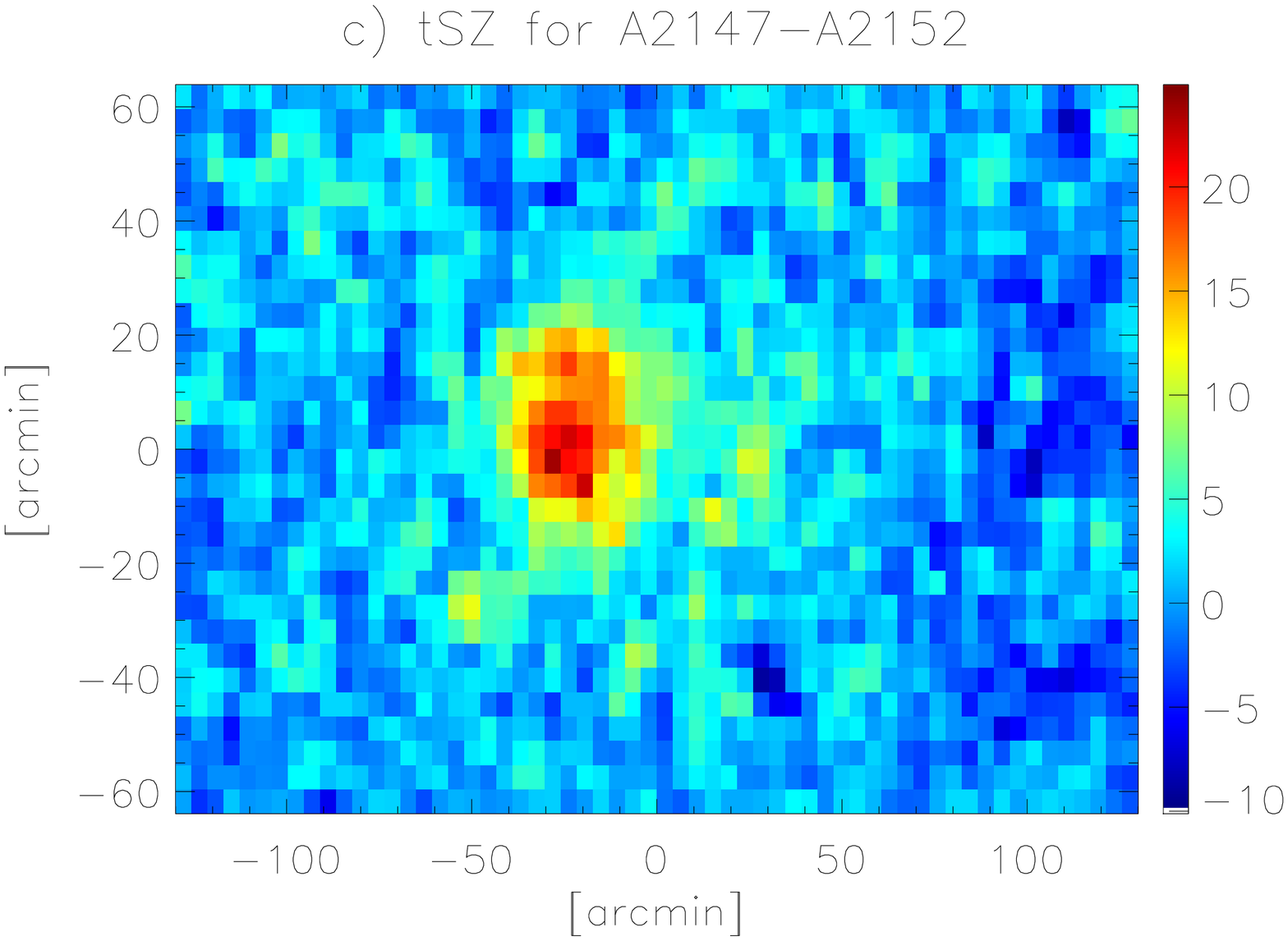}  
       \includegraphics[width=4cm,height=8cm,angle=90]{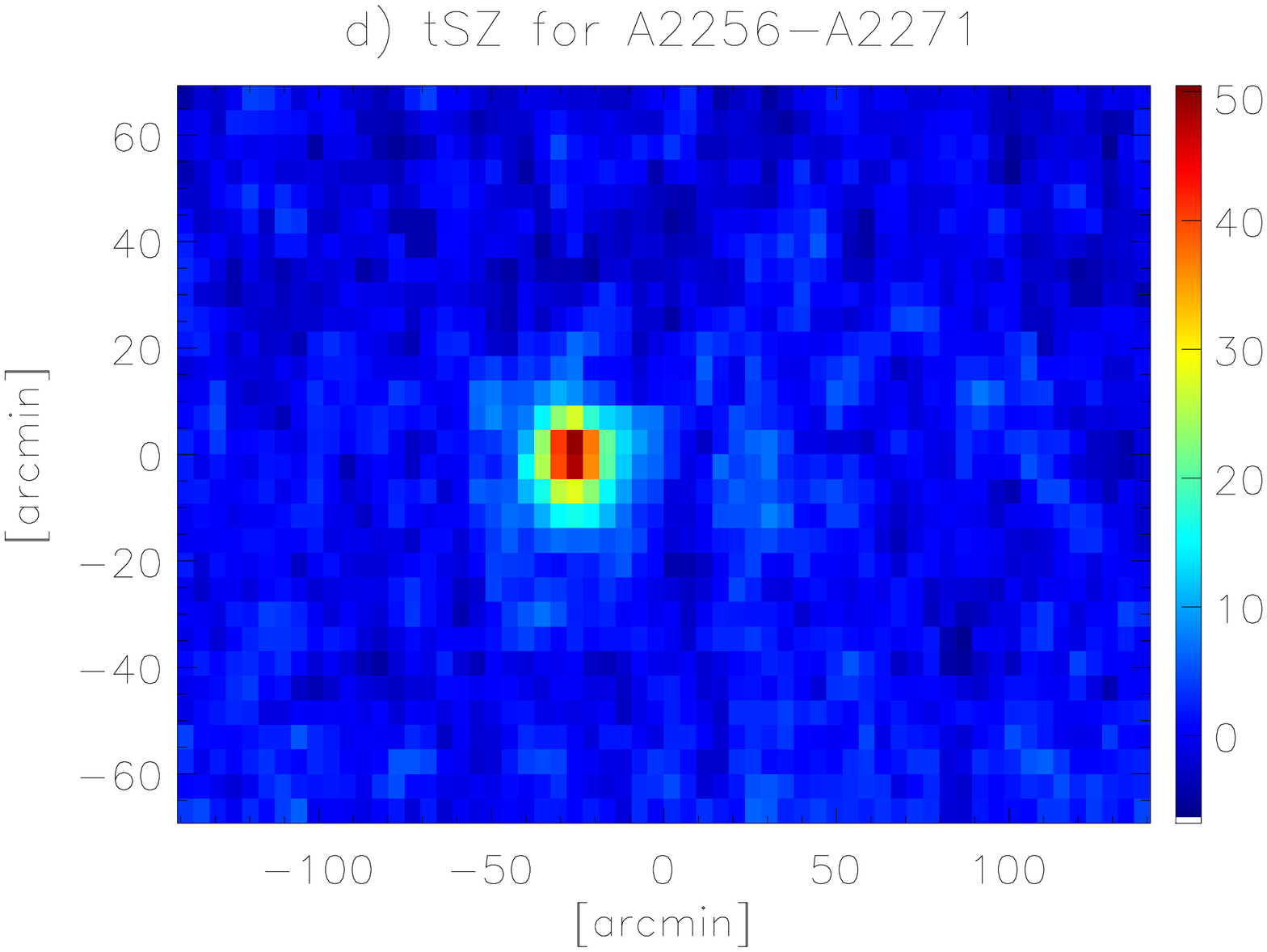} 
       \includegraphics[width=4cm,height=8cm,angle=90]{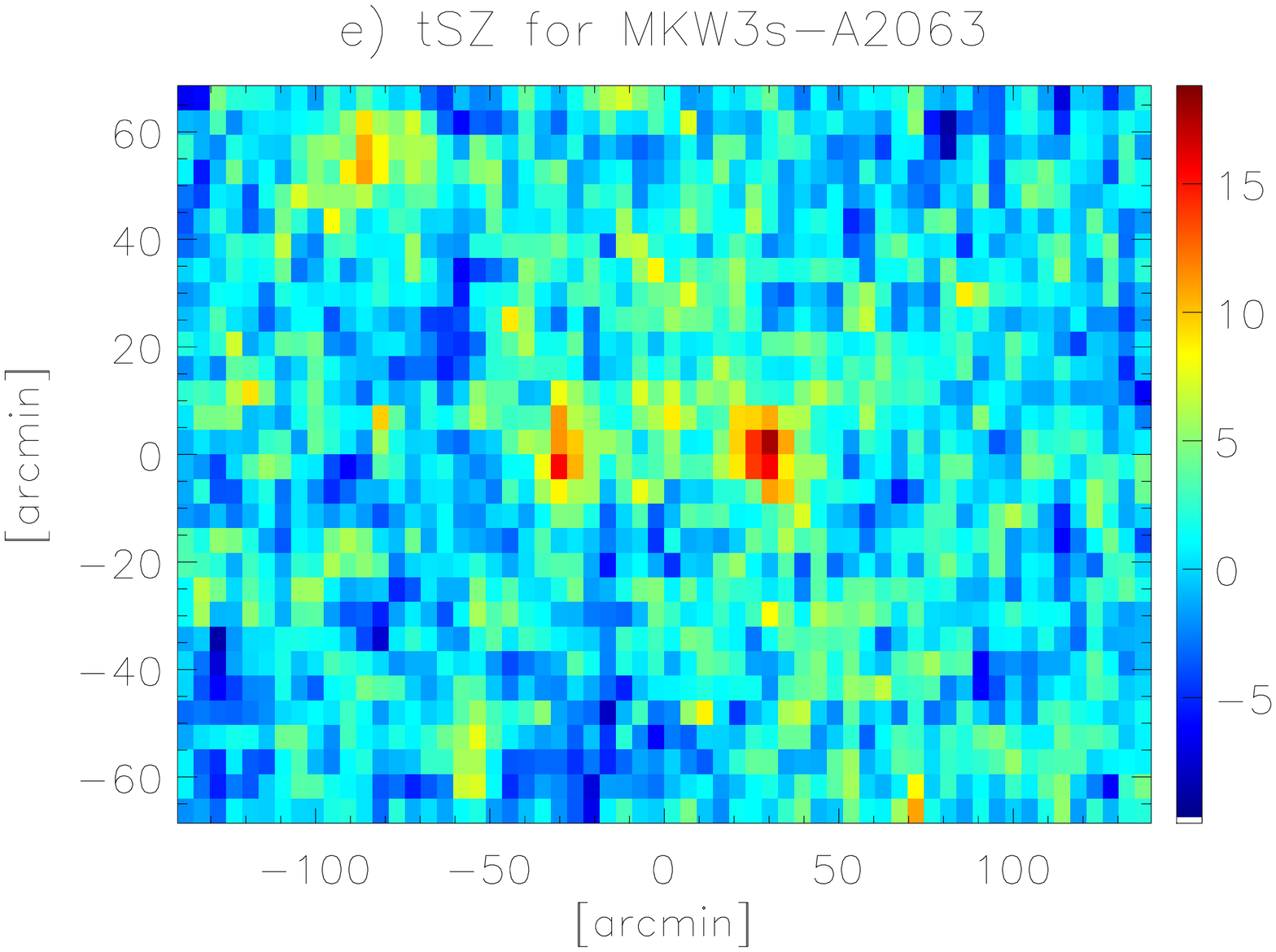}  
        \includegraphics[width=4cm,height=8cm,angle=90]{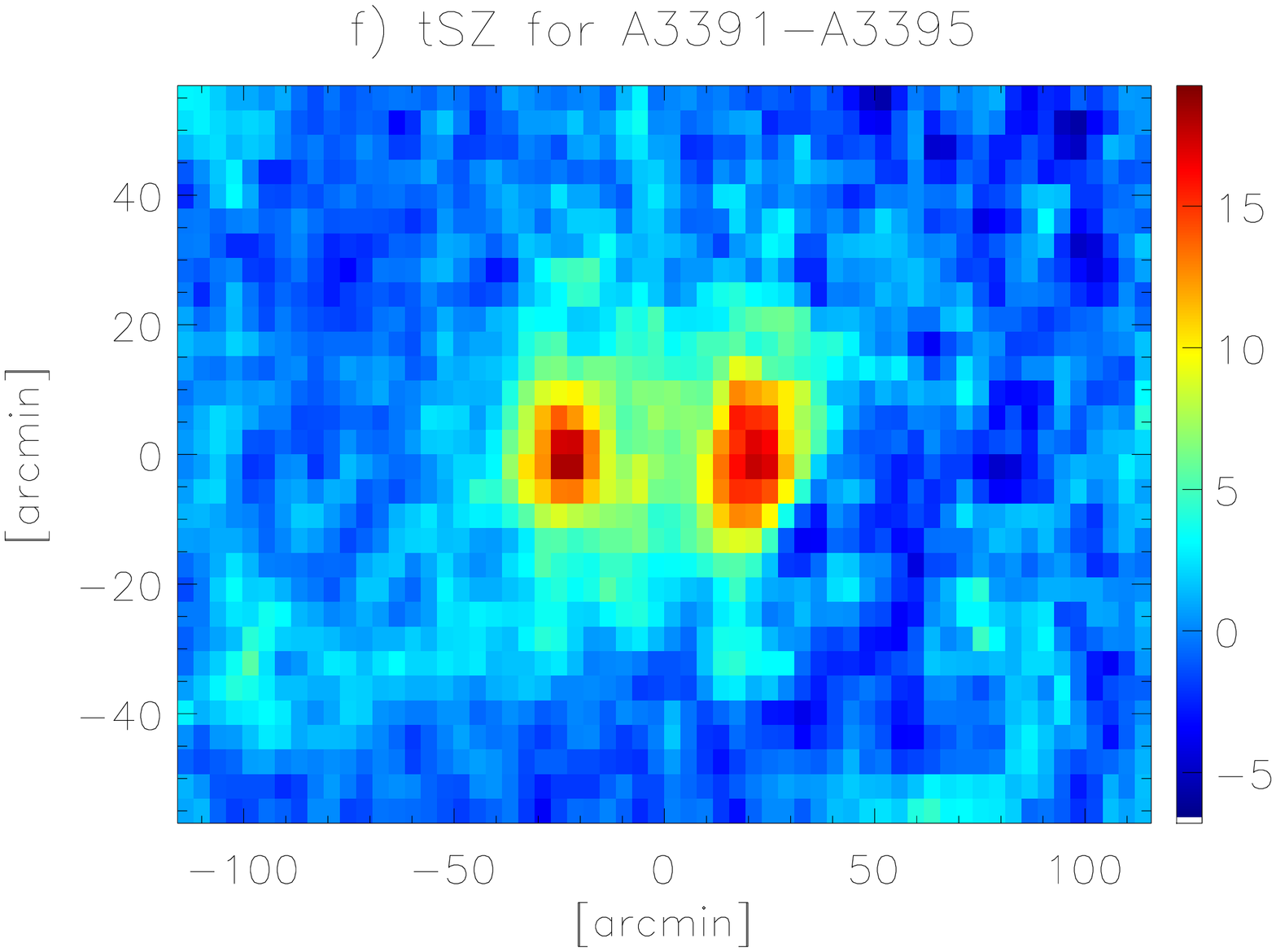}
        \includegraphics[width=4cm,height=8cm,angle=90]{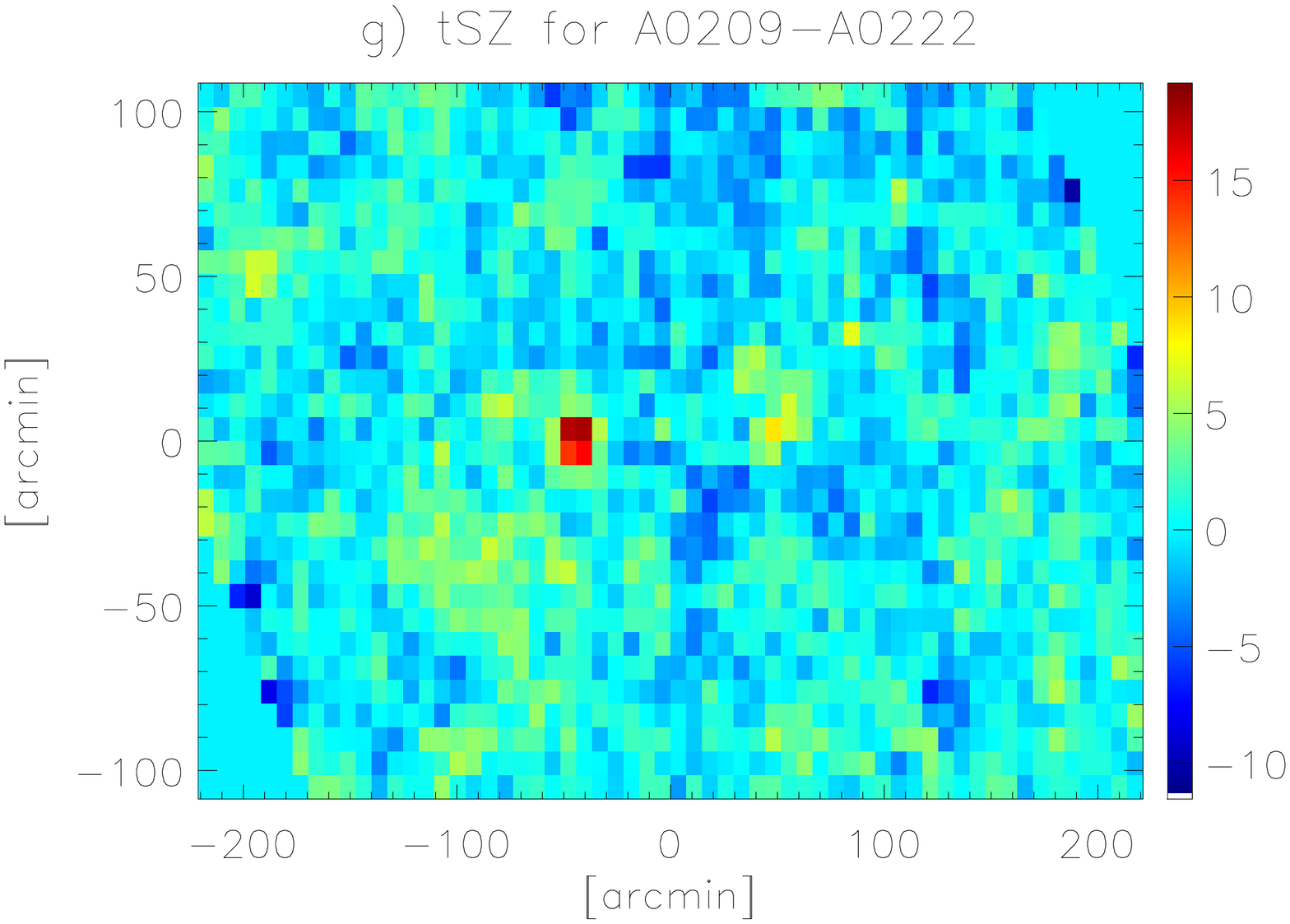}  

     \caption{MILCA maps of the Compton parameter $y \times 10^6$ for the selected pairs of clusters. From left to right and from top to bottom we show
      	the pairs of clusters a) A0399-A0401, b) A2029-A2033, c) A2147-A2152, d) A2256-A2271, e) MKW 3s-A2063, f) A3391-A3395 and g) A0209-A0222.
	}
         \label{selectionmaps}     
   \end{figure*}

In the region trapped between interacting pairs of clusters we expect a hotter and denser phase of the WHIM (see Section~\ref{introduction}) that might produce enough SZ emission to be detected by the \Planck\ satellite. The combined full sky capabilities, lack of atmospheric fluctuations, and wide frequency coverage makes \Planck\ a unique instrument to study these peculiar objects.  \\

To identify pairs of merger clusters in the \Planck\ maps we used the MCXC catalogue of clusters, \cite{Piffaretti2011}, which provides the main physical
parameters of X-ray detected clusters of galaxies (position, redshift, angular extension $\theta_{500}$, and mass $M_{500}$).
We selected a sample of cluster pairs whose difference in redshift is smaller than
0.01 and whose angular distance is between 10 and 120$\arcm$. For all selected pairs we constructed maps of the tSZ (tSZ hereafter) emission from the \Planck\ HFI frequency maps at a resolution
of 7.18$\arcm$ using different component separation techniques: MILCA (maximum internal linear component analysis, \cite{milca}), NILC (needlet internal linear combination, \cite{nilc}) and GMCA (generalized morphological component analysis, \cite{gmca}).  
A detailed discussion of the relative performance of these component separation techniques can be found in \cite{ppp_paper} and \cite{Melin2012}.
Below we take the MILCA maps as a reference to simplify the discussion. The tSZ maps are centred on the barycentre of the X-ray position of
the two clusters and extend the distance between the two clusters (as given in the MCXC catalogue) by up to five times. The error distributions on the final SZ maps were computed using the \Planck\ jack-knife
maps described in \cite{2011A&A...536A...6P}.  \\

Using these tSZ maps, we finally selected those pairs of clusters for which at least one of the clusters has a signal-to-noise ratio higher than five.
Table~\ref{tbl_sel} shows the position (in Galactic coordinates), the redshift, and the angular diameter (in $\theta_{500}$ units) of the two clusters
as well as the angular distance  between them (in arcminutes) for each of the selected pairs of clusters. We divided the selected clusters
into three different sets (separated by horizontal lines in Table~\ref{tbl_sel}). The first set corresponds to the pairs of clusters for which both clusters
are significantly detected in SZ by \Planck. The second set corresponds to the pairs for which one of the clusters is only marginally
detected in SZ by \Planck. The third set consists of pairs of clusters separated by less than 30 \arcm. After a first assessment we concluded 
that for the last two sets the modelling of the clusters in the pair is too complex, given the lower significance of the data. This makes it difficult to extract reliable information on the intercluster region, therefore we do not consider them further in the analysis presented in the next sections. \\

\begin{figure*}
   \centering
   \includegraphics[width=8cm]{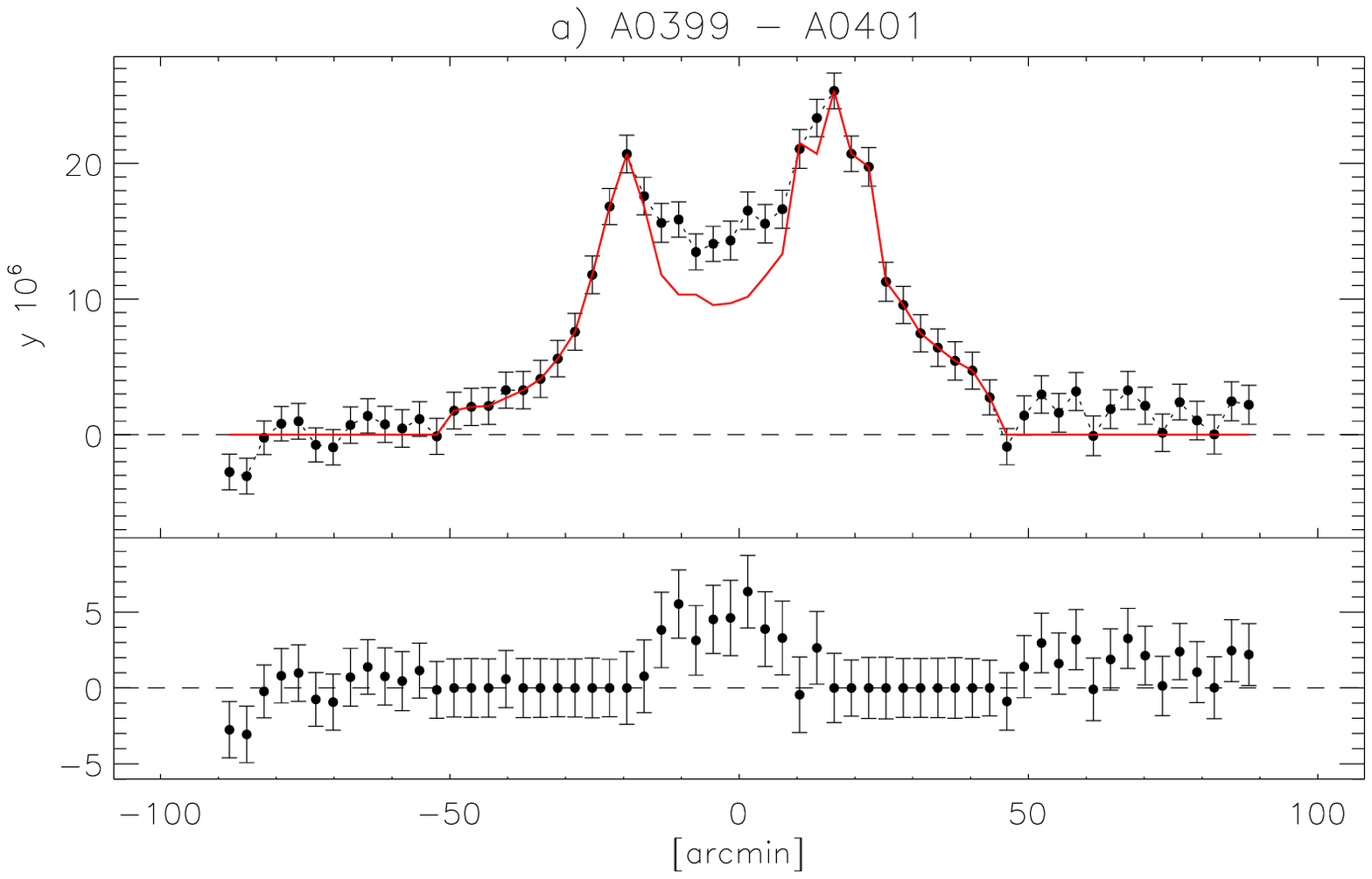}
\includegraphics[width=8cm]{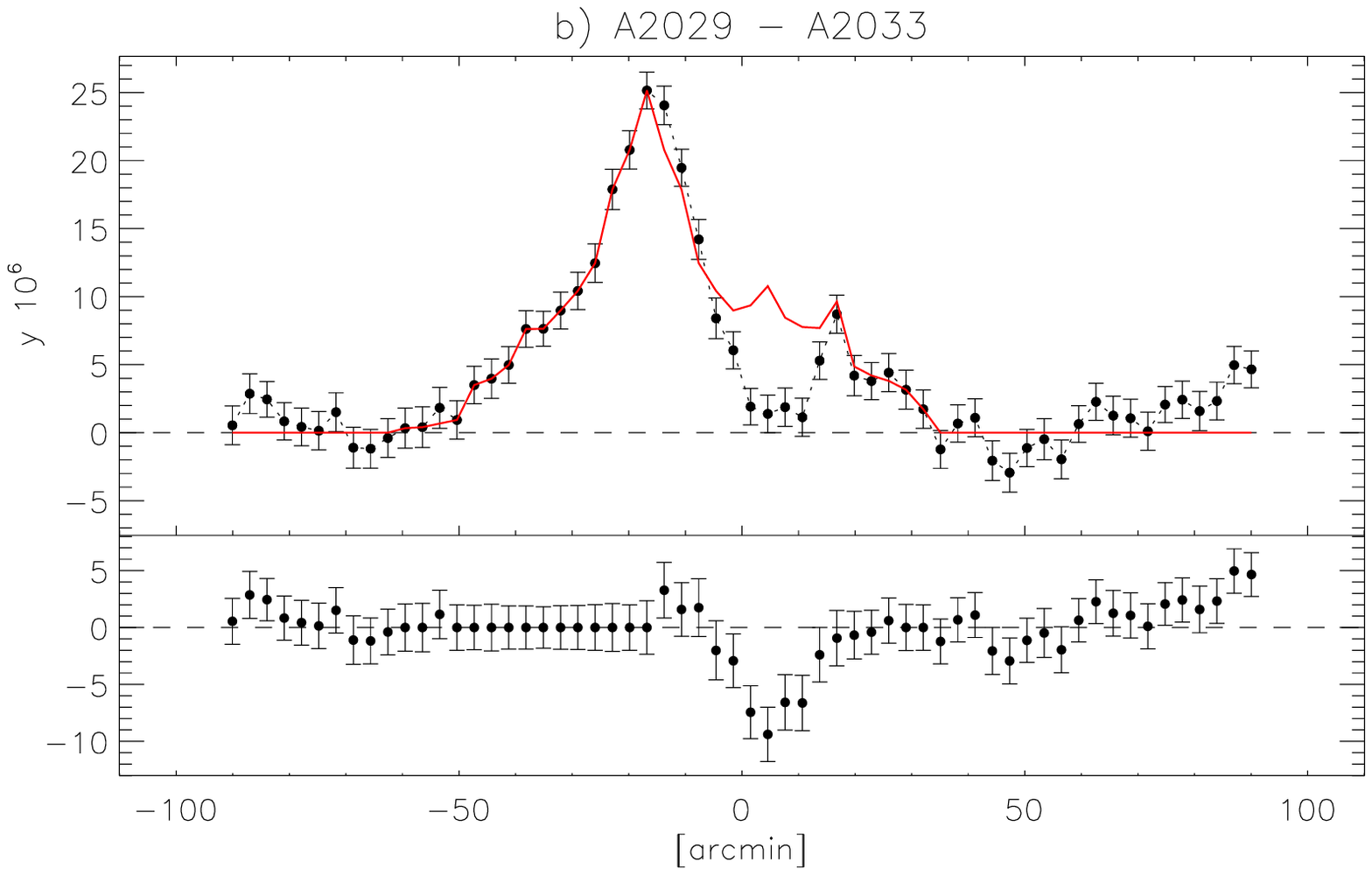}
 \includegraphics[width=8cm]{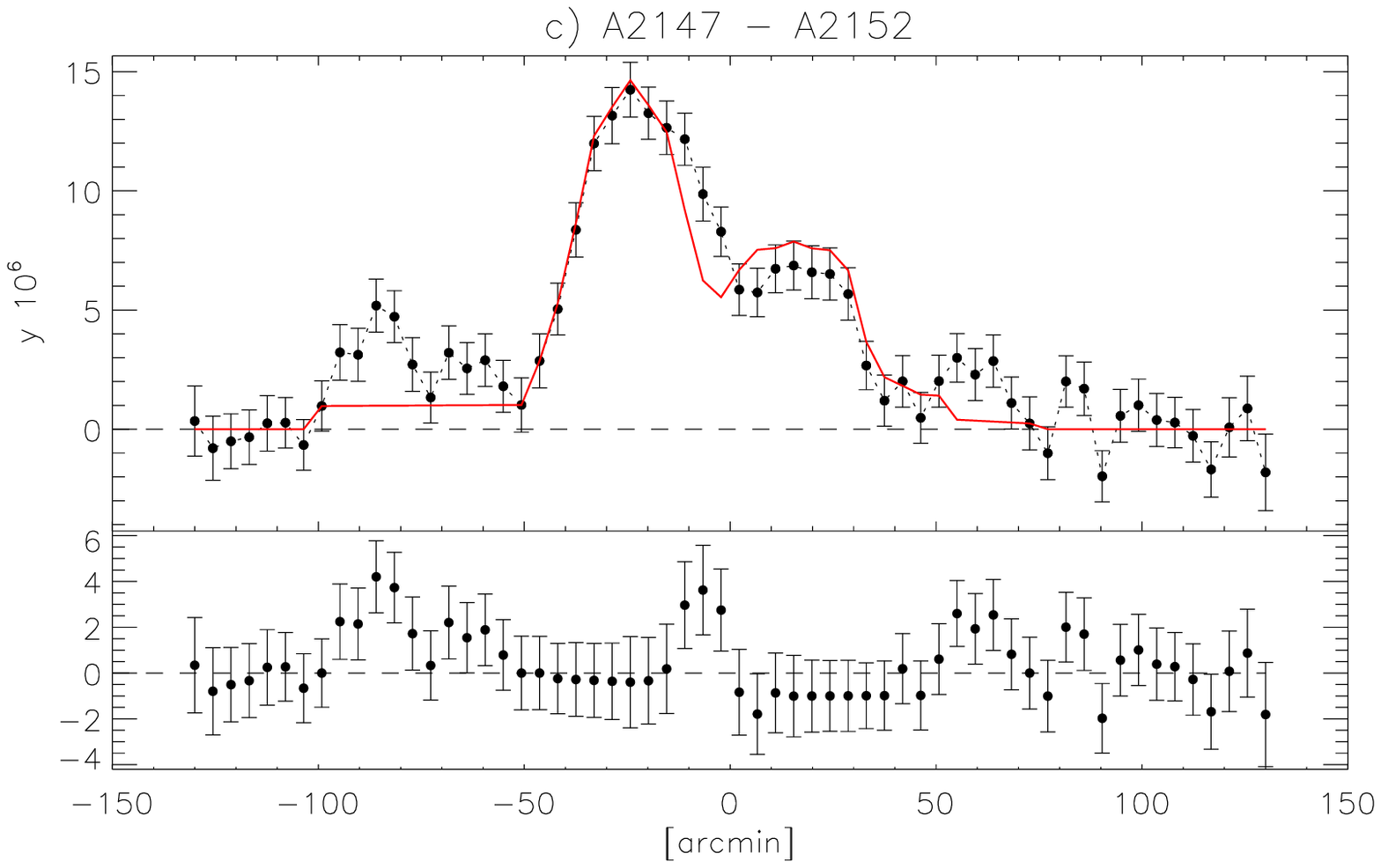}
 \includegraphics[width=8cm]{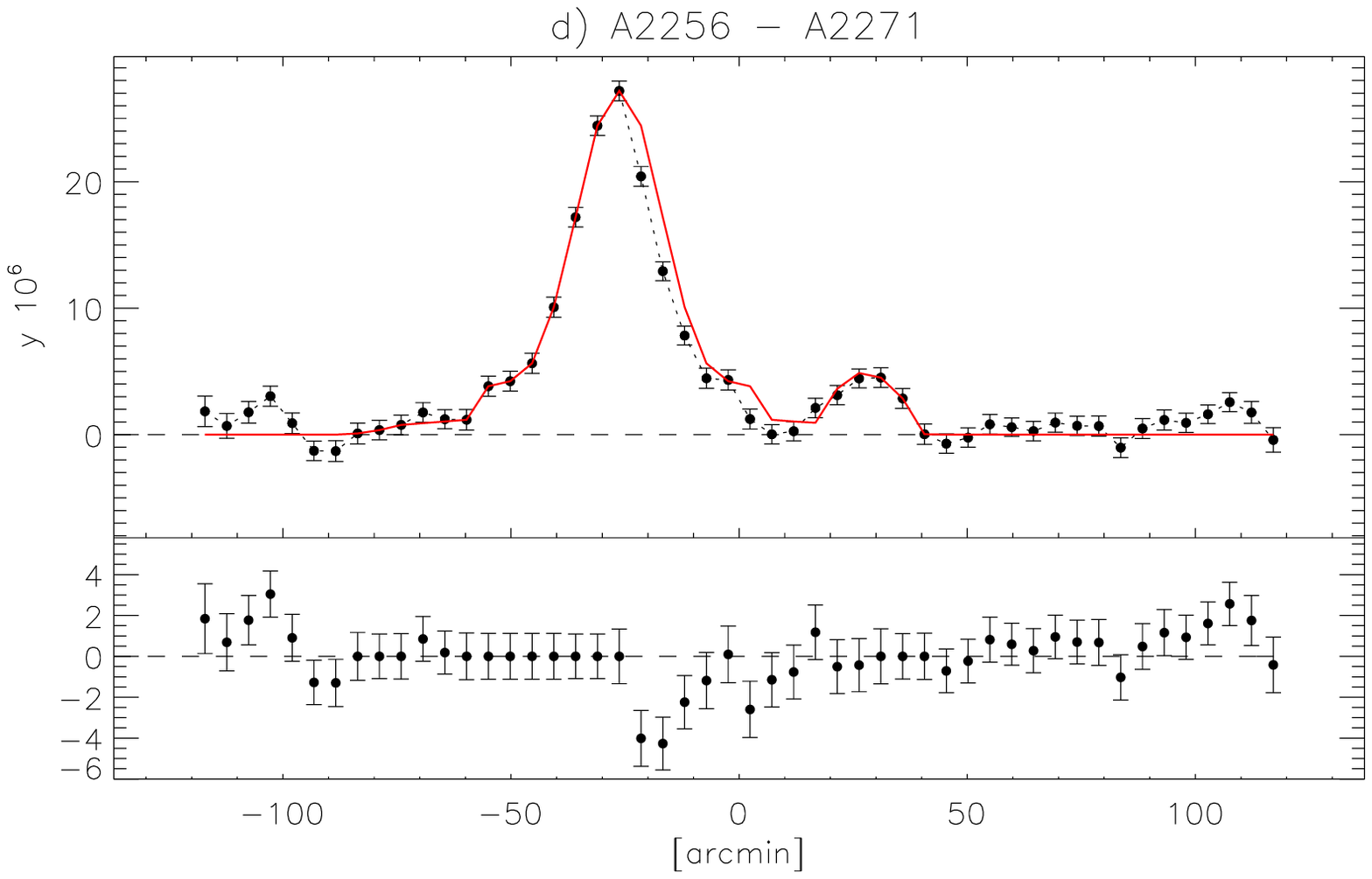}
 \includegraphics[width=8cm]{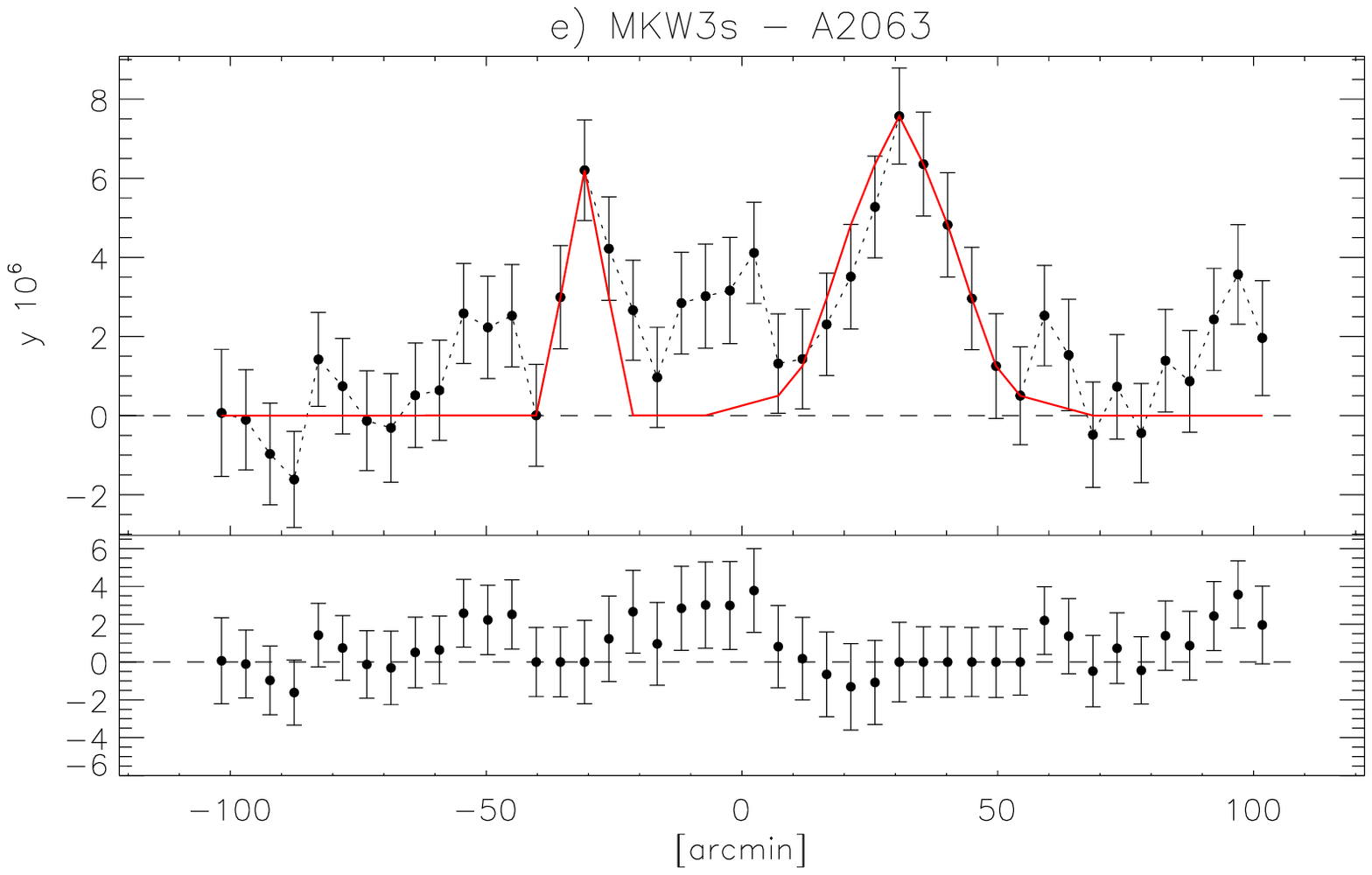}
  \includegraphics[width=8cm]{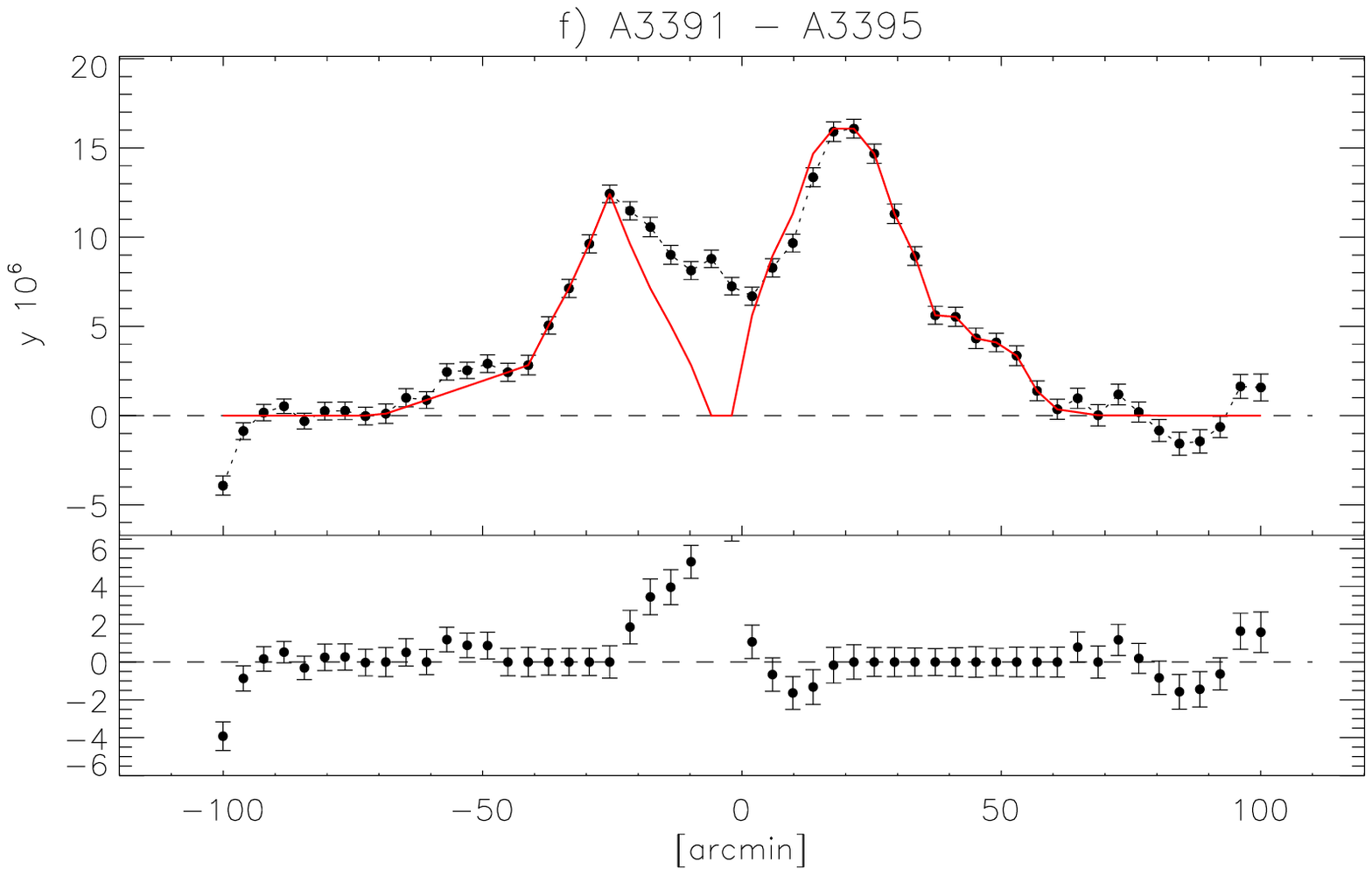}
 \includegraphics[width=8cm]{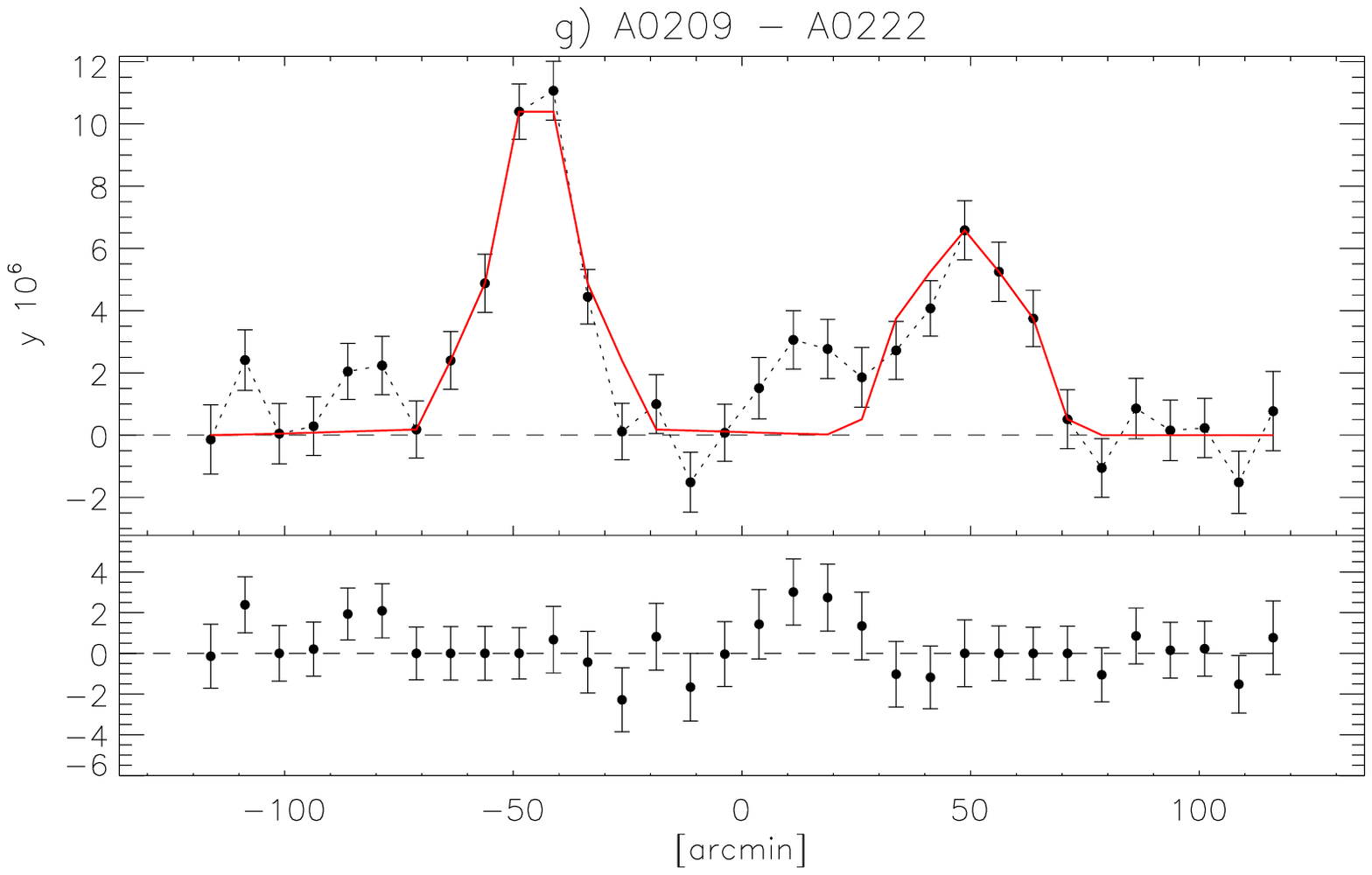}

      \caption{Like in figure \ref{selectionmaps}. From left to right and from top to bottom, 1D tSZ longitudinal profiles and residuals after subtracting the contribution from the clusters (see text for details) for the pairs of clusters a) A399-A401, b) A2029-A2033, c) A2147-A2152, d) A2256-A2271,  e)  MKW 3s-A2063,  f) A3391-A3395, and g) A0209-A222. }
         \label{sym_profile_interpolation}     
   \end{figure*}

Figure~\ref{selectionmaps} shows maps of the Compton parameter $y \times 10^6$ for the six pairs of clusters in the first group in Table~\ref{tbl_sel} (in decreasing order of $y$ from a) to f) ). These maps are rotated so that the virtual line connecting the centre of the clusters is horizontal  and at the middle of the figures. The direction defined
by this line is called longitudinal hereafter, and the direction perpendicular to it is called radial. The pixel size in the longitudinal and radial directions is defined by the angular
distance between the centre of the two clusters, $d$, as follows:
$\theta_{pix} =  5 \times d/60$. From visual inspection of these maps we infer that there might be significant SZ emission
in the intercluster region of the pairs of clusters A399-A401, A2147-A2152, MKW 3s-A2063, and A3391-A3395. To confirm these findings we computed 1D longitudinal profiles by stacking the above maps in the radial direction. Figure~\ref{sym_profile_interpolation} shows the 1D profiles for the six pairs of clusters (notice that the errors are correlated).
The figure also shows the residuals after subtracting an estimate of the tSZ emission from the clusters. In the intercluster region we simply used a symmetric interpolation (in red in the figure) of  the 
external profile of each of the clusters. When there was an obvious contribution from extra clusters (see for example the shoulders on the A3391-A3395 pair, Figure 2e) and the peak on the
 A2029-2033 pair, Figure 2b), we excluded the affected region from the analysis. The error bars on the residuals are increased by a factor $\surd 2$ in the outskirts and $\surd 3$ in the intercluster region to account
for the interpolation, symmetrization, and subtraction. We observe clear extra emission for two of the pairs:  A399-A401 (Figure 2a) and A3391-A3395, (Figure 2e). In the intercluster region,  $\chi^2$ for the null hypothesis (no extra tSZ signal) is
20  for  A399-A401 and 174 for A3391-A3395, for  eight degrees of freedom (11 data samples minus 3 parameters). We also found high $\chi^2$ values for the
A2029-A2033 (Figure 2b) and A2256-A2271 (Figure 2d) pairs of clusters. However, for these two pairs we observe a deficit of tSZ signal in the intercluster region that is induced either by an over-estimation of the tSZ emission of the clusters themselves or by their asymmetric geometry. Therefore, we only considered the A399-A401 and A3391-A3395 pairs of clusters for further analysis because they show the most prominent tSZ signal in the intercluster region.
To check that the excess of tSZ effect observed cannot be explained by contamination by foreground emissions (Galactic), we estimated the dust temperature  using \Planck\ and \IRAS\ data and a simple modified blackbody SED that we fitted to the \Planck+\IRAS\ data. No strong deviations in the temperature  that could compromise the component separation process were observed in the field of view.

\section{X-ray data for the selected  pairs}
\label{xrays}

\begin{figure*}
   \centering
      \includegraphics[width=7cm,height=8cm]{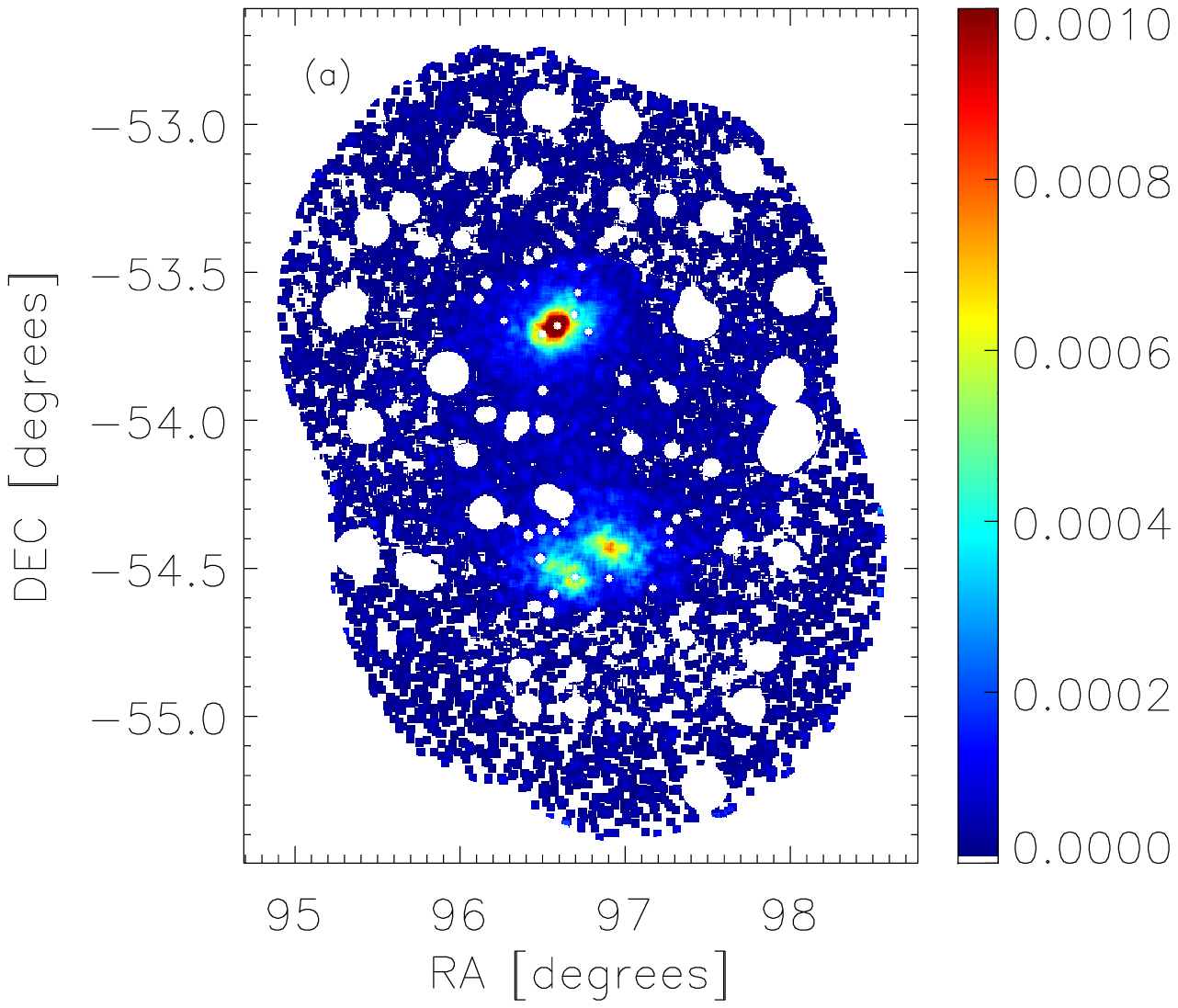}
     \includegraphics[width=7cm,height=7cm]{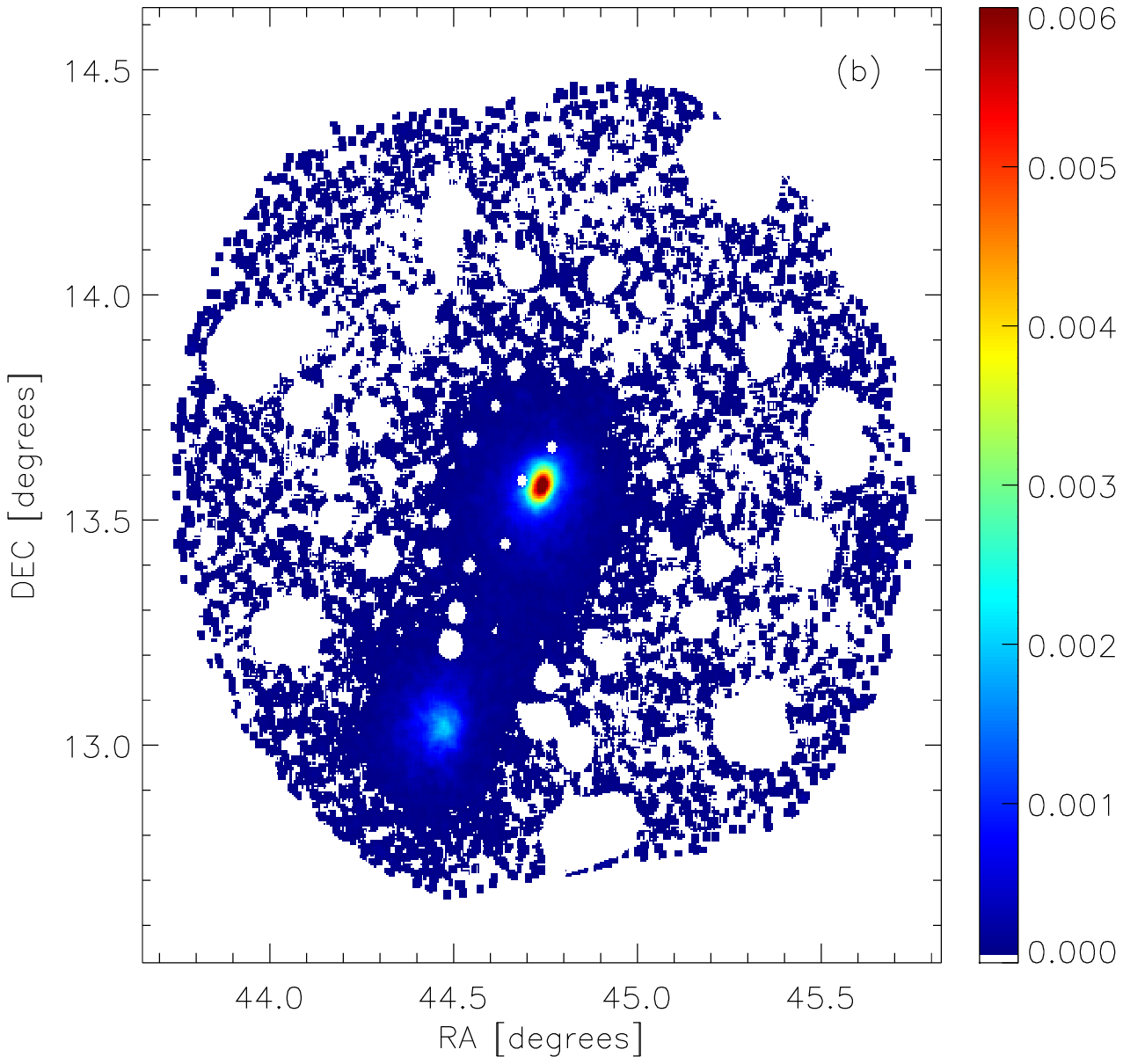}
      \caption{Maps of the X-ray emission (in counts per second) for the cluster pairs a) A3391-A3395 and b)A399-A401.}
         \label{xraysmaps}     
   \end{figure*}

To improve the quality of our analysis we complemented the \Planck\ data with X-ray observations
retrieved from the \ROSAT\ archive. Our choice of using \ROSAT/PSPC
 is motivated by i) its ability to detect the faint surface
brightness emission (i.e the cluster emission at large radii
\cite{vikhlinin99, eckert12} and the intercluster region in
superclusters \citep{kull99}), ii) its large field of view ($\sim$2 \,deg$^2$) and iii) the low
instrumental background. These factors make \ROSAT/PSPC the X-ray instrument that is  most
sensitive to diffuse low surface brightness emission. 
On the negative side, its limited bandpass and poor spectral
resolution is not best-suited for measuring plasma temperatures. \\

We used the \ROSAT\ extended source analysis software (ESAS, \cite{snowden94})
for data reduction, following the procedure in \citet{eckert12}. 
We produced images of the counts in the R37 \ROSAT\ band, i.e. in the energy range
$0.42-2.01$ \,keV, and corrected for vignetting effects through the
exposure map produced with the ESAS software. 
Following \cite{eckert12}, we detected point sources 
up to a constant flux threshold, to resolve
the same fraction of the CXB on the entire field of view (FOV). We estimated 
the sky background components in the fraction of the FOV not contaminated
by cluster emission ($r>1.3r_{200}$) and then subtracted this constant
value from the image. Error bars for the count rate were estimated by
propagating Poissonian errors of the original source and background
count images.\\

The \ROSAT/PSPC point spread function (PSF) strongly depends on 
the position in the FOV, ranging from 15$\arcs$ on axis to 2$\arcm$ in the
outer parts of the FOV. We considered an average PSF representative of our data sets as computed assuming a position 
1$\arcm$ off axis and having 1 \,keV energy. We checked that 
variations within reasonable limits ($E = 0.5-2$\,keV and off axis = 0-10 arcmin) do not introduce significant changes in our results.
To reduce statistical noise we smoothed the X-ray data with a 2$\arcm$ Gaussian kernel.
The scale of this smoothing is significantly smaller than \Planck's angular resolution, so it does not compromise 
the results derived from X-ray data. This also helps to make the uncertainties in the \ROSAT\ PSF a second-order effect. \\

Since we used PSPC archive data and covered regions that might be larger than the FOV of \ROSAT, 
we needed to combine different pointings into a single mosaic. 
The pair A399-A401 (Figure 3a) is contained in a single PSPC pointing, therefore in principle it is not necessary to perfomr a mosaic here. 
However, we still combined the neighbouring pointing 
rp800182n00 and rp80235n00, to increase the statistics. The total combined exposure time is $\sim 14$ \,ks.
For A3391-A3395 (Figure 3b), the larger separation in the sky between the clusters, forced us to combine at least two pointings. 
We combined the two available observations for this pair: rp800079n00 ($2.5$ \,ks centred on A3395) and  rp800080n00 (6 \,ks centred on A3391). 
The final X-ray maps in counts per second are presented in Figure~\ref{xraysmaps}.


\section{Modelling of the tSZ and X-ray emission from the clusters}
\label{Models}

One of the key problems of the analysis presented in this paper is estimating the tSZ and X-ray emission of the clusters themselves because  
this estimate is later used to remove the contribution of the clusters from the intercluster region.
In this section we present a detailed description of modelling the cluster emission for the A399-A401 and A3391-A3395 cluster pairs. \\

\subsection{tSZ  and X-ray emission from clusters}
When CMB photons cross a galaxy cluster, some of the photons will probably interact with the free electrons in the hot plasma through inverse Compton scattering. 
The temperature decrement (increment for frequencies $\nu > 217$ \,GHz) observed in the direction $\theta$ can be described as 
\begin{equation}                             
\Delta T(\theta) = T_o \int n_e(l) T(l) dl,      
\label{eq_DeltaT}                                                               
\end{equation}                                                                  
where $T_o$ contains all relevant constants, including the frequency dependence ($g_x = x(e^x + 1)/(e^x - 1) - 4$ with $x=h\nu/kT$), 
$n_e$ is the electron density and $T$ is the electron temperature. The integral is performed along the line of sight. 
The quantity $P(l) = n_e(l) T(l) $ is often referred to as the pressure produced by the plasma of thermal electrons along the line of sight.\\

We can also consider the thermal  X-ray emission from the hot ionized plasma including bremsstrahlung and line- and recombination emission from metals given by
\begin{equation}                                                                
S_x(\theta) = S_o \frac{\int n_e^2  \lambda(T) dl}{D_l(z)^2}, 
\label{eq_Sx}                                                                   
\end{equation}                                                                                        
where $D_l(z)$ is the luminosity distance. The quantity $S_o$ contains all relevant constants and corrections (including the band- and k-corrections). By combining X-ray and SZE observations 
it is in principle possible to break the degeneracy between different models because of their different dependency on $T$ and specially on $n_e$. 

\subsection{Pressure profile models}
The tSZ and X-ray emission of clusters can be computed from the density and temperature profiles of the thermal electrons in the cluster.
Below we describe the models of the electron pressure profiles for clusters that were used to estimate and subtract the cluster contributions from the pairs of clusters.

\subsection*{$\beta$-model}
The isothermal $\beta$-model is historically the most widely used in the literature in the context of galaxy clusters. Even if cluster observations have shown that this model is over-simplified and unable to describe the details of the intercluster medium (ICM) distribution, it allows us to use a limited number of parameters to produce a useful first approximation of the radial behaviour of the gas. Indeed, we are not interested in the details of the single objects, but in removing the main signal coming from the two clusters in the system to study the SZ excess signal observed between them by \Planck.
In the isothermal $\beta$-model, the electron temperature is assumed to be constant and its density is given by
\begin{equation}
	n_{e}(r)=n_{e0}\left(1+\frac{r^{2}}{r_{c}^{2}}\right)^{\frac{-3\beta}{2}},
\end{equation}
where $n_{e0}$ is the central electron density and $r_c$ is the core radius.

\subsection*{Generalized Navarro Frenk and White pressure profile}
Given the characteristic pressure P$_{500}$, defined as
\begin{equation}
P_{500}=1.65\cdot 10^{-3}h(z)^{8/3}\left[\frac{M_{500}}{3\cdot 10^{14}h_{70}^{-1}}M_{\odot}\right]^{2/3}h_{70}^{2}\,keV\,cm^{3},
\end{equation}
the dimensionless universal (scaled) pressure profile can be written as \citep{Nagai2007,Arnaud2010}
\begin{equation}
\mathds{P}(x)=\frac{p_{o}}{\left(c_{500}x\right)^{\gamma}\left[1+\left(c_{500}x\right)^{\alpha}\right]^{\left(\beta-\gamma\right)/\alpha}},
\label{Pr_Arnaud}
\end{equation}
in which $x=r/R_{500}$, 
$c_{500}=R_{500}/r_{s}$, and $\gamma$, $\alpha$ and $\beta$ are the central ($r << r_{s}$), intermediate ($r \sim r_{s}$) and outer ($r >> r_{s}$) slopes. 
The radial pressure can be the expressed as 
\begin{equation}
P(r)=P_{500} \mathds{P}(x), 
\end{equation}
with $[p_{o}, c_{500},\gamma,\alpha,\beta]=[8.403 h_{70}^{-3/2}, 1.117, 0.3081, 1.0510, 5.4905]$, as derived by \cite{Arnaud2010}. For the sake of simplicity, we redefine the normalization as $P_{0}=  P_{500}*p_{o}$, also in units of  $keV\, cm^{3}$. 
As a result of the anticorrelation of the density and temperature profiles, at cluster cores the pressure exhibits less scatter than the electron density and temperature, and is accordingly better suited to define a universal profile.
However, in the core region, the dispersion about the average profile found by \cite{Arnaud2010} is still significant ($\sim$ 80 \% at 0.03 $R_{500}$), while it becomes less than 30\% beyond 0.2 $R_{500}$.

 \begin{figure*}
  \centering
  \includegraphics[width=4cm,height=8cm,angle=90]{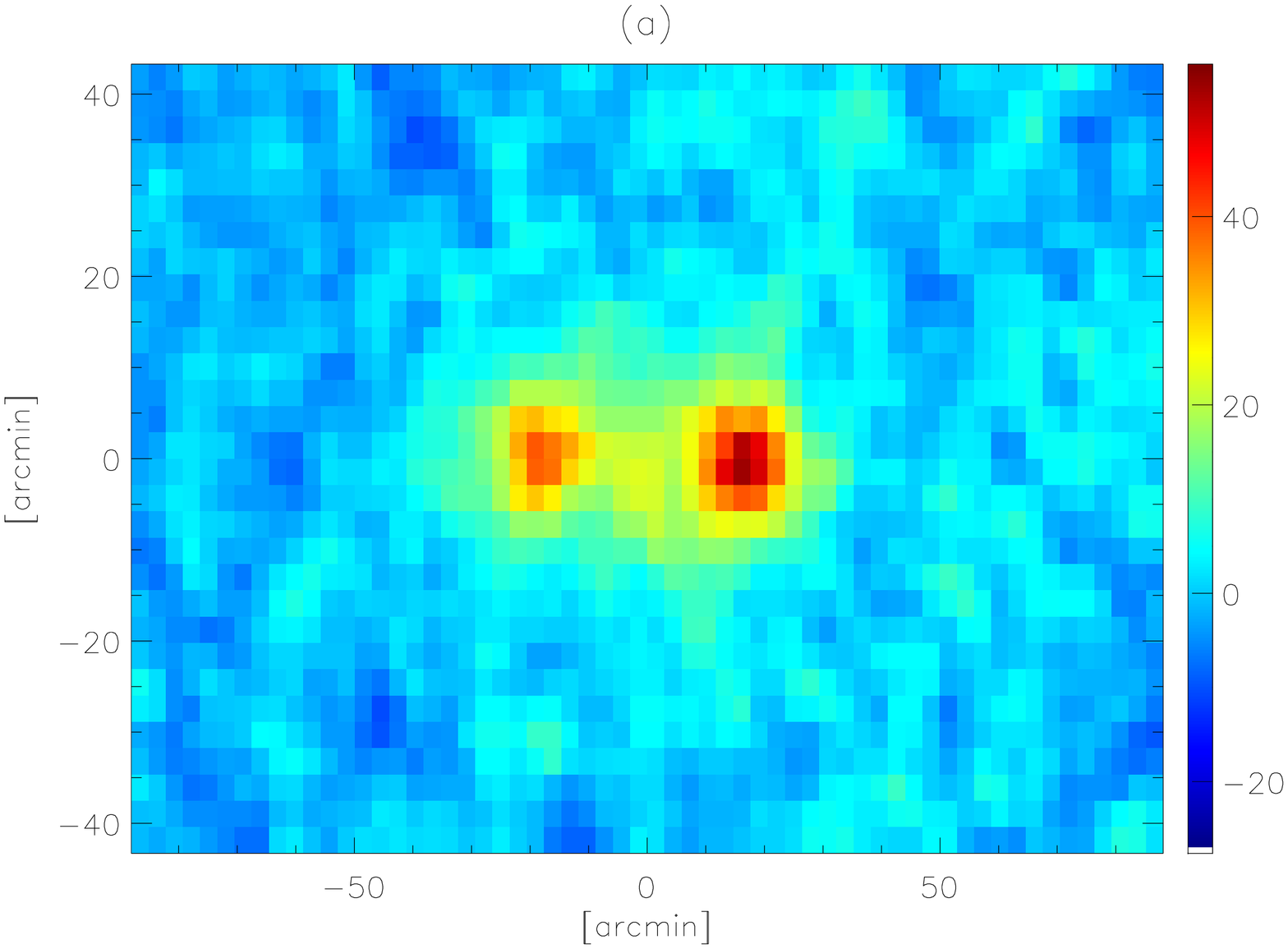}
   \includegraphics[width=4cm,height=8cm,angle=90]{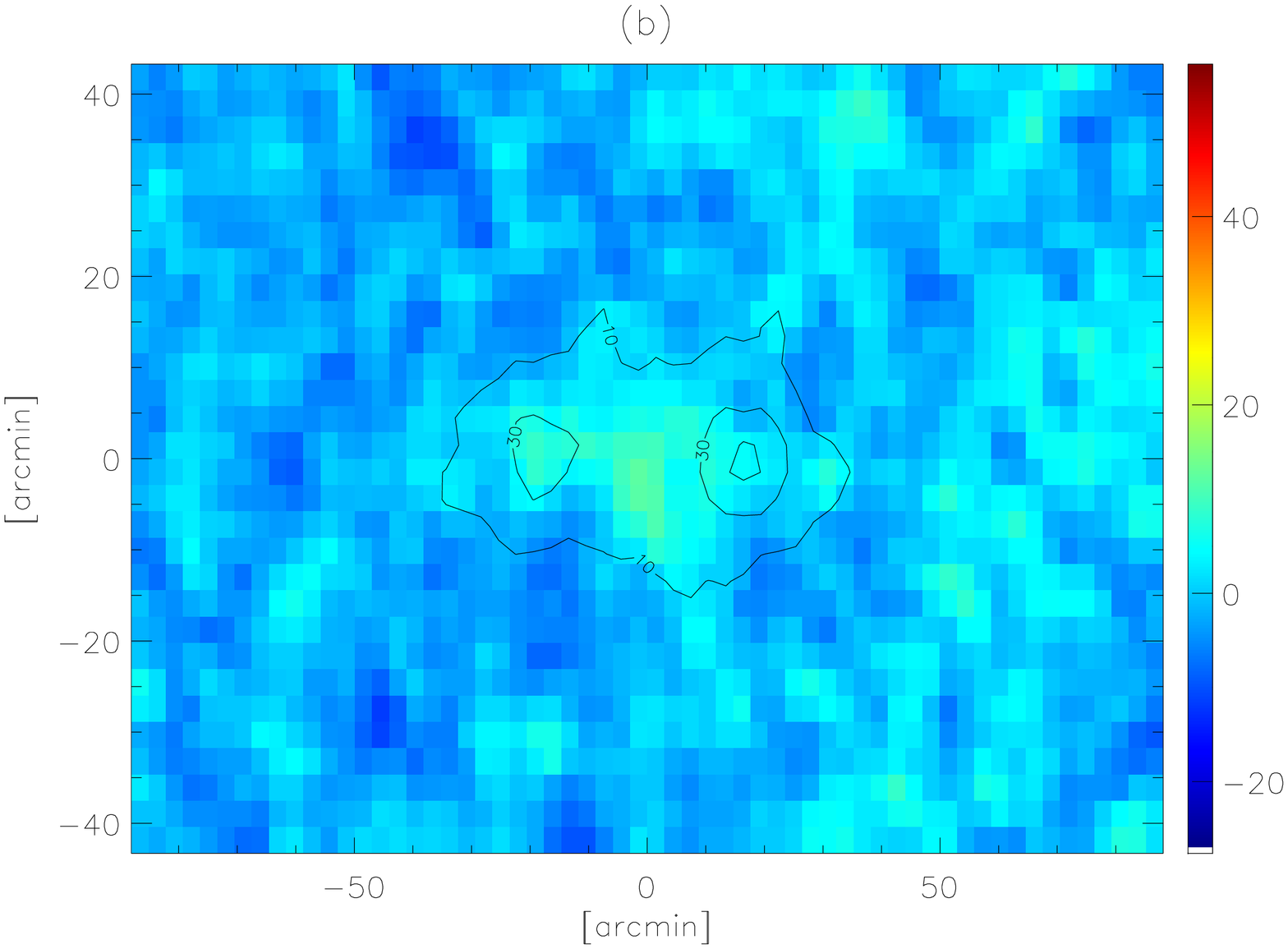}
   \includegraphics[width=4cm,height=8cm,angle=90]{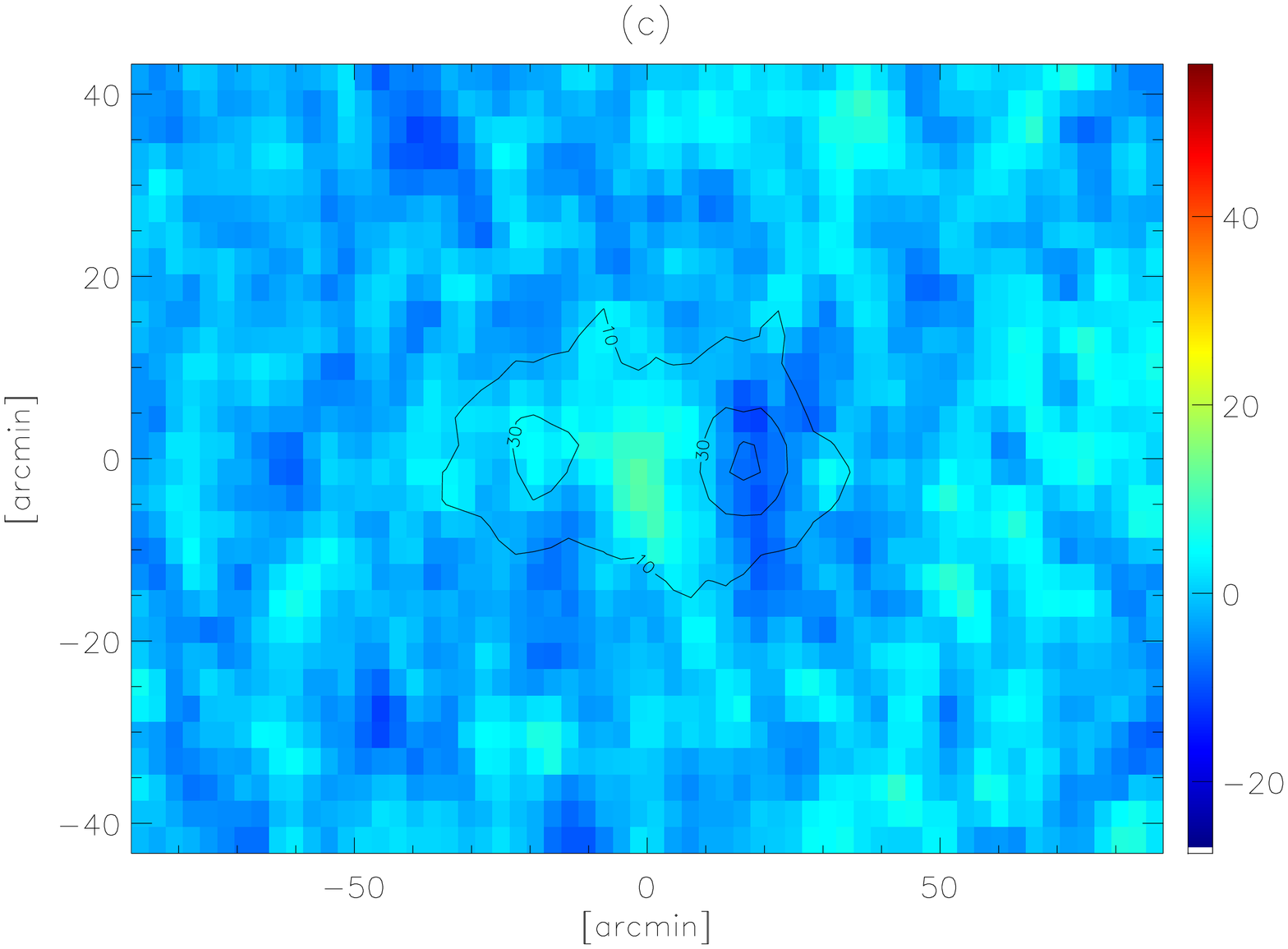}
   \includegraphics[width=4cm,height=8cm,angle=90]{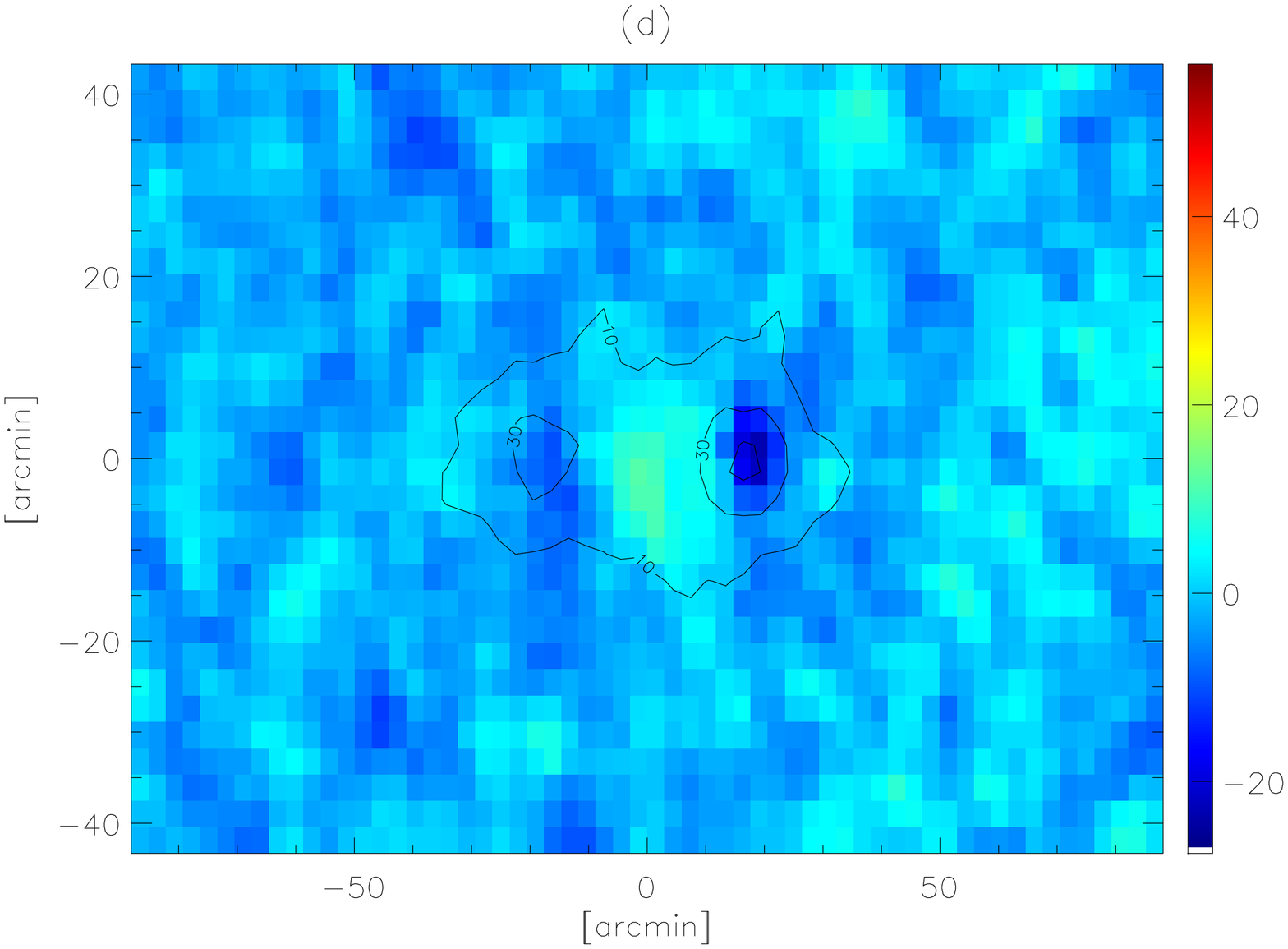}
   
      \caption{ a) \Planck\ tSZ Compton parameter map ($y  \times 10^6$); b) residuals after subtracting of the GNFW2 model; 
                c) residuals after subtracting of the GNFW1 model; and d) residuals after subtracting of the $\beta$ model
                of the clusters as described in the text.}
         \label{BestFitClusterMapsA399A401}     
  \end{figure*}

 \begin{figure*}
  \centering
  \includegraphics[width=4cm,height=8cm,angle=90]{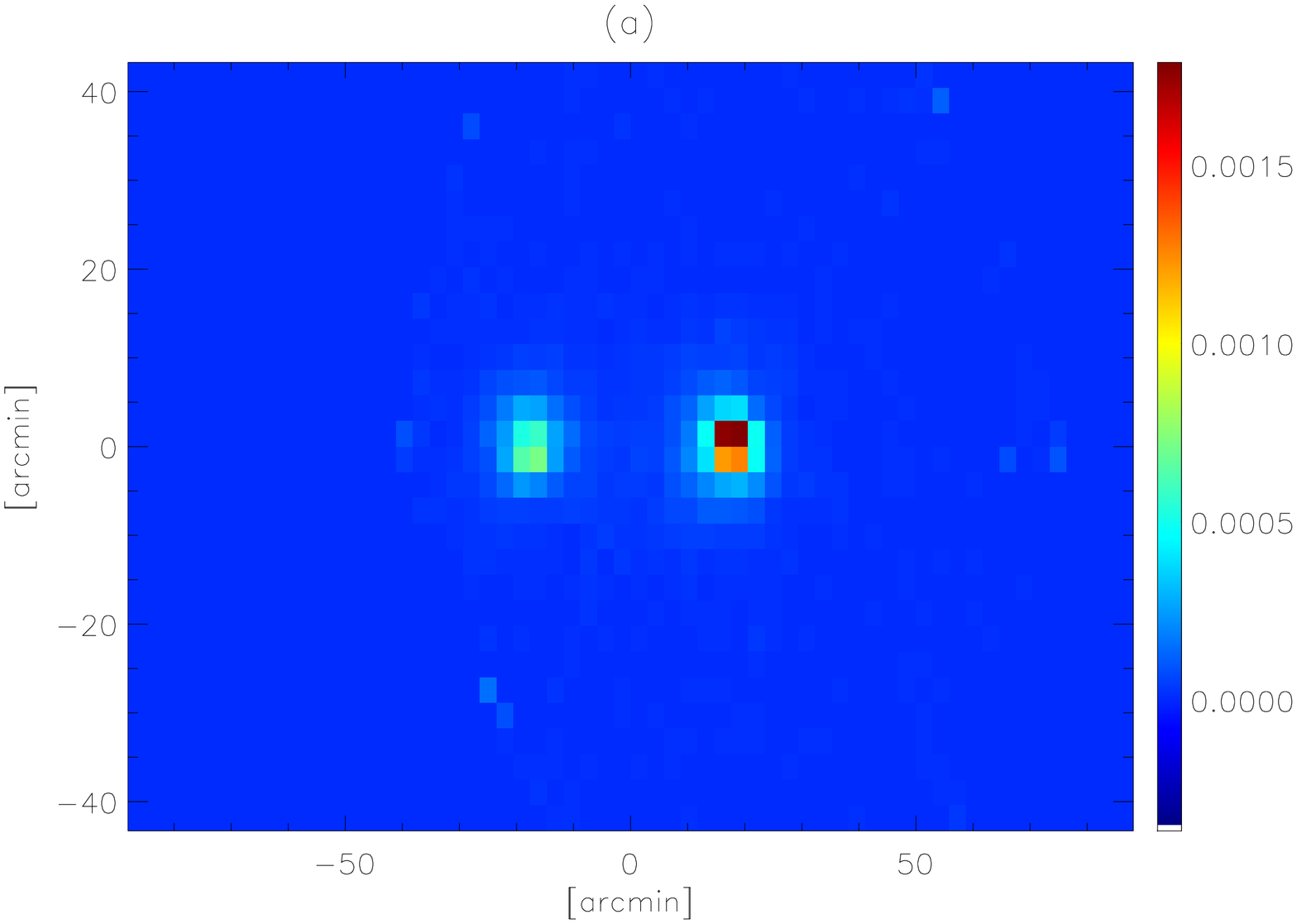}
   \includegraphics[width=4cm,height=8cm,angle=90]{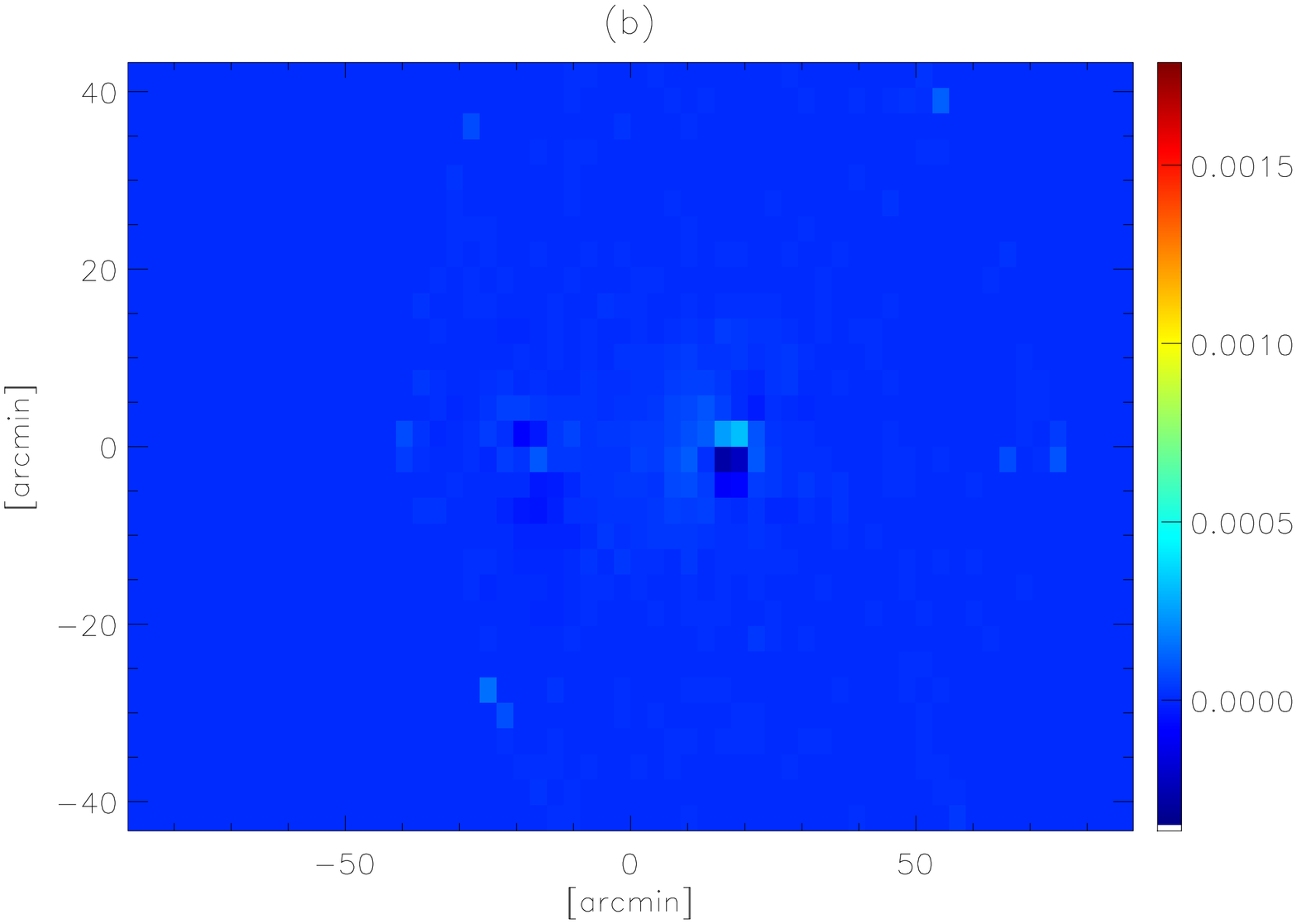}
  \includegraphics[width=4cm,height=8cm,angle=90]{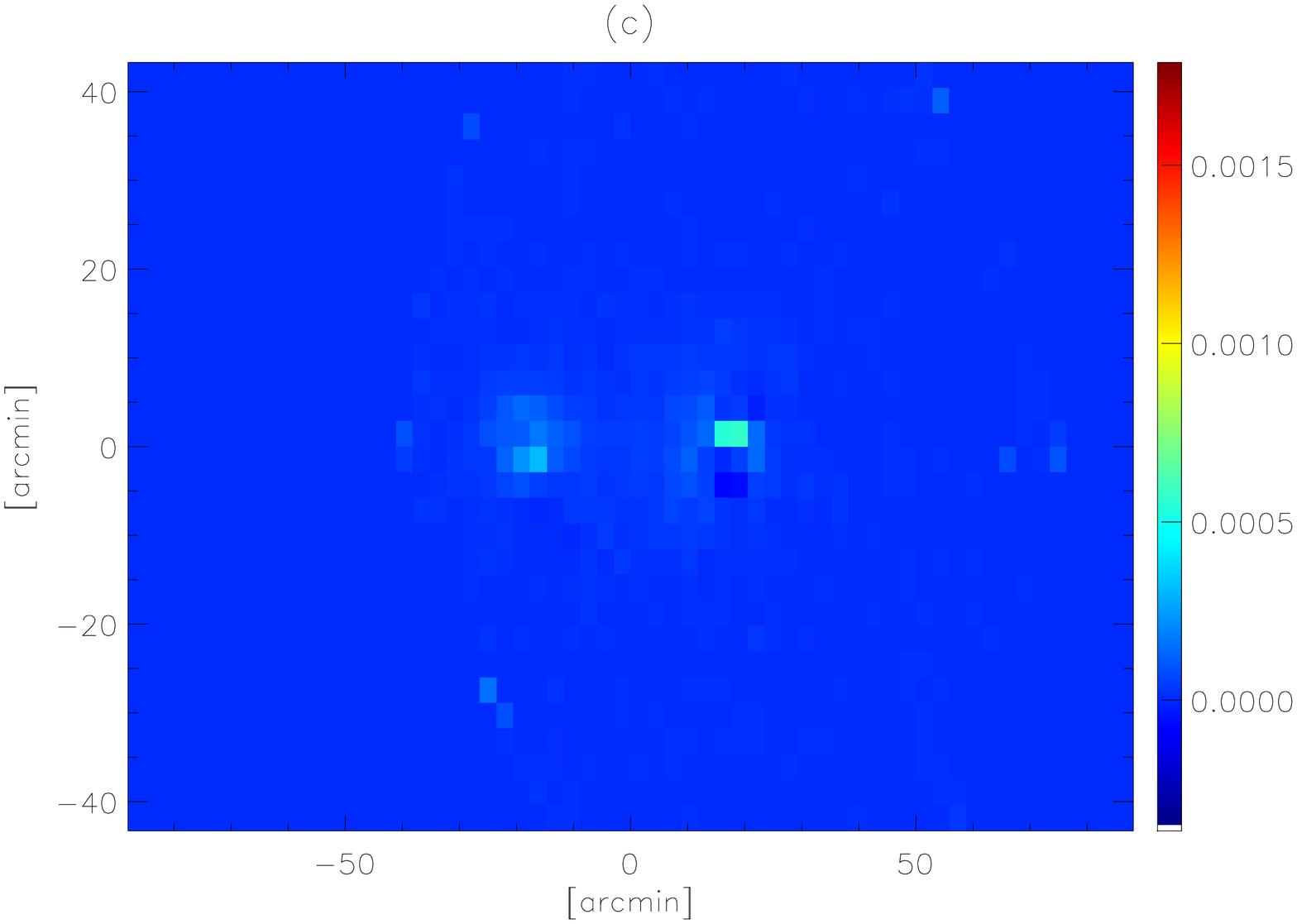}
   \includegraphics[width=4cm,height=8cm,angle=90]{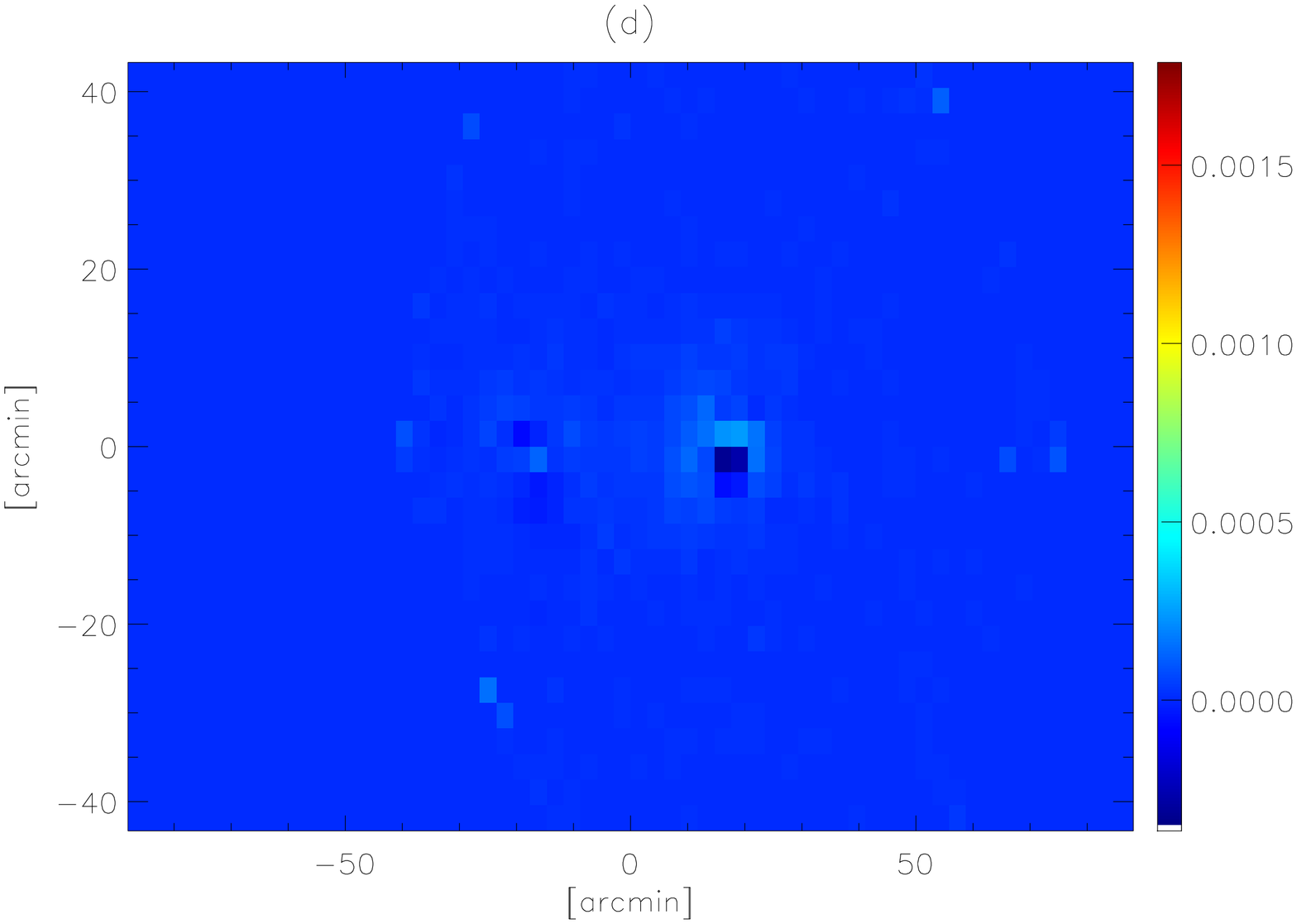}
    
      \caption{X-ray data and residuals of the best-fit cluster model for the A399 and A401 pair of clusters. 
      	a) \ROSAT\ map; b) residuals for the GNFW2 model; c) residuals for the GNFW1 model; 
        and d) residuals for the $\beta$ model.} 
         \label{BestFitClusterXrayMapsA399A401}     
  \end{figure*}

 \begin{figure*}
  \centering
  \includegraphics[width=5cm,angle=90]{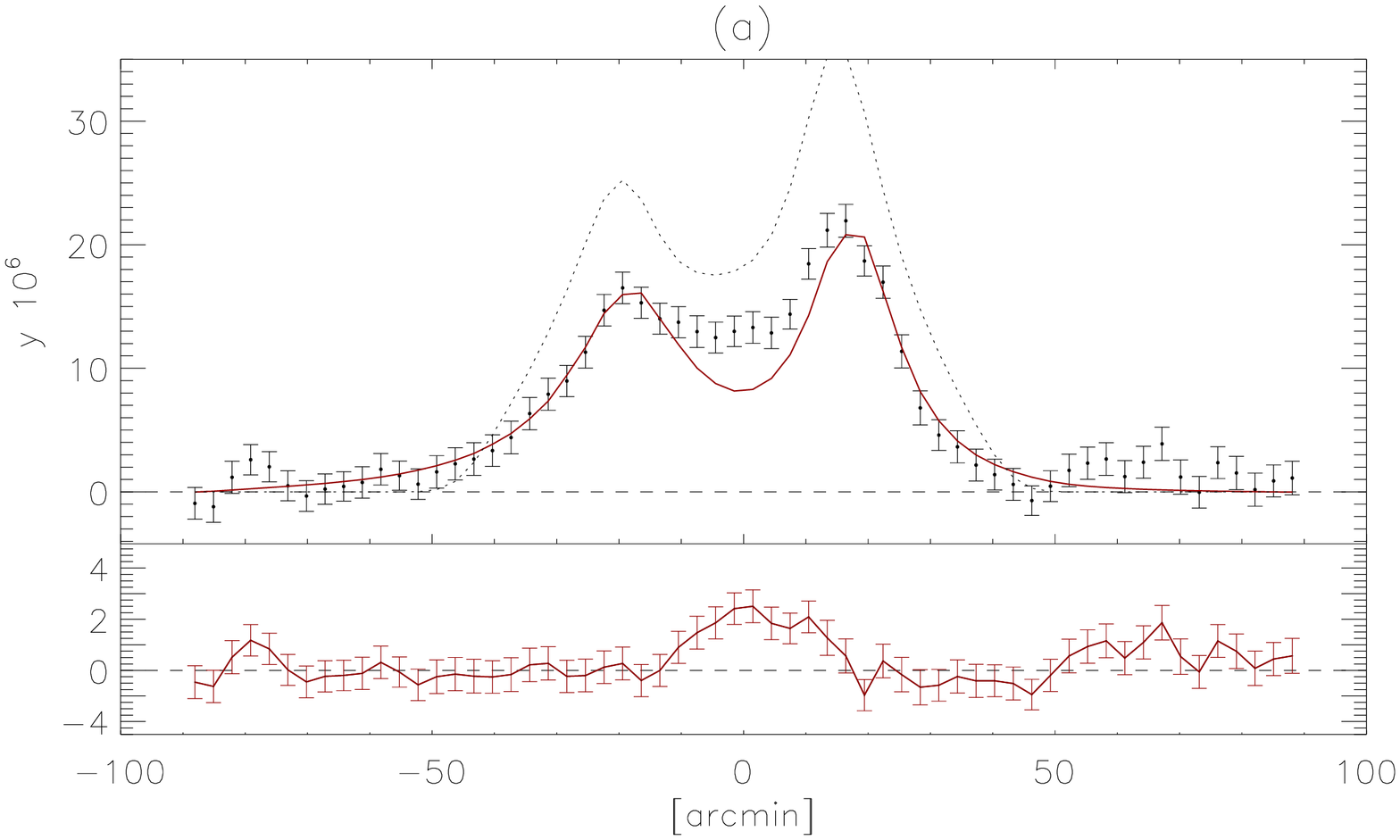}
  \includegraphics[width=5cm,angle=90]{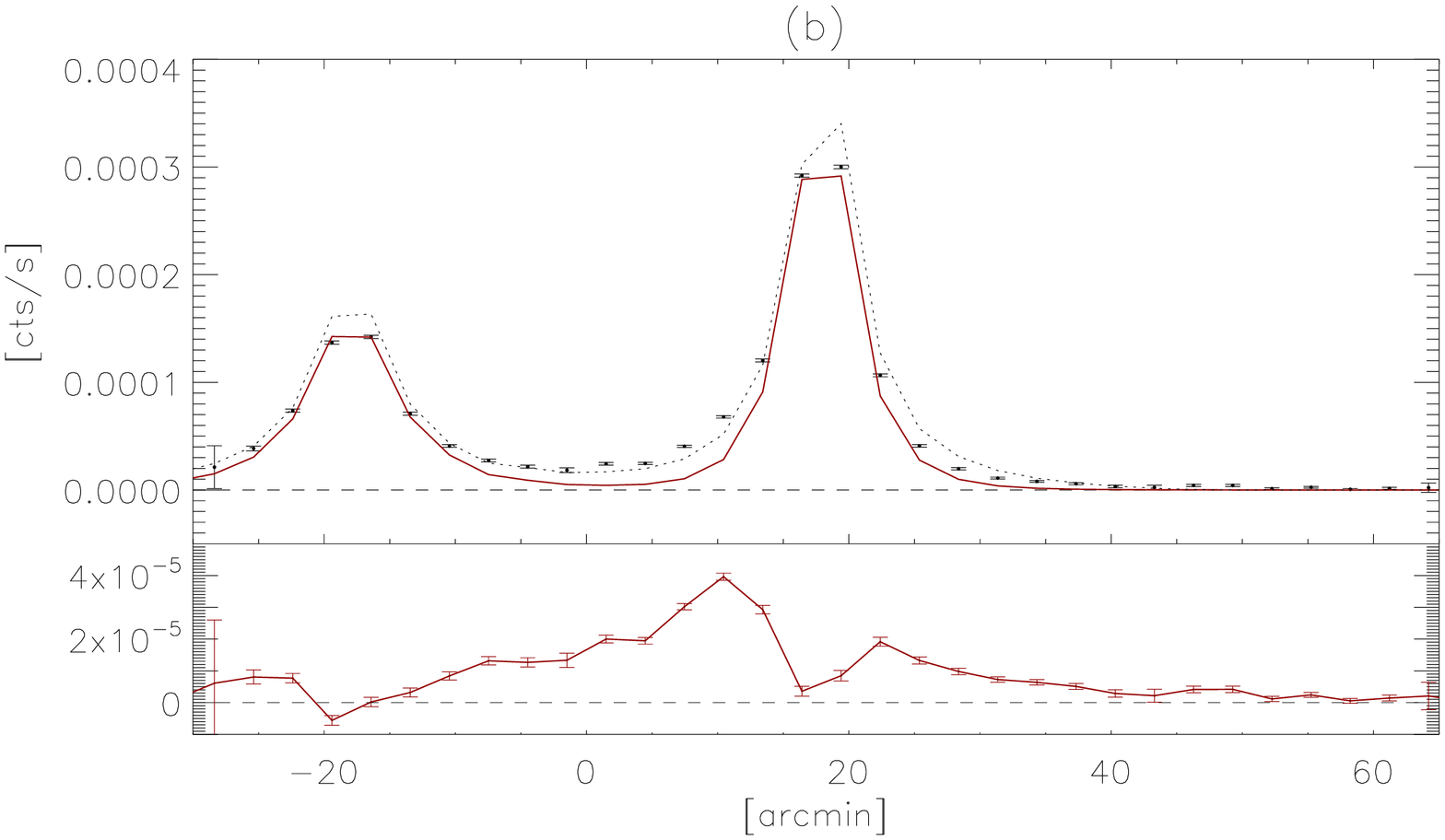}
   
      \caption{1D a) tSZ and b) X-ray profiles and residuals for the best-fit models of the A399 and A401 clusters. The red curve corresponds to the
      GNFW2 model. The dotted lines are the model from \cite{Sakelliou2004}.
         \label{BestFitClusterProfilesA399A401}     }
  \end{figure*}

\subsection{Fitting the tSZ and X-ray emissions from the clusters}

We used the $\beta$ and generalized Navarro Frenk and White (or GNFW hereafter) pressure profile
models to fit the tSZ and X-ray emissions of the clusters in the A399-A401 and A3391-A3395 pairs.
For the $\beta$-model we varied the central density, $n_o$, the slope $\beta$, and the core radius $r_c$ of each cluster.
For the GNFW model we considered two different sets of models: GNFW1 and GNFW2. For the first one, we fixed $\beta=3.5$ and fitted the other parameters.
For the second one,  $c_{500}$, $\gamma$ and $\alpha$ were fixed to the \cite{Arnaud2010} values, leaving $P_{0}$ and $\beta$ free to vary.  \\

We fitted each cluster individually excluding the intercluster region
from the fit. For the tSZ emission we integrated the pressure profile along the line of sight up to $5 \times R_{500}$ (unless stated otherwise) following equation~\ref{eq_DeltaT}.
The resulting map was convolved with a Gaussian beam of 7.18$\arcm$ to match the resolution of the MILCA SZ map.
For the X-ray emission we considered an isothermal scenario and computed the electron density from the pressure profile.
We then used equation \ref{eq_Sx} to compute the surface brightness of the clusters.
An X-ray map in counts per second was obtained from the latter using the MEKAL model \citep{mewe85,mewe86,kaas92,lied95} for cluster emission and the WABS
model \citep{morr83} for absorption of the neutral hydrogen along the line-of-sight. The MEKAL model is a function of the square of the electron density, the temperature, 
and the redshift of the cluster. The WABS model uses column density maps~\citep{Dickey1990,Kalberla2005}.
We convolved the final X-ray map with a Gaussian beam of 2$\arcm$ to match the resolution of the  2$\arcm$ resolution (degraded) \ROSAT\ map. \\

We performed different fits including X-ray data only, SZ data only, and both X-ray and SZ data. 
The fits were performed directly on the 2-D maps presented above and correlated errors were accounted for
in the likelihood analysis. For the joint tSZ and X-ray fits the likelihood functions were normalized by their volume and then 
multiplied to obtain the best-fit parameters and errors. \\

\subsection*{Results for the A399-A401 pair}

\begin{table*}[tmb]                 
\begingroup
\newdimen\tblskip \tblskip=5pt
\caption{Best-fit parameters for the pressure profile $\beta$-model. 
For this analysis KT and $R_{max}$ were fixed to the values shown.}
\label{tbl_beta}    
\nointerlineskip
\vskip -3mm
\footnotesize
\setbox\tablebox=\vbox{
   \newdimen\digitwidth 
   \setbox0=\hbox{\rm 0} 
   \digitwidth=\wd0 
   \catcode`*=\active 
   \def*{\kern\digitwidth}
   \newdimen\signwidth 
   \setbox0=\hbox{+} 
   \signwidth=\wd0 
   \catcode`!=\active 
   \def!{\kern\signwidth}
%
{\tabskip=3em                                 
\halign{\hfil#\hfil&                       
        \hfil#\hfil&
        \hfil#\hfil&
        \hfil#\hfil&
        \hfil#\hfil&
        \hfil#\hfil \cr
\noalign{\vskip 3pt\hrule\vskip 1.5pt\hrule\vskip 5pt}
Cluster & $\beta$ & $r_{c}$ (kpc)  & n$_0$ (cm$^-3$)           & KT (keV)  &   $R_{max}$  (kpc)\cr
\noalign{\vskip 3pt\hrule\vskip 5pt}
A399    &  0.928    &  432.8              & $2.37 \times 10^{-3}$  & 7.23           &   2160 \cr
A401    &  0.928    &  298.4              & $5.50 \times 10^{-3}$  & 8.70           &   2340 \cr
\noalign{\vskip 3pt\hrule\vskip 5pt}
A3391        &  0.620    &  213.0              & $2.61 \times 10^{-3}$  & 6.00     &   1500\cr
A3395SW     &  0.673    &  328.0              & $1.17 \times 10^{-3}$  & 4.80     &   1400\cr
A3395E     &  0.726    &  356.6              & $1.17 \times 10^{-3}$  & 5.00     &   1200\cr
\noalign{\vskip 5pt\hrule\vskip 3pt}}}  
}                                   
\endPlancktablewide                 
\endgroup
\end{table*}                        

\begin{table*}[tmb]                 
\begingroup
\newdimen\tblskip \tblskip=5pt
\caption{Best-fit parameters for the GNFW1 pressure profile model. For this analysis
 $\beta$  and KT were fixed to the values shown.}
\label{tbl_gnfw1}    
\nointerlineskip
\vskip -3mm
\footnotesize
\setbox\tablebox=\vbox{
   \newdimen\digitwidth 
   \setbox0=\hbox{\rm 0} 
   \digitwidth=\wd0 
   \catcode`*=\active 
   \def*{\kern\digitwidth}
   \newdimen\signwidth 
   \setbox0=\hbox{+} 
   \signwidth=\wd0 
   \catcode`!=\active 
   \def!{\kern\signwidth}
%
{\tabskip=2em                                 
\halign{\hfil#\hfil&                       
        \hfil#\hfil&
        \hfil#\hfil&
        \hfil#\hfil&
        \hfil#\hfil&
        \hfil#\hfil&
        \hfil#\hfil&
        \hfil#\hfil \cr
\noalign{\vskip 3pt\hrule\vskip 1.5pt\hrule\vskip 5pt}
Cluster & c$_{500}$  & $\gamma$  &  $\alpha$   & $\beta$    &   r$_{500}$ (Mpc)   & P$_{0}$ (keV / cm$^3$)   &  kT (keV) \cr
\noalign{\vskip 3pt\hrule\vskip 5pt}
A399       &  1.80    &  0.000   &  1.25    &  3.5   &  1.11   &  2.04  $\times 10^{-2}$  &  7.23 \cr
A401       &  2.58    &  0.016   &  1.59    &  3.5   &  1.24   &  3.76  $\times 10^{-2}$  &  8.47 \cr
\noalign{\vskip 3pt\hrule\vskip 5pt}
A3391      &  0.69    &  0.033   &   0.72  &   3.5   &  0.90   &  3.08  $\times 10^{-2}$  &  6.00 \cr
A3395SW    &  2.65    &  0.066   &   2.00  &   3.5   &  1.10   &  0.70  $\times 10^{-2}$  &  4.80 \cr
A3395E     &  0.3     &  0.000   &   0.58  &   3.5   &  0.90   &  1.89  $\times 10^{-2}$  &  5.00 \cr
\noalign{\vskip 5pt\hrule\vskip 3pt}}}  
}                                   
\endPlancktablewide                 
\endgroup
\end{table*}                        

%
\begin{table*}[tmb]                 
\begingroup
\newdimen\tblskip \tblskip=5pt
\caption{Best-fit parameters for the GNFW2 pressure profile model. For this model
c$_{500}$, $\gamma$,  $\alpha$ and  r$_{500}$ were fixed to the values shown.}
\label{tbl_gnfw2}    
\nointerlineskip
\vskip -3mm
\footnotesize
\setbox\tablebox=\vbox{
   \newdimen\digitwidth 
   \setbox0=\hbox{\rm 0} 
   \digitwidth=\wd0 
   \catcode`*=\active 
   \def*{\kern\digitwidth}
   \newdimen\signwidth 
   \setbox0=\hbox{+} 
   \signwidth=\wd0 
   \catcode`!=\active 
   \def!{\kern\signwidth}
%
{\tabskip=2em                                 
\halign{\hfil#\hfil&                       
        \hfil#\hfil&
        \hfil#\hfil&
        \hfil#\hfil&
        \hfil#\hfil&
        \hfil#\hfil&
        \hfil#\hfil&
        \hfil#\hfil \cr
\noalign{\vskip 3pt\hrule\vskip 1.5pt\hrule\vskip 5pt}
Cluster & c$_{500}$  & $\gamma$  &  $\alpha$   & $\beta$    &   r$_{500}$ (Mpc)   & P$_{0}$ (keV cm$^3$)   &  kT (keV)\cr   
\noalign{\vskip 3pt\hrule\vskip 5pt}
A399    &  1.18   &  0.308      &  1.05   &  3.5  &  1.11  & 0.97 $\times 10^{-2}$ &  5.0 \cr
A401    &  1.18   &  0.308      &  1.05   &  5.0  &  1.24  & 2.40 $\times 10^{-2}$ &  6.5 \cr
\noalign{\vskip 5pt\hrule\vskip 3pt}
A3391        & 1.18  &  0.308    &  1.05   &  3.5     & 0.89   &  0.75  $\times 10^{-2}$  &  6.0   \cr
A3395SW     &  1.18 &  0.308    &   1.05  &    3.5   &  0.93   &  0.40      $\times 10^{-2}$   &   4.8  \cr
A3395E     &  1.18   &  0.308  &   1.05  &   3.5    &   0.93  &  0.40   $\times 10^{-2}$    &    5.0  \cr
\noalign{\vskip 5pt\hrule\vskip 3pt}}}  
}                                   
\endPlancktablewide                 
\endgroup
\end{table*}                        

We converted the density profile into \ROSAT\ PSPC count rates, using an
absorbed MEKAL model within XSPEC \citep{Arnaud96}. The conversion rate
was computed using the temperatures in \cite{Sakelliou2004}, a
metal abundance $Z=0.2$ solar, the Galactic column density in the
direction of the pair $nH=1.09\times10^{21}\,\rm{cm}^{-2}$ \citep{Dickey1990}
and the redshifts $z=0.0737$ for A401 and $z=0.0724$ for
A399. We fixed the truncation radius in the 3-D integrals to the
$r_{200}$ value in \cite{Sakelliou2004} (this choice is not mandatory provided the radius is reasonably large).

Before model fitting, we used three different methods to estimate the background or zero level of the tSZ Compton parameter maps. The results are consistent between the three methods. The first method computes the mean signal in an area around the cluster pairs that excludes the pairs. This region does not show any other source that needs to be masked out. The second method computes the mean signal of the averaged profiles of the clusters beyond R200 and in the directions opposite to the other cluster. The third method computes a histogram of the background region defined in the first method and finds the median of the distribution. We adopted method one for the final results.  \\

Tables~\ref{tbl_beta}, \ref{tbl_gnfw1}, and \ref{tbl_gnfw2} present the best-fit parameters of the A399 and A401 clusters for the $\beta$, GNFW1, and GNFW2 pressure profile models.
The residual tSZ maps for these best-fit cluster models are shown in Figure~\ref{BestFitClusterMapsA399A401}. We note that for the GNFW2 pressure profile model (Figure 4d) the derived temperatures are lower than those derived from X-ray-only data and in particular with the results from \XMM\ by \cite{Sakelliou2004}. For the other two models the temperatures were fixed to the \cite{Sakelliou2004} value.
We show the \Planck\ tSZ map (Figure 4a) and the residuals after subtracting the best-fit $\beta$ model (Figure 4b) and the best-fit GNFW model for the two sets described above (Figure 4c and 4d). 
In the residuals we clearly observe an excess of tSZ emission with respect to the background. This can also be observed in Figure~\ref{BestFitClusterProfilesA399A401}a) where we present the 1D tSZ longitudinal profile and residuals for the GNFW2 model described before and for the \cite{Sakelliou2004} best-fit model . \\

In Figure~\ref{BestFitClusterXrayMapsA399A401} we show the \ROSAT\ X-ray maps for the A399 and A401 clusters and the residuals after subtracting the $\beta$, GNFW1 and GNFW2 models.
We observe that the models slightly under-estimate the signal at the centre of the clusters. This is also seen in Figure~\ref{BestFitClusterProfilesA399A401}b) where we present the 1D x-ray longitudinal profile and residuals for the three pressure profile models. A similar behaviour was observed when fitting \XMM\ data in \cite{Sakelliou2004}. Quoting \cite{Sakelliou2004}, {\it ``neither of the two central galaxies is known to host an active nucleus, whose presence could be invoked to explain the requirement for an extra, but small, central component''.}
For illustration we also plot  (dotted lines) the best-fit model of the clusters found by~\cite{Sakelliou2004}.
This model clearly over-predicts the tSZ cluster emission. However, this discrepancy is not surprising, given the different dependencies of the SZ and X-ray signals on the ICM density and temperature. Indeed, as pointed out by \cite{Hallman07}, the inadequacy of the isothermal assumptions affects the parameter determination performed on SZ and X-ray data in a different way, therefore X and SZ profiles cannot be represented properly using the same parameter values for an isothermal $\beta$-model. Isothermal $\beta$-models derived from X-ray observations, in general, overpredict the SZ effect and steeper profiles are needed to simultaneously fit SZ and XR data (e.g. \citep{diego2010}). Other factors that affect the SZ and XR  in different ways are for instance clumpiness and triaxiality that can be invoked to explain part of the discrepancy, as we discuss below. 
The different sensitivity to the details of temperature and gas density distribution is still relevant even when comparing integrated quantities such as $Y$, the integrated Comptonization parameter, and the proxy 
$Y_{X}=k_{B}TM_{gas}$, \citep{Kravtsov06}, whose ratio has been found to be lower than 1 by several different authors (\cite{Arnaud2010,2011A&A...536A..10P, Andersson10, Rozo12}). This agrees with our findings and confirms the importance of a joint X-ray/SZ analysis, which is able to break the degeneracy between models.  
Nevertheless, despite the uncertainties in the central part of the clusters, we observe an excess of X-ray and tSZ signal in the intercluster region. This excess is discussed in section~\ref{analysis}.

 \begin{figure*}
  \centering
  \includegraphics[width=5cm,height=8cm,angle=90]{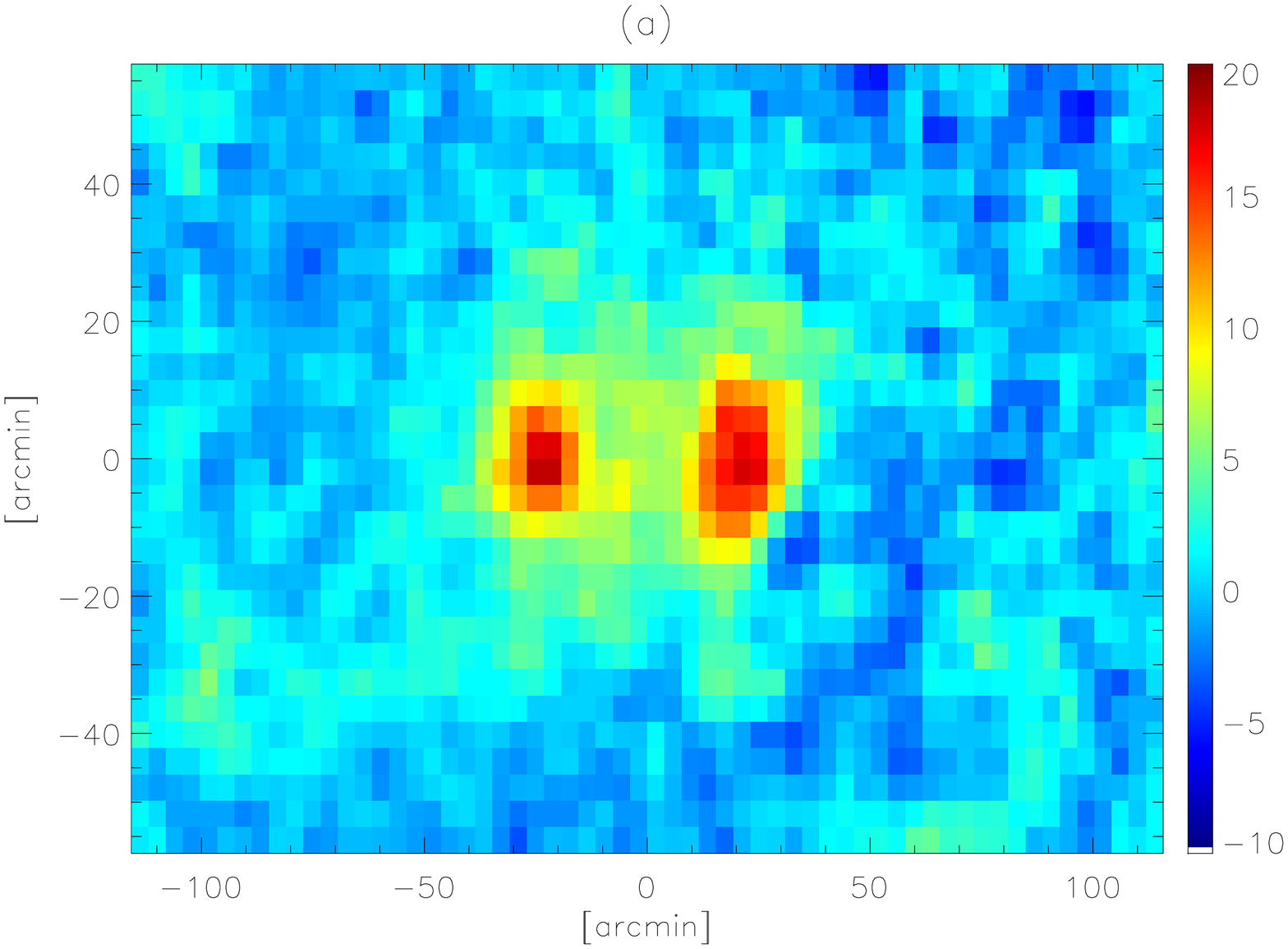}
   \includegraphics[width=5cm,height=8cm,angle=90]{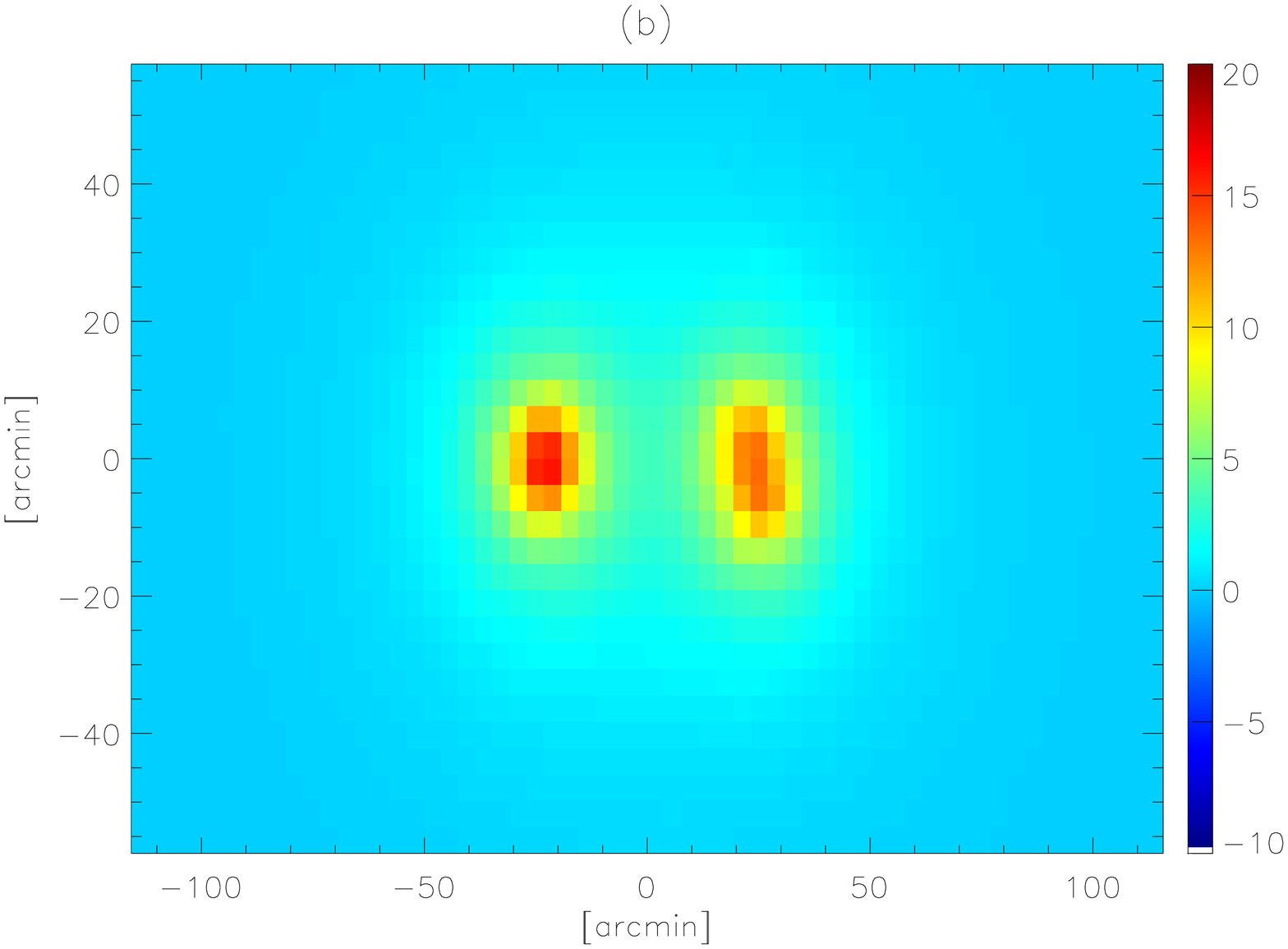}
   \includegraphics[width=5cm,height=8cm,angle=90]{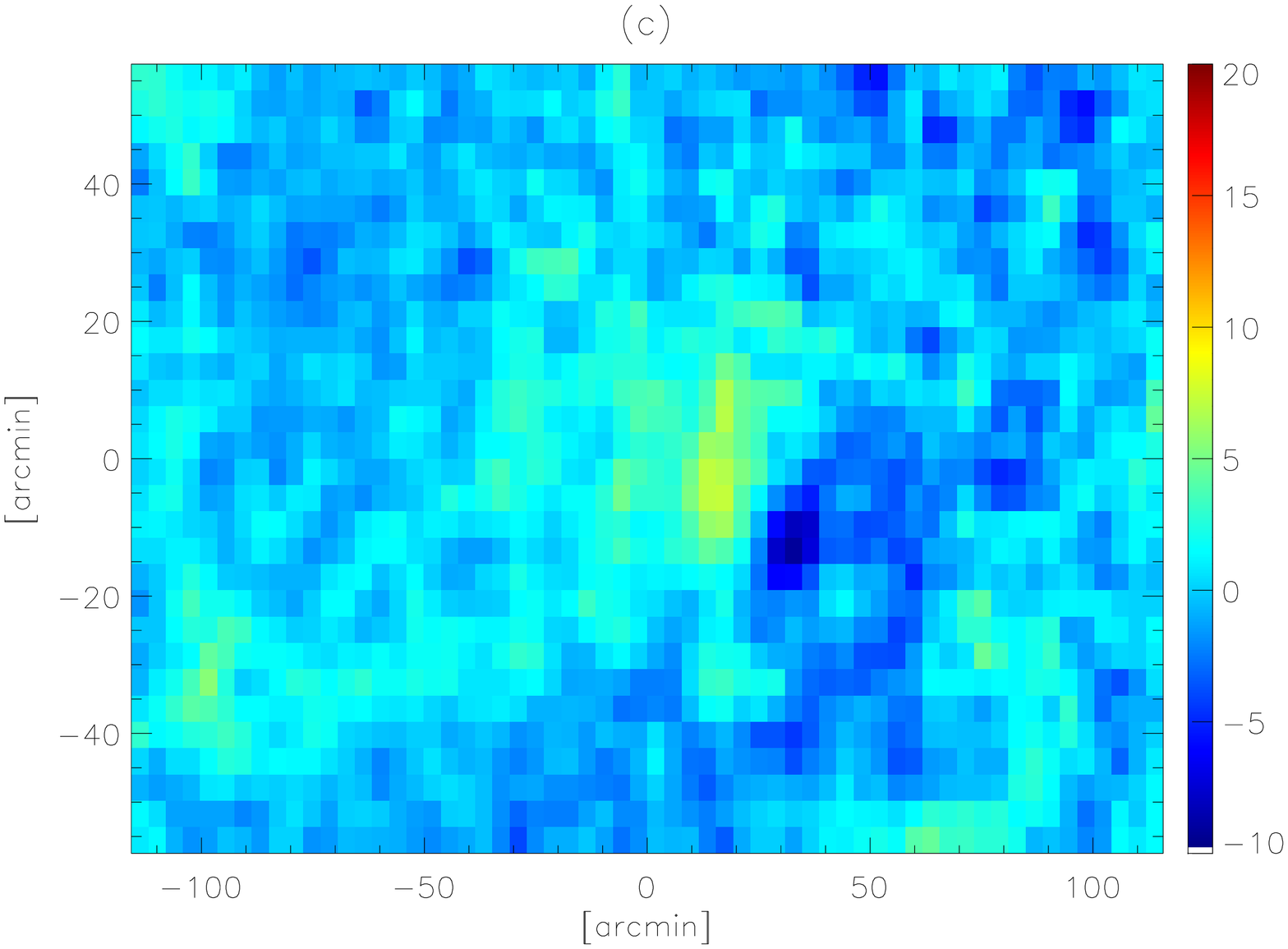}
   
         \caption{ a) tSZ Compton parameter map ($y  \times 10^6$), 
                   b) best-fit model of the clusters, and c) residuals after subtracting the best fit model for the A3391 and A3395 pair of clusters.  }
         \label{BestFitClusterMapsA3391A3395}     
  \end{figure*}

\subsection*{Results for the A3391-A3395 pair}

Modelling the A3391 and A3395 pair is a challenging task. A3395 is a multiple system formed of at least three identified clusters (see \cite{Tittley2001}).
For the purpose of this paper we considered the two clusters most important in X-rays (SW and E), which can be clearly identified in Figure~\ref{xraysmaps}. 
The quality of the X-ray data is significantly poorer than in the case of A399-A401 (less than half the amount of exposure time), making it difficult to distinguish between background and diffuse X-rays in the intercluster region. Moreover, the clusters that form the system A3395 are clearly elongated (this is even clearer when studying \XMM\ and \CHANDRA\  images of the cluster, see for example
\cite{Tittley2001}), which adds another complication  to the model. For the SZ part of the data, even though there seems to be a signal in the intercluster region, 
the modelling of the SZ signal from A3395 is affected because \Planck\ cannot 
resolve the individual subsystems in A3395. Despite all these complications, we are able to perform a basic fit to the SZ and X-ray data.
The best-fit parameters for the three pressure profile models are presented in the last three rows of tables~\ref{tbl_beta}, \ref{tbl_gnfw1}, and \ref{tbl_gnfw2}.
We show in Figure~\ref{BestFitClusterMapsA3391A3395} a) the tSZ Planck map for the A3391-A3395 pair, b) the best-fit GNFW2 model, and c) the residuals after subtracting this
model. We clearly observe in the residuals a significant excess of signal in the region near the A3395 system. 
Indeed, the integrated Compton parameter in the intercluster region is $Y = (4.5 \pm 0.7) \ 10^{-3}$ (arcminutes)$^2$ --of the same order of magnitude as that for the A399-A401 pair  (see below).
However, this residual is suspiciously close to the NW component of the A3395 cluster
that was discussed in~\cite{Tittley2001} (and not included in our model), indicating that a far more complex model is necessary to subtract the cluster components. 
We can also notice in the figure the contribution from a radio point-source that appears as a decrement in the reconstructed Compton parameter map. The area contaminated by this
point source was excluded from the analysis. Due to the uncertainties in the model and in the foreground subtraction (radio point source contribution), the analysis of the intercluster region gives unreliable constraints on the density and temperature of this region and is not considered in the following analysis.

 \section{Analysis of the intercluster residuals}
 \label{analysis}
 
 \subsection{Models}
 
 To model the intercluster region of the pair of clusters we considered either an extra background cluster or a filament-like structure described by a parametric model.
 
 \subsection*{Extra cluster}
 
In this case we assumed a background cluster in the intercluster region of the pair. To simplify the modelling, we
considered that the cluster properties are fully defined by its redshift and $M_{500}$ mass, which are the free parameters of the model.
From  M$_{500} $ we computed $R_{500}$ and used scaling relations to compute the temperature.
We finally assumed that the pressure profile of the cluster is well described by a universal profile and used the parameters obtained
by \cite{Arnaud2010} for cool cores discussed in the previous section.

\subsection*{Filament-like structure}
In this case we assumed that the intercluster region can be described by a tube-like filament orientated perpendicularly to the line of sight.
We considered an isothermal filament of temperature T.  For the electron density we considered two cases, one in which the density remains constant inside the tube and another in which the density falls from a maximum to zero in the border of the tube. The main physical parameters are the normalization density, the temperature of the filament, and its radius. 
For the latter the electron density profile is defined in the radial direction in cylindrical coordinates as follows
\begin{equation}
n_e(r) =   \frac{n_e(0)}{\left(1+(\frac{r}{r_c})^2 \right)^{\frac{3}{2} \beta}},
\end{equation}
with $\beta = 2/3$ and $r_c=10$$\arcm$. The free parameters of the model are the temperature, $T$, and the central electron density, $n_e(0)$.

\subsection{A399-A401}
\begin{figure}
   \centering
   \includegraphics[width=9cm]{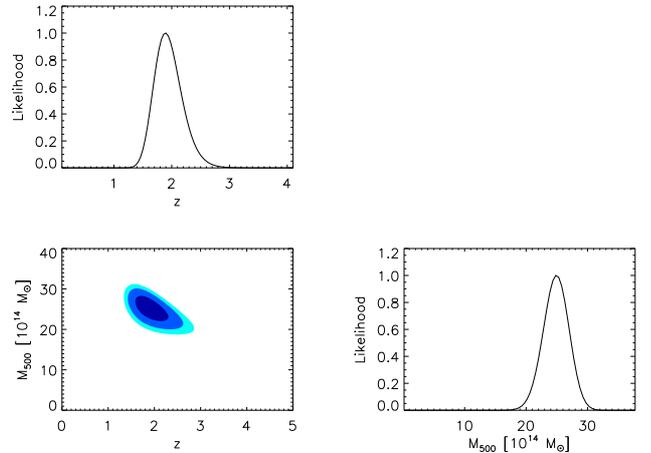}
      \caption{Constraints  on the redshift and $M_{500}$ for an extra cluster in the intercluster region of the A399-A401 system. }
         \label{FigA399constraintsExtraCluster}
   \end{figure}

From the three models considered in the previous section, we computed the residual signal in tSZ and X-ray. The amount 
of signal in the intercluster region was then used to constrain the parameters of the extra cluster or the filament.  
For the case of an extra cluster behind the intercluster region we obtain the results in Figures~\ref{FigA399constraintsExtraCluster} a) and c)
where we trace the 1D  and (Figure 8b) 2-D likelihood  contours  at 68\%, 95.5\%, and 99\%  confidence level for z and $M_{500}$. 
Only clusters at high redshift  $z=1.95$ and with a large mass $M_{500} = 2.4 \times 10^{15} M_{\circ}$ will be capable of producing a strong enough 
SZ signal with the strength of our \Planck\ residual but with X-rays that do not exceed the \ROSAT\ signal. These results are
not consistent with existing constraints on structure formation, which predict that massive clusters are formed at low
redshifts (see for example Figure 2 in \cite{HarrisonColes}.

\begin{figure}
   \centering
   \includegraphics[width=9cm]{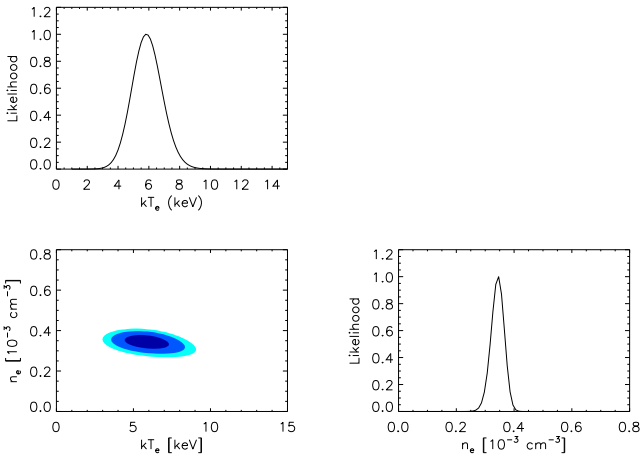}
      \caption{Constraints on the temperature and density of the filament in the intercluster region of the A399-A401 system. A high temperature $\sim 7$ keV is also favoured by the XMM data 
      \citep{Sakelliou2004}.}
         \label{FigA399constraints}
   \end{figure}

\begin{figure*}
   \centering
   \includegraphics[width=5cm,angle=90]{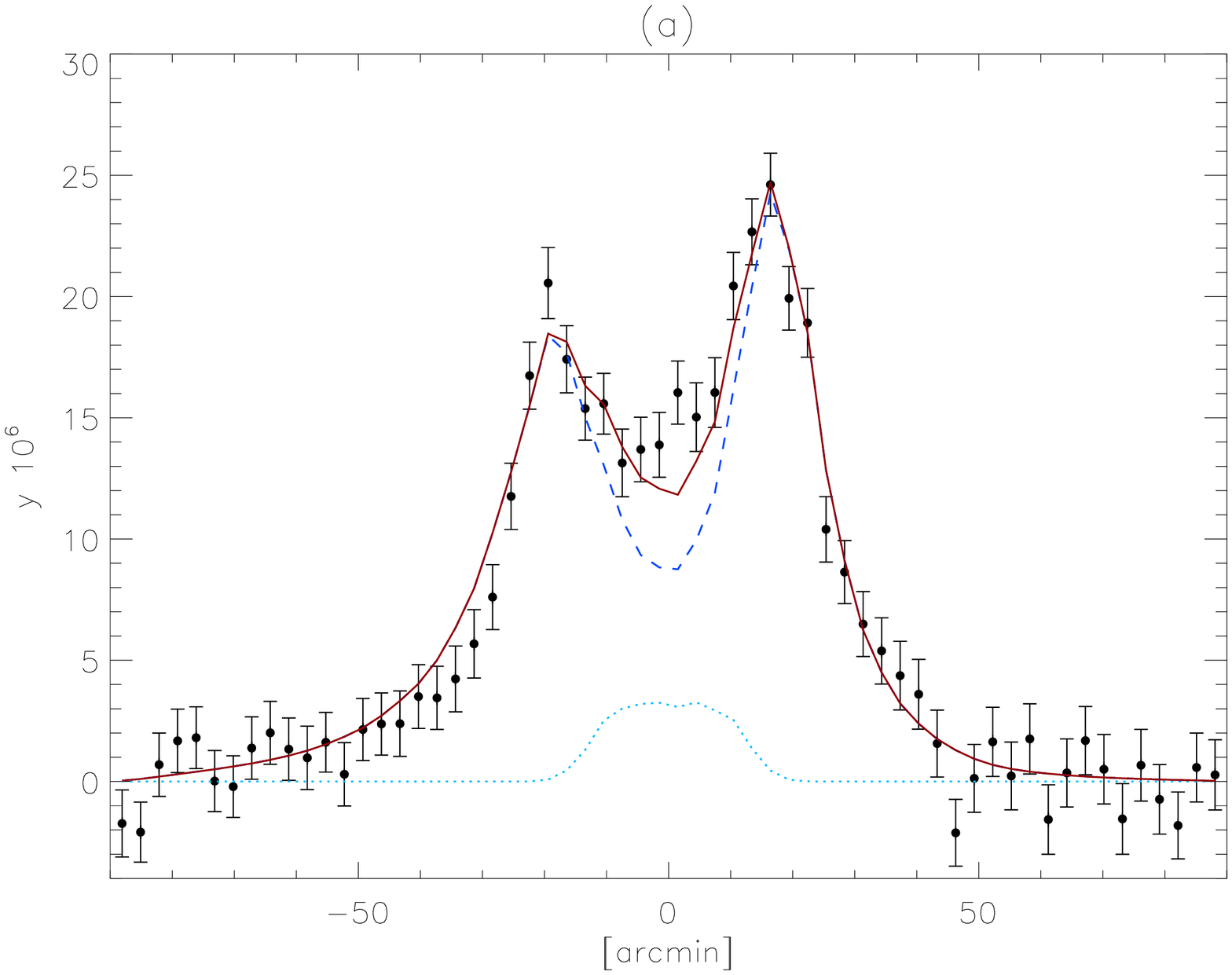}
   \includegraphics[width=5cm,angle=90]{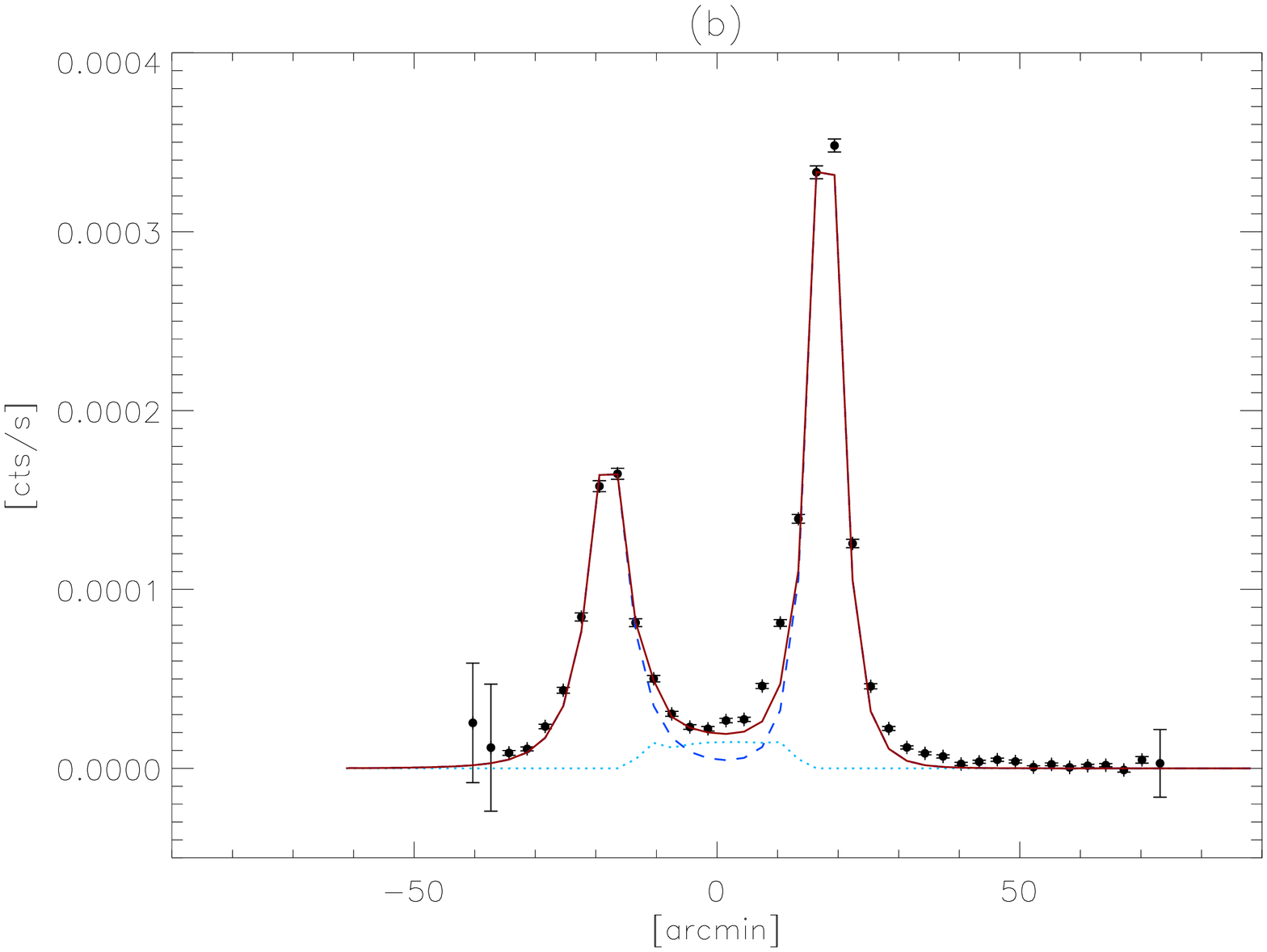}
   
      \caption{ a) tSZ longitudinal profile and b) X-ray longitudinal profile. Data from \Planck\  and \ROSAT\ (black points), from the total model (red line), the PXCC model only (light blue line), and the filament model only (dark blue line). }
         \label{FigA399profile}
   \end{figure*}

The best-fit parameters in the filament case are presented in Figure~\ref{FigA399constraints} where we show the 1D and 2-D likelihood contours
at  68\%, 95.5\%, and 99\%  confidence level for T and $n_e(0)$. We find that the temperature of the filament is $(7.08 \pm 0.85)$ \,keV  and 
the central electron density is $(3.72 \pm 0.17) \ 10^{-4}$ \,cm$^{-3}$. These results are consistent with previous estimates by 
\cite{Sakelliou2004}. It is interesting to look at the residuals for the best-fit clusters (we used the GNFW2 model here)
and the filament model. Figure~\ref{FigA399profile} shows the 1D
longitudinal profiles and residuals for the tSZ effect and X-ray emission after subtracting the clusters and filament models.
We observe that the excess of tSZ effect in the intercluster region is well described by the filament model. The integrated Compton parameter
in the intercluster region is $Y = (6.1 \pm 0.7) \ 10^{-3} $ \,arcmin$^2$. 

\section{Comparison with hydrodynamical simulations}
\label{hydrosimulations}
We applied the full analysis
described in the previous sections to hydrodynamical simulations
of a supercluster-like region ~\citep{dolag2006} that mimic our cluster pairs. 
Unlike the original simulations presented in ~\cite{dolag2006}, this simulation
was carried out including the treatment of radiative cooling, heating
by a uniform UV background and star formation feedback processes
based on a subresolution model for the multiphase structure of the
interstellar medium \citep{Springel2003}. The simulation also
follows the pattern of metal production from the past history of cosmic
star formation \citep{2004MNRAS.349L..19T,2007MNRAS.382.1050T}. This is done by computing
the contributions from both type-II and type-Ia supernovae. Energy
is fed back and metals are released gradually in time, according to
the appropriate lifetimes of the different stellar populations. This
treatment also includes, in a self-consistent way, the dependence
of the gas cooling on the local metallicity. The feedback scheme
assumes a Salpeter initial mass function (IMF) \citep{1955ApJ...121..161S} and the parameters
were fixed to obtain a wind velocity of $\approx 480 \,km\,s^{-1}$. \\
\begin{figure*}
   \centering
   \includegraphics[width=4cm,height=8cm,angle=90]{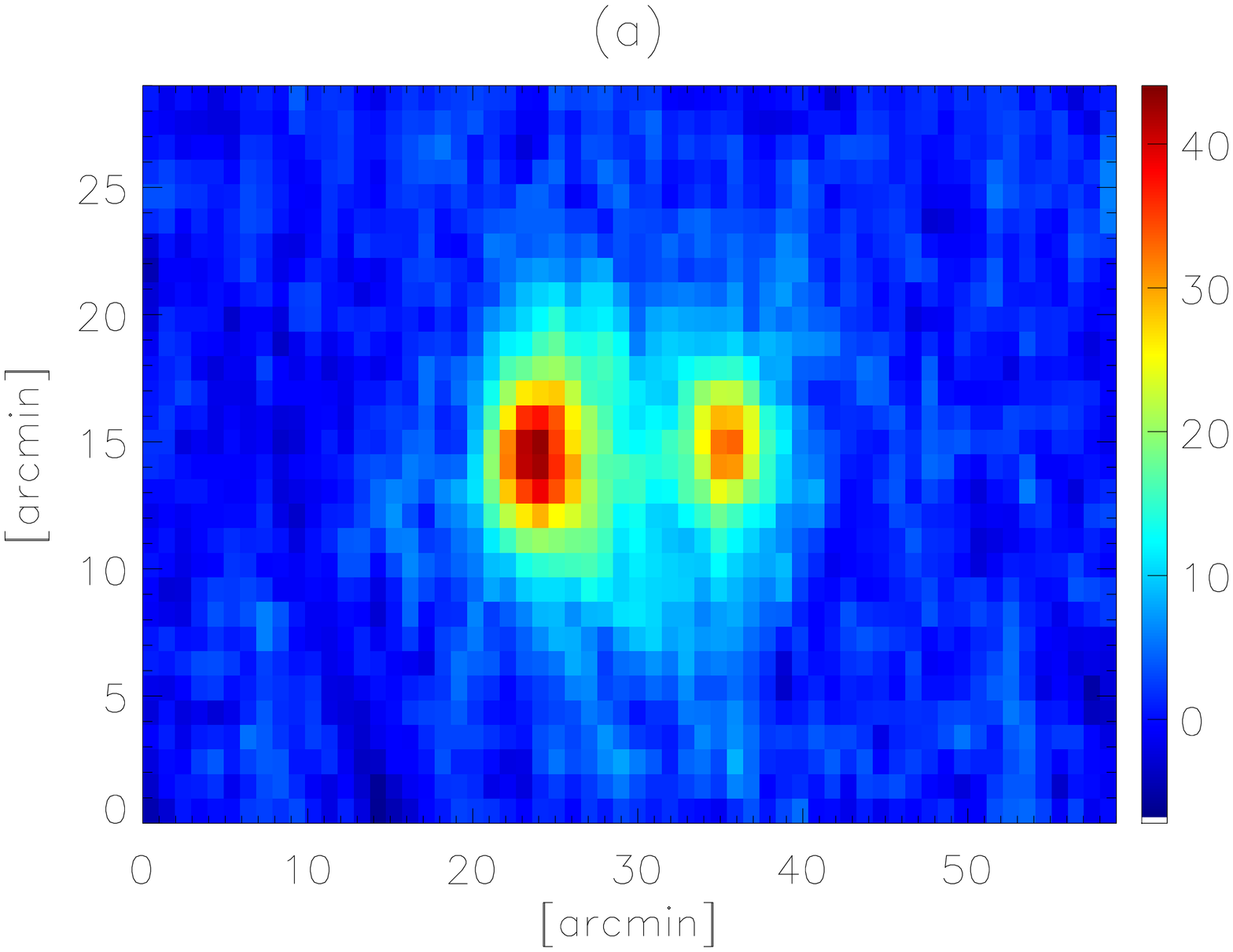}
   \includegraphics[width=4cm,height=8cm,angle=90]{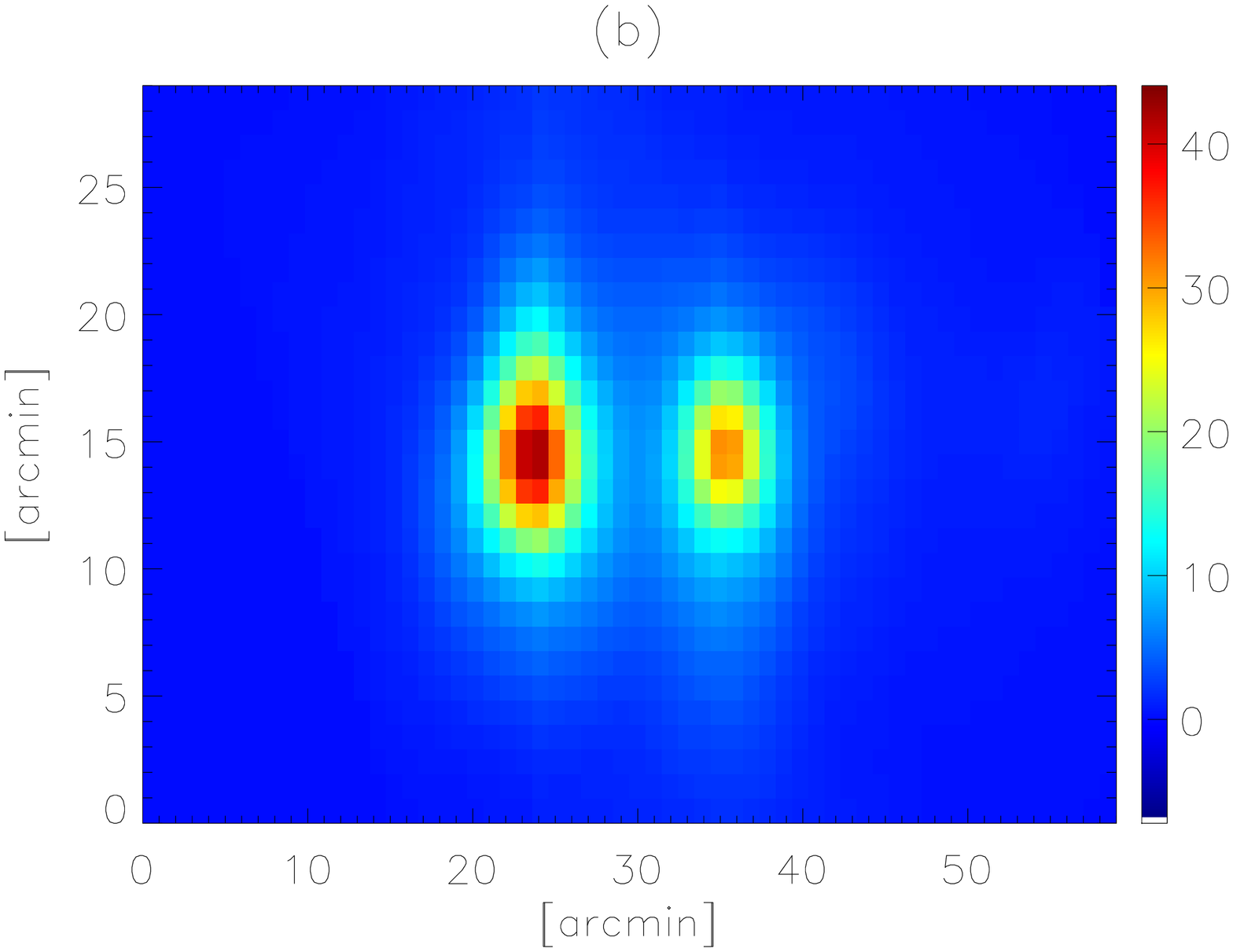}
   \includegraphics[width=4cm,height=8cm,angle=90]{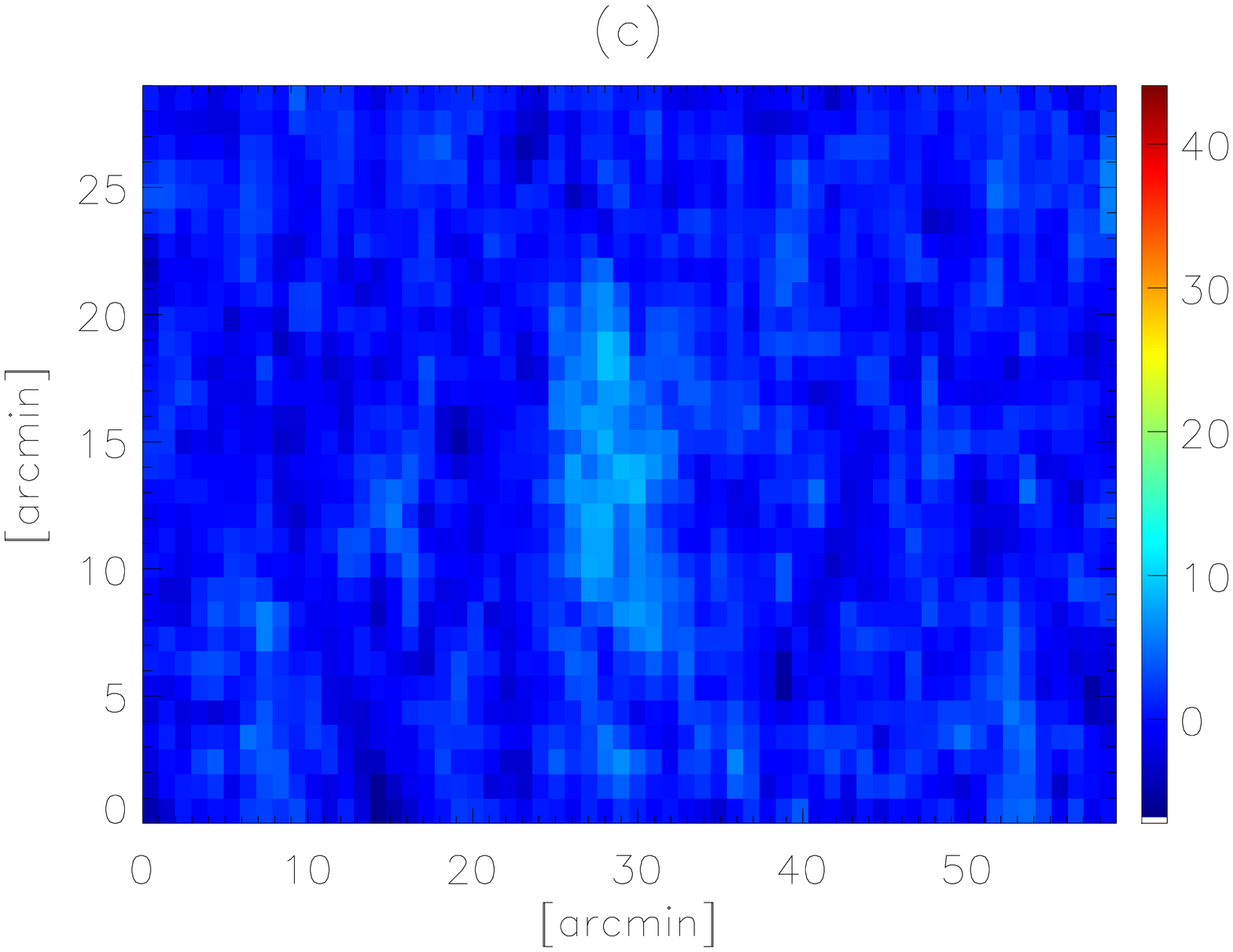}
   
      \caption{ a) tSZ Compton parameter map ($y  \times 10^6$) for Halo b in the hydrodynamical simulations,
     b) best-fit model for the clusters in the pair, and c) residuals after subtracting the latter.
         \label{FigSimuSZMaps}}
   \end{figure*}

We concentrated our analysis on a system of two merging clusters 
with characteristics similar to the A399-A401 system.
In the simulation, a cluster with a virial mass of $6.5 \times 10^{14}$ $M_{\odot}$/h
(Halo d) is merging with a cluster with a mass of $1.1 \times 10^{15}$ $M_{\odot}$/h
(Halo b). At $z=0.07$ the two systems are physically separated
by 4\,Mpc/h (2.9\,Mpc/h in the projection shown in figure 12), which
compares remarkably well with the A399-A401 system, where the projected
separation of the systems is $\sim 3$ \,Mpc. More details on the
appearance and the geometry of the simulated system can be found
in ~\cite{dolag2006}. Figure~\ref{FigSimuSZMaps} shows the Compton parameter maps (multiplied by 10$^6$).
After subtracting the best-fit model for the clusters (Figure 11b), we fitted the residuals (Figure 11c)) in the 
intercluster region with the filament model described above. 
For the best-fit model of the filament the integrated
Compton parameter is $Y = (3.5 \pm 0.7) \times 10^{-3}$ \,arcmin$^2$ and agrees well 
with the input $Y_{input} = 3.2 \times 10^{-3}$ \,arcmin$^2$. It is also important
to notice that the error bars obtained are consistent with those obtained for the cluster pair A399-A401. 

  \begin{figure*}
    \centering
    \includegraphics[scale=0.8]{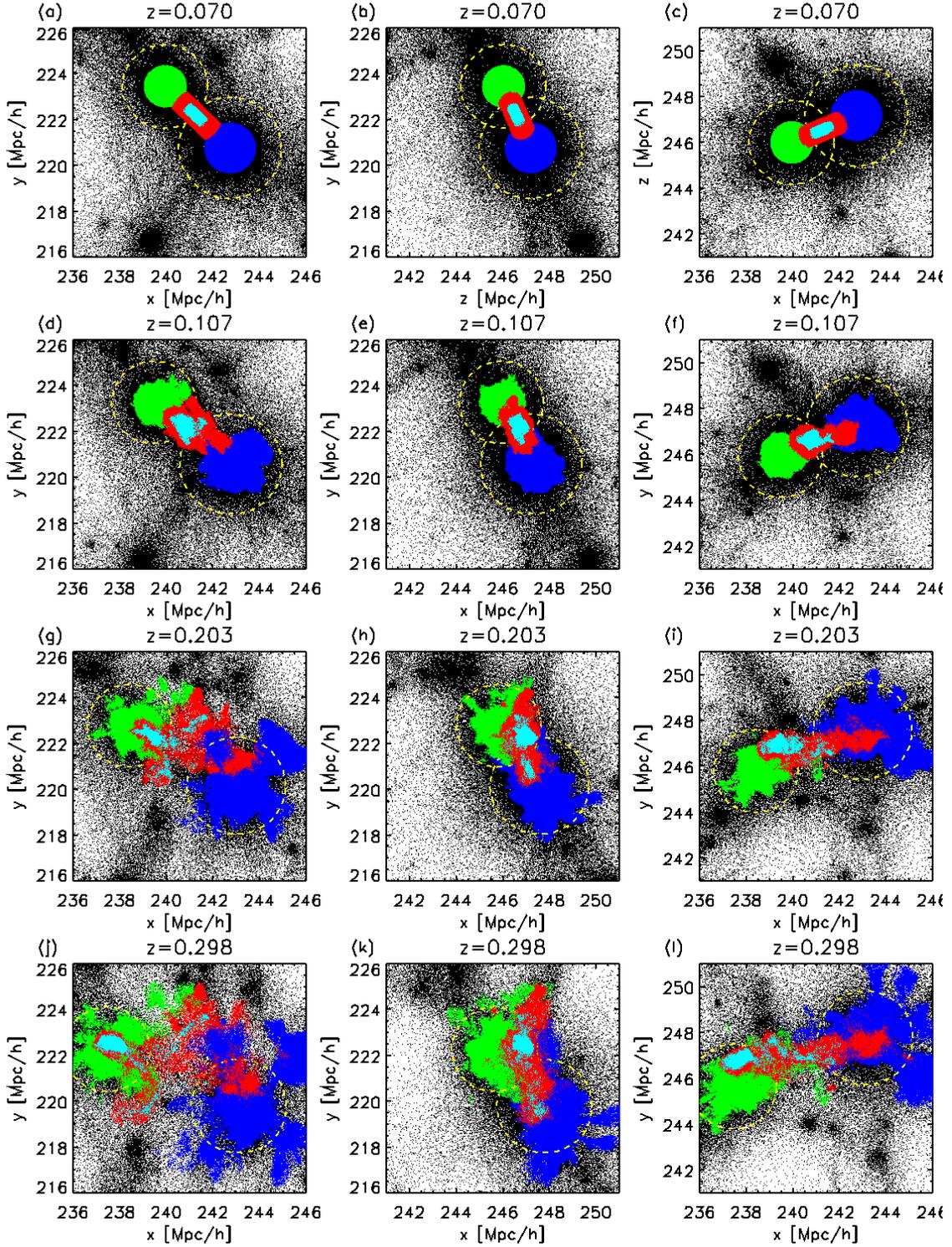}
        \caption{Spatial projections (along x,y and z axis) of the particles in the simulation.
The  rows show the simulation at z=0.07, z=0.1,
z=0.2 and z=0.3 respectively. We colour code the particles depending on their location:
clusters (green/blue), the filamentary region in between (red) and the central part
of this filament (cyan). See text for a more detailed definition. Additionally the yellow
circles mark the virial radius of the two clusters.}
        \label{FigDiscussion1}
  \end{figure*}

  \begin{figure*}
    \centering
    \includegraphics[scale=0.6]{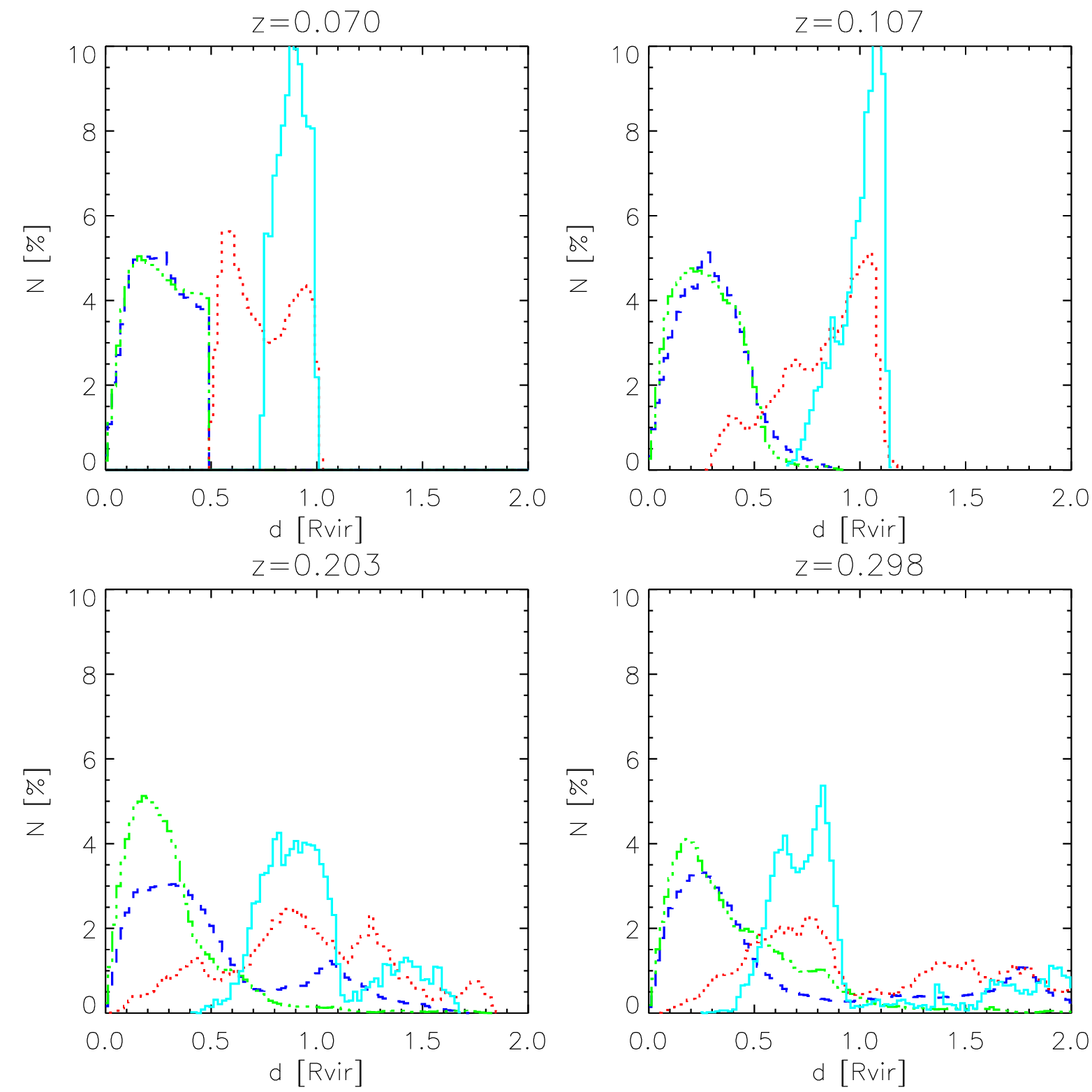}
        \caption{Evolution of the radial distribution of the particles with respect to the
centre of the closest of the two clusters. The colour code is the same as in figure~\ref{FigDiscussion1}. 
The dashed and dot-dashed lines correspond to the blue and green particles (clusters) while the dotted 
and solid lines orrespond to the red and cyan particles (filament).}
        \label{FigDiscussion3}
  \end{figure*}

We traced the origin of the particles
causing the signal within the region between the two clusters.
To do this, we distinguished between particles belonging to the main part
of one of the two galaxy clusters (e.g. within 0.5 R$_{vir}$ \footnote{
We define R$_{vir}$ based on the over-density from the top hat spherical
collapse model.}) and the particles belonging to the filament, selected
as those particles within a cylinder of 1Mpc/h in diameter between the two clusters and
lying outside 0.5 R$_{vir}$. Additionally, we defined particles
in the central part of the filament as those within a cylinder of
0.5\,Mpc/h in diameter between the two clusters and  lying outside 
0.75 R$_{vir}$. Figures~\ref{FigDiscussion1} a-c show the
different location of the particles colour-coded in blue and green
for particles belonging to the clusters, red  for particles in the filament, and cyan 
for those in the inner part of the filament.
The yellow circles mark the virial radius of the two clusters.
Going back in time to z=0.1, z=0.2, and z=0.3 (corresponding to 0.4, 1.5
and 2.5 Gyr) reveals the spatial origin of these particles.
We can observe that according to the origin of these particles, the intercluser 
region at z=0.07 is populated by two separate
components. One component is coming from material that belonged to the
outer atmosphere of one of the two clusters at z=0.3, and the other
one is coming from a very diffuse structure, outside but connecting
both clusters. This structure has a sheet-like geometry at early
times. When we trace back  the origin of the particles in the very central part 
of the filament, most of the particles were farther than 0.5 R$_{vir}$ from one of the main clusters within the last 2.5\,Gyr. 
This is made more evident in Figure~\ref{FigDiscussion3}, which shows
the spatial distribution of the particles of the different components
more quantitatively. Here we sorted each particle by 
its shortest distance to either cluster.
When we concentrate our attention on the  filament's core  (light blue particles), only a very small fraction of the particles in this region were closer than 0.5 R$_{vir}$ to one of the two main halos at z=0.3. More than half of
them are coming from regions with  0.5 R$_{vir} <$ R $<$ R$_{vir}$ and a considerable fraction is coming from regions beyond R$_{vir}$. 

This simulation shows that the intercluster region might be populated by a mix of 
components from both the clusters and the intergalactic medium.

\section{Discussion}
\label{discussions}
 
The significant SZ signal between pairs of clusters combined with the lack of significant X-ray signal suggestes that merging events may have occurred, leaving filaments of matter between these interacting pairs.  Merging of the clusters would contribute to the increased pressure that boosts the SZ signal.  These results qualitatively agree with hydrodynamical simulations in which
the intercluster region between an interacting pair of clusters can show significant SZ signal \citep{dolag2006}. \\

Regarding the cluster pair A399-A401, earlier studies showed that the intercluster medium in this pair is compressed by the merging process (\cite{Fujita1996}, \cite{Sakelliou2004}). The increased pressure would increase the SZ signal. \cite{Akahori2008} studied the non-equilibrium state of this system with simulated data and found that there might be significant shock layers at the edge of the linked region between the clusters that could explain a boost in the SZ signal. Also, \Suzaku\  observations found that the intercluster medium has a relatively high metallicity of 0.2 solar (\cite{Fujita2008}). These works estimated that the filamentary bridge would have an electron density of $n_e \sim 10^{-4}$\,cm$^{-3}$ (\cite{Fujita1996}, \cite{Sakelliou2004}, \cite{Fujita2008}).  \\
Given that the angular separation between the two clusters is $\sim 3$ Mpc and that both clusters are at slightly different redshifts (0.0724 and 0.0737),  we can assume 
that the clusters are not exactly on the same plane of the sky and that their true separation is larger than 3 Mpc. 
That is, the clusters would be separated by more than their respective virial radii. 
Consequently we can conclude that the signal (or at least a significant percentage of it) seen by \Planck\ in the intercluster region corresponds 
to baryons outside the clusters.

Our results show that there is evidence for a filamentary structure between the pair A399-A401 and outside the clusters, but the results also raise some questions about the origin of this gas. The uncertainties in estimating the cluster contributions around their virial radii makes it difficult to distinguish between the different scenarios. In a pre-merger scenario a filament could have been trapped in between the clusters, with its material being reprocessed and compressed as the clusters approach each other. In a post-merger scenario (by post-merger we mean not direct crossing of the cluster cores but a gravitational interaction as the two approaching clusters orbit their common centre of mass), the intercluster signal could be just the result of the overlapping tails of the disturbed clusters.

These clusters could be elongated due to their gravitational interaction, or a bridge of matter may have formed between the clusters after an interaction. The fits to SZ and X-Ray data reveals that there are some apparently conflicting results, which make it harder to reconcile the SZ and X-ray data with spherical, standard models. A good example is the best-fitting model of \cite{Sakelliou2004} that over-predicts the SZ data. This could be, for instance, an indication that spherical models may not be the most appropriate for describing these clusters. The failure of the \cite{Sakelliou2004} model to properly describe the Planck data shows how Planck data can be used to add information on the third dimension.
\\

Non-sphericity is expected (and indeed observed) in X-rays and could introduce corrections to the best-fitting models, particularly if the elongations of the clusters point towards each other (as is the case for A401 from the X-ray images). In this case, the gas density in the intercluster region would be  enhanced due to this elongation. 
\cite{sereno2006} showed how clusters seem to favour prolate geometries. 
A prolate model would increase the X-ray signal for the same number of electrons (or SZ), or vice versa, the same X-ray signal requires fewer electrons (or SZ). 
We estimated that a ratio of $\sim 1.3$ between the axis might be sufficient to reduce the SZ signal by $\sim 30\%$ with a fixed X-ray observation (i.e, fixing the other two axes and the total X-ray signal). \\

Clumpiness is another factor that might be introducing a bias in the models that best subtract the cluster components. \cite{1999ApJ...520L..21M} used simulations to estimate the mean mass-weighted clumping factor $C=<n^2>/<n>^2$. They found typical values for $C$ between 1.3 and 1.4 within a density contrast of 500. Since for an isothermal model $C$ can be seen as the ratio of X-ray to SZ signal (squared), as in the case of prolate clusters discussed above, clumpiness of the gas would increase the X-ray signal for the same number of electrons (or SZ) or vice-versa, the same X-ray signal would require fewer electrons (or SZ). 

\begin{figure}
   \centering
   \includegraphics[width=8cm]{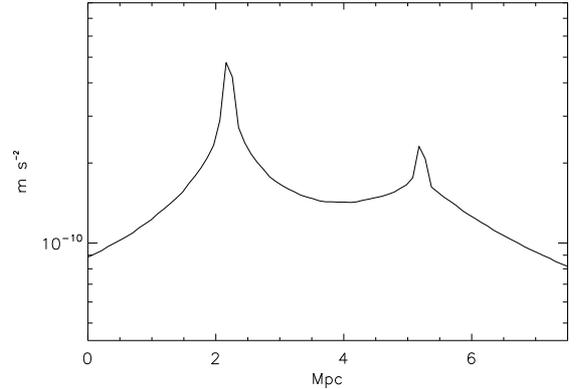}
      \caption{Gravitational acceleration across the line connecting the two cluster centres (A401 left and A399 right). The acceleration is computed assuming hydrostatic equilibrium. 
               The model GNFW1 was assumed for the computation of the hydrostatic equilibrium equation}
         \label{FigGravtA1}
   \end{figure}

One cluster will have an effect on the other cluster and in the intercluster region through its gravitational potential. 
Figure \ref{FigGravtA1} shows the resulting gravitational field across the line intersecting the two cluster centres. 
The gravitational field is computed from the hydrostatic equilibrium equation assuming a profile of the gas density (and a constant temperature). We use one of our best-fitting models described earlier (universal profile) for this purpose. The intercluster region shows a significant enhancement in the gravitational potential. 
The effect is more significant in A399 where the level in the intercluster region is higher with 
respect to the maximum acceleration at the centre of the cluster. Therefore we should expect the gravitational attraction experienced by the gas in the clusters to be stronger (compared with the peak of the potential) over A399 than over A401. 
The gravitational field in the intercluster region increases as the two clusters approach, creating a pulling effect over the gas of the clusters towards this region. In this scenario, since the 
gas would be moving from gravitational fields with similar intensity, it would not undergo an adiabatic expansion and hence would retain its original temperature. This would explain 
the high temperature found in the intercluster region. This scenario would also explain the high metallicity in the intercluster area (Z=0.2, Fujita el al. 2007, although the constraints on the metallicity are not very strong and should be take with caution). If the gas in the intercluster region was originally in a filament, it would be difficult to explain this high metallicity 
(if confirmed), but not if the gas originally comes from the clusters.  \\

The gravitational pull of the intercluster region over the outer parts of the cluster may possibily explain how the intercluster region can be partially 
populated with metal-rich gas from the clusters. As the simulation 
presented in section \ref{hydrosimulations} shows, this is a reasonable scenario, but it also
shows that the intercluster region can be populated by intergalactic material. Thus 
the intercluster region might be a mix of cluster and intergalactic material.

\section{Conclusions.}
\label{conclusions}

Using \Planck\ and \ROSAT\ data, we have studied the tSZ and X-ray maps of 25 pairs of clusters of galaxies. After modelling (assuming a
spherical symmetric model) and subtracting the
contribution of each individual cluster, we detected significant tSZ residuals in at least two of these pairs: A399-A401 and
A3391-A3395. In the case of the A399-A401 pair, these residuals are compatible with an intercluster filament of hot, 7 keV (in agreement with \cite{Sakelliou2004}),
and diffuse, $3.7 \times 10^{-4}$ \,cm$^{-3}$, gas connecting the two clusters. 
A chance coincidence of a background cluster is ruled out for canonical scaling relations because it would have to be a very massive cluster ($M_{500} = 2.4 \times 10^{15} \, M_{\odot}$) at a very high
redshift ($z=1.9$).  The signal detected by \Planck\ is significant independently of the cluster model.  
Hydrodynamical simulations show that the intercluster signal in A399-A401 is  compatible with a scenario where the intercluster region is populated with a mixture 
of material from the clusters and the intergalactic medium, indicating that there might be a bridge of matter connecting the two clusters.
If the measured signal of the merging cluster pair can be interpreted
in terms of spherically symmetric individual clusters, evidence remains
for an intercluster SZ signal detected by \Planck. It is consistent with
simulated data and may constitute the first detection of the tSZ effect between clusters.
Under this interpretation, the signal is unambiguous in the sense that
it is detected with high significance (as shown by Figure 2), is not
caused by any known artefact, is clearly resolved by Planck, and is
located in what one would consider the external region of a standard
cluster. The exact interpretation of the origin of the signal is more
open to speculation.
The analysis presented in this work shows the potential of \Planck\ data for studying these yet unexplored regions.
Better angular resolution observations of the tSZ would improve the modelling of the  clusters and reduce the uncertainties in the estimation of the signal excess in the intercluster region.

\section{Acknowledgements}
Based on observations obtained with \Planck\ (http://www.esa.int/Planck), an ESA science mission with instruments and contributions directly funded by ESA Member States, NASA, and Canada.

The development of \Planck\ has been supported by: ESA; CNES and CNRS/INSU-IN2P3-INP (France); ASI, CNR, and INAF (Italy); NASA and DoE (USA); STFC and UKSA (UK); CSIC, MICINN and JA (Spain); Tekes, AoF and CSC (Finland); DLR and MPG (Germany); CSA (Canada); DTU Space (Denmark); SER/SSO (Switzerland); RCN (Norway); SFI (Ireland); FCT/MCTES (Portugal); and The development of \Planck\ has been supported by: ESA; CNES and CNRS/INSU-IN2P3-INP (France); ASI, CNR, and INAF (Italy); NASA and DoE (USA); STFC and UKSA (UK); CSIC, MICINN and JA (Spain); Tekes, AoF and CSC (Finland); DLR and MPG (Germany); CSA (Canada); DTU Space (Denmark); SER/SSO (Switzerland); RCN (Norway); SFI (Ireland); FCT/MCTES (Portugal); and PRACE (EU). We acknowledge the use of the Healpix software \citep{healpix} and \small{SEXTRACTOR} \citep{sextractor}.

\bibliographystyle{aa} 
\bibliography{biblio} 

\end{document}

%% file: Planck.tex
\def\setsymbol#1#2{\expandafter\def\csname #1\endcsname{#2}}
\def\getsymbol#1{\csname #1\endcsname}

\def\Planck{\textit{Planck}}
\def\ASCA{\textit{ASCA}}
\def\ROSAT{\textit{ROSAT}}
\def\IRAS{\textit{IRAS}}
\def\XMM{\textit{XMM-Newton}}
\def\CHANDRA{\textit{Chandra}}
\def\Suzaku{\textit{Suzaku}}

\newbox\tablebox    \newdimen\tablewidth
\def\leaderfil{\leaders\hbox to 5pt{\hss.\hss}\hfil}
\def\endPlancktablewide{\tablewidth=\textwidth 
    $$\hss\copy\tablebox\hss$$
    \vskip-\lastskip\vskip -2pt}
\def\tablenote#1 #2\par{\begingroup \parindent=0.8em
    \abovedisplayshortskip=0pt\belowdisplayshortskip=0pt
    \noindent
    $$\hss\vbox{\hsize\tablewidth \hangindent=\parindent \hangafter=1 \noindent
    \hbox to \parindent{$^#1$\hss}\strut#2\strut\par}\hss$$
    \endgroup}

\def\L2{\ifmmode L_2\else $L_2$\fi}

\def\DeltaT{\ifmmode \Delta T\else $\Delta T$\fi}
\def\deltat{\ifmmode \Delta t\else $\Delta t$\fi}
\def\fknee{\ifmmode f_{\rm knee}\else $f_{\rm knee}$\fi}
\def\Fmax{\ifmmode F_{\rm max}\else $F_{\rm max}$\fi}
\def\solar{\ifmmode{\rm M}_{\mathord\odot}\else${\rm M}_{\mathord\odot}$\fi}
\def\Msolar{\ifmmode{\rm M}_{\mathord\odot}\else${\rm M}_{\mathord\odot}$\fi}
\def\Lsolar{\ifmmode{\rm L}_{\mathord\odot}\else${\rm L}_{\mathord\odot}$\fi}

\def\inv{\ifmmode^{-1}\else$^{-1}$\fi}
\def\mo{\ifmmode^{-1}\else$^{-1}$\fi}
\def\sup#1{\ifmmode ^{\rm #1}\else $^{\rm #1}$\fi}
\def\expo#1{\ifmmode \times 10^{#1}\else $\times 10^{#1}$\fi}
\def\,{\thinspace}
\def\lsim{\mathrel{\raise .4ex\hbox{\rlap{$<$}\lower 1.2ex\hbox{$\sim$}}}}
\def\gsim{\mathrel{\raise .4ex\hbox{\rlap{$>$}\lower 1.2ex\hbox{$\sim$}}}}

\def\simprop{\mathrel{\raise .4ex\hbox{\rlap{$\propto$}\lower 1.2ex\hbox{$\sim$}}}}
\def\deg{\ifmmode^\circ\else$^\circ$\fi}
\def\pdeg{\ifmmode $\setbox0=\hbox{$^{\circ}$}\rlap{\hskip.11\wd0 .}$^{\circ}
          \else \setbox0=\hbox{$^{\circ}$}\rlap{\hskip.11\wd0 .}$^{\circ}$\fi}
\def\arcs{\ifmmode {^{\scriptstyle\prime\prime}}
          \else $^{\scriptstyle\prime\prime}$\fi}
\def\arcm{\ifmmode {^{\scriptstyle\prime}}
          \else $^{\scriptstyle\prime}$\fi}
\newdimen\sa  \newdimen\sb
\def\parcs{\sa=.07em \sb=.03em
     \ifmmode \hbox{\rlap{.}}^{\scriptstyle\prime\kern -\sb\prime}\hbox{\kern -\sa}
     \else \rlap{.}$^{\scriptstyle\prime\kern -\sb\prime}$\kern -\sa\fi}
\def\parcm{\sa=.08em \sb=.03em
     \ifmmode \hbox{\rlap{.}\kern\sa}^{\scriptstyle\prime}\hbox{\kern-\sb}
     \else \rlap{.}\kern\sa$^{\scriptstyle\prime}$\kern-\sb\fi}
\def\ra[#1 #2 #3.#4]{#1\sup{h}#2\sup{m}#3\sup{s}\llap.#4}
\def\dec[#1 #2 #3.#4]{#1\deg#2\arcm#3\arcs\llap.#4}
\def\deco[#1 #2 #3]{#1\deg#2\arcm#3\arcs}
\def\rra[#1 #2]{#1\sup{h}#2\sup{m}}

\def\dots{\relax\ifmmode \ldots\else $\ldots$\fi}
\def\WHzsr{\ifmmode $W\,Hz\mo\,sr\mo$\else W\,Hz\mo\,sr\mo\fi}
\def\mHz{\ifmmode $\,mHz$\else \,mHz\fi}
\def\GHz{\ifmmode $\,GHz$\else \,GHz\fi}
\def\mKs{\ifmmode $\,mK\,s$^{1/2}\else \,mK\,s$^{1/2}$\fi}
\def\muKs{\ifmmode \,\mu$K\,s$^{1/2}\else \,$\mu$K\,s$^{1/2}$\fi}
\def\muKRJs{\ifmmode \,\mu$K$_{\rm RJ}$\,s$^{1/2}\else \,$\mu$K$_{\rm RJ}$\,s$^{1/2}$\fi}
\def\muKHz{\ifmmode \,\mu$K\,Hz$^{-1/2}\else \,$\mu$K\,Hz$^{-1/2}$\fi}
\def\MJysr{\ifmmode \,$MJy\,sr\mo$\else \,MJy\,sr\mo\fi}
\def\MJysrmK{\ifmmode \,$MJy\,sr\mo$\,mK$_{\rm CMB}\mo\else \,MJy\,sr\mo\,mK$_{\rm CMB}\mo$\fi}
\def\microns{\ifmmode \,\mu$m$\else \,$\mu$m\fi}

\def\muK{\ifmmode \,\mu$K$\else \,$\mu$\hbox{K}\fi}
\def\microK{\ifmmode \,\mu$K$\else \,$\mu$\hbox{K}\fi}
\def\muW{\ifmmode \,\mu$W$\else \,$\mu$\hbox{W}\fi}
\def\kms{\ifmmode $\,km\,s$^{-1}\else \,km\,s$^{-1}$\fi}
\def\kmsMpc{\ifmmode $\,\kms\,Mpc\mo$\else \,\kms\,Mpc\mo\fi}
\setsymbol{LFI:center:frequency:70GHz:units}{70.3\,GHz}
\setsymbol{LFI:center:frequency:44GHz:units}{44.1\,GHz}
\setsymbol{LFI:center:frequency:30GHz:units}{28.5\,GHz}

\setsymbol{LFI:center:frequency:70GHz}{70.3}
\setsymbol{LFI:center:frequency:44GHz}{44.1}
\setsymbol{LFI:center:frequency:30GHz}{28.5}

\setsymbol{LFI:center:frequency:LFI18:Rad:M:units}{71.7\GHz}
\setsymbol{LFI:center:frequency:LFI19:Rad:M:units}{67.5\GHz}
\setsymbol{LFI:center:frequency:LFI20:Rad:M:units}{69.2\GHz}
\setsymbol{LFI:center:frequency:LFI21:Rad:M:units}{70.4\GHz}
\setsymbol{LFI:center:frequency:LFI22:Rad:M:units}{71.5\GHz}
\setsymbol{LFI:center:frequency:LFI23:Rad:M:units}{70.8\GHz}
\setsymbol{LFI:center:frequency:LFI24:Rad:M:units}{44.4\GHz}
\setsymbol{LFI:center:frequency:LFI25:Rad:M:units}{44.0\GHz}
\setsymbol{LFI:center:frequency:LFI26:Rad:M:units}{43.9\GHz}
\setsymbol{LFI:center:frequency:LFI27:Rad:M:units}{28.3\GHz}
\setsymbol{LFI:center:frequency:LFI28:Rad:M:units}{28.8\GHz}
\setsymbol{LFI:center:frequency:LFI18:Rad:S:units}{70.1\GHz}
\setsymbol{LFI:center:frequency:LFI19:Rad:S:units}{69.6\GHz}
\setsymbol{LFI:center:frequency:LFI20:Rad:S:units}{69.5\GHz}
\setsymbol{LFI:center:frequency:LFI21:Rad:S:units}{69.5\GHz}
\setsymbol{LFI:center:frequency:LFI22:Rad:S:units}{72.8\GHz}
\setsymbol{LFI:center:frequency:LFI23:Rad:S:units}{71.3\GHz}
\setsymbol{LFI:center:frequency:LFI24:Rad:S:units}{44.1\GHz}
\setsymbol{LFI:center:frequency:LFI25:Rad:S:units}{44.1\GHz}
\setsymbol{LFI:center:frequency:LFI26:Rad:S:units}{44.1\GHz}
\setsymbol{LFI:center:frequency:LFI27:Rad:S:units}{28.5\GHz}
\setsymbol{LFI:center:frequency:LFI28:Rad:S:units}{28.2\GHz}

\setsymbol{LFI:center:frequency:LFI18:Rad:M}{71.7}
\setsymbol{LFI:center:frequency:LFI19:Rad:M}{67.5}
\setsymbol{LFI:center:frequency:LFI20:Rad:M}{69.2}
\setsymbol{LFI:center:frequency:LFI21:Rad:M}{70.4}
\setsymbol{LFI:center:frequency:LFI22:Rad:M}{71.5}
\setsymbol{LFI:center:frequency:LFI23:Rad:M}{70.8}
\setsymbol{LFI:center:frequency:LFI24:Rad:M}{44.4}
\setsymbol{LFI:center:frequency:LFI25:Rad:M}{44.0}
\setsymbol{LFI:center:frequency:LFI26:Rad:M}{43.9}
\setsymbol{LFI:center:frequency:LFI27:Rad:M}{28.3}
\setsymbol{LFI:center:frequency:LFI28:Rad:M}{28.8}
\setsymbol{LFI:center:frequency:LFI18:Rad:S}{70.1}
\setsymbol{LFI:center:frequency:LFI19:Rad:S}{69.6}
\setsymbol{LFI:center:frequency:LFI20:Rad:S}{69.5}
\setsymbol{LFI:center:frequency:LFI21:Rad:S}{69.5}
\setsymbol{LFI:center:frequency:LFI22:Rad:S}{72.8}
\setsymbol{LFI:center:frequency:LFI23:Rad:S}{71.3}
\setsymbol{LFI:center:frequency:LFI24:Rad:S}{44.1}
\setsymbol{LFI:center:frequency:LFI25:Rad:S}{44.1}
\setsymbol{LFI:center:frequency:LFI26:Rad:S}{44.1}
\setsymbol{LFI:center:frequency:LFI27:Rad:S}{28.5}
\setsymbol{LFI:center:frequency:LFI28:Rad:S}{28.2}

\setsymbol{LFI:white:noise:sensitivity:70GHz:units}{134.7\muKs}
\setsymbol{LFI:white:noise:sensitivity:44GHz:units}{164.7\muKs}
\setsymbol{LFI:white:noise:sensitivity:30GHz:units}{143.4\muKs}

\setsymbol{LFI:white:noise:sensitivity:70GHz}{134.7}
\setsymbol{LFI:white:noise:sensitivity:44GHz}{164.7}
\setsymbol{LFI:white:noise:sensitivity:30GHz}{143.4}

\setsymbol{LFI:white:noise:sensitivity:LFI18:Rad:M:units}{512.0\muKs}
\setsymbol{LFI:white:noise:sensitivity:LFI19:Rad:M:units}{581.4\muKs}
\setsymbol{LFI:white:noise:sensitivity:LFI20:Rad:M:units}{590.8\muKs}
\setsymbol{LFI:white:noise:sensitivity:LFI21:Rad:M:units}{455.2\muKs}
\setsymbol{LFI:white:noise:sensitivity:LFI22:Rad:M:units}{492.0\muKs}
\setsymbol{LFI:white:noise:sensitivity:LFI23:Rad:M:units}{507.7\muKs}
\setsymbol{LFI:white:noise:sensitivity:LFI24:Rad:M:units}{462.2\muKs}
\setsymbol{LFI:white:noise:sensitivity:LFI25:Rad:M:units}{413.6\muKs}
\setsymbol{LFI:white:noise:sensitivity:LFI26:Rad:M:units}{478.6\muKs}
\setsymbol{LFI:white:noise:sensitivity:LFI27:Rad:M:units}{277.7\muKs}
\setsymbol{LFI:white:noise:sensitivity:LFI28:Rad:M:units}{312.3\muKs}
\setsymbol{LFI:white:noise:sensitivity:LFI18:Rad:S:units}{465.7\muKs}
\setsymbol{LFI:white:noise:sensitivity:LFI19:Rad:S:units}{555.6\muKs}
\setsymbol{LFI:white:noise:sensitivity:LFI20:Rad:S:units}{623.2\muKs}
\setsymbol{LFI:white:noise:sensitivity:LFI21:Rad:S:units}{564.1\muKs}
\setsymbol{LFI:white:noise:sensitivity:LFI22:Rad:S:units}{534.4\muKs}
\setsymbol{LFI:white:noise:sensitivity:LFI23:Rad:S:units}{542.4\muKs}
\setsymbol{LFI:white:noise:sensitivity:LFI24:Rad:S:units}{399.2\muKs}
\setsymbol{LFI:white:noise:sensitivity:LFI25:Rad:S:units}{392.6\muKs}
\setsymbol{LFI:white:noise:sensitivity:LFI26:Rad:S:units}{418.6\muKs}
\setsymbol{LFI:white:noise:sensitivity:LFI27:Rad:S:units}{302.9\muKs}
\setsymbol{LFI:white:noise:sensitivity:LFI28:Rad:S:units}{285.3\muKs}

\setsymbol{LFI:white:noise:sensitivity:LFI18:Rad:M}{512.0}
\setsymbol{LFI:white:noise:sensitivity:LFI19:Rad:M}{581.4}
\setsymbol{LFI:white:noise:sensitivity:LFI20:Rad:M}{590.8}
\setsymbol{LFI:white:noise:sensitivity:LFI21:Rad:M}{455.2}
\setsymbol{LFI:white:noise:sensitivity:LFI22:Rad:M}{492.0}
\setsymbol{LFI:white:noise:sensitivity:LFI23:Rad:M}{507.7}
\setsymbol{LFI:white:noise:sensitivity:LFI24:Rad:M}{462.2}
\setsymbol{LFI:white:noise:sensitivity:LFI25:Rad:M}{413.6}
\setsymbol{LFI:white:noise:sensitivity:LFI26:Rad:M}{478.6}
\setsymbol{LFI:white:noise:sensitivity:LFI27:Rad:M}{277.7}
\setsymbol{LFI:white:noise:sensitivity:LFI28:Rad:M}{312.3}
\setsymbol{LFI:white:noise:sensitivity:LFI18:Rad:S}{465.7}
\setsymbol{LFI:white:noise:sensitivity:LFI19:Rad:S}{555.6}
\setsymbol{LFI:white:noise:sensitivity:LFI20:Rad:S}{623.2}
\setsymbol{LFI:white:noise:sensitivity:LFI21:Rad:S}{564.1}
\setsymbol{LFI:white:noise:sensitivity:LFI22:Rad:S}{534.4}
\setsymbol{LFI:white:noise:sensitivity:LFI23:Rad:S}{542.4}
\setsymbol{LFI:white:noise:sensitivity:LFI24:Rad:S}{399.2}
\setsymbol{LFI:white:noise:sensitivity:LFI25:Rad:S}{392.6}
\setsymbol{LFI:white:noise:sensitivity:LFI26:Rad:S}{418.6}
\setsymbol{LFI:white:noise:sensitivity:LFI27:Rad:S}{302.9}
\setsymbol{LFI:white:noise:sensitivity:LFI28:Rad:S}{285.3}

\setsymbol{LFI:knee:frequency:70GHz:units}{29.5\mHz}
\setsymbol{LFI:knee:frequency:44GHz:units}{56.2\mHz}
\setsymbol{LFI:knee:frequency:30GHz:units}{113.7\mHz}

\setsymbol{LFI:knee:frequency:70GHz}{29.5}
\setsymbol{LFI:knee:frequency:44GHz}{56.2}
\setsymbol{LFI:knee:frequency:30GHz}{113.7}

\setsymbol{LFI:knee:frequency:LFI18:Rad:M:units}{16.3\mHz}
\setsymbol{LFI:knee:frequency:LFI19:Rad:M:units}{15.1\mHz}
\setsymbol{LFI:knee:frequency:LFI20:Rad:M:units}{18.7\mHz}
\setsymbol{LFI:knee:frequency:LFI21:Rad:M:units}{37.2\mHz}
\setsymbol{LFI:knee:frequency:LFI22:Rad:M:units}{12.7\mHz}
\setsymbol{LFI:knee:frequency:LFI23:Rad:M:units}{34.6\mHz}
\setsymbol{LFI:knee:frequency:LFI24:Rad:M:units}{46.2\mHz}
\setsymbol{LFI:knee:frequency:LFI25:Rad:M:units}{24.9\mHz}
\setsymbol{LFI:knee:frequency:LFI26:Rad:M:units}{67.6\mHz}
\setsymbol{LFI:knee:frequency:LFI27:Rad:M:units}{187.4\mHz}
\setsymbol{LFI:knee:frequency:LFI28:Rad:M:units}{122.2\mHz}
\setsymbol{LFI:knee:frequency:LFI18:Rad:S:units}{17.7\mHz}
\setsymbol{LFI:knee:frequency:LFI19:Rad:S:units}{22.0\mHz}
\setsymbol{LFI:knee:frequency:LFI20:Rad:S:units}{8.7\mHz}
\setsymbol{LFI:knee:frequency:LFI21:Rad:S:units}{25.9\mHz}
\setsymbol{LFI:knee:frequency:LFI22:Rad:S:units}{15.8\mHz}
\setsymbol{LFI:knee:frequency:LFI23:Rad:S:units}{129.8\mHz}
\setsymbol{LFI:knee:frequency:LFI24:Rad:S:units}{100.9\mHz}
\setsymbol{LFI:knee:frequency:LFI25:Rad:S:units}{38.9\mHz}
\setsymbol{LFI:knee:frequency:LFI26:Rad:S:units}{58.9\mHz}
\setsymbol{LFI:knee:frequency:LFI27:Rad:S:units}{104.4\mHz}
\setsymbol{LFI:knee:frequency:LFI28:Rad:S:units}{40.7\mHz}

\setsymbol{LFI:knee:frequency:LFI18:Rad:M}{16.3}
\setsymbol{LFI:knee:frequency:LFI19:Rad:M}{15.1}
\setsymbol{LFI:knee:frequency:LFI20:Rad:M}{18.7}
\setsymbol{LFI:knee:frequency:LFI21:Rad:M}{37.2}
\setsymbol{LFI:knee:frequency:LFI22:Rad:M}{12.7}
\setsymbol{LFI:knee:frequency:LFI23:Rad:M}{34.6}
\setsymbol{LFI:knee:frequency:LFI24:Rad:M}{46.2}
\setsymbol{LFI:knee:frequency:LFI25:Rad:M}{24.9}
\setsymbol{LFI:knee:frequency:LFI26:Rad:M}{67.6}
\setsymbol{LFI:knee:frequency:LFI27:Rad:M}{187.4}
\setsymbol{LFI:knee:frequency:LFI28:Rad:M}{122.2}
\setsymbol{LFI:knee:frequency:LFI18:Rad:S}{17.7}
\setsymbol{LFI:knee:frequency:LFI19:Rad:S}{22.0}
\setsymbol{LFI:knee:frequency:LFI20:Rad:S}{8.7}
\setsymbol{LFI:knee:frequency:LFI21:Rad:S}{25.9}
\setsymbol{LFI:knee:frequency:LFI22:Rad:S}{15.8}
\setsymbol{LFI:knee:frequency:LFI23:Rad:S}{129.8}
\setsymbol{LFI:knee:frequency:LFI24:Rad:S}{100.9}
\setsymbol{LFI:knee:frequency:LFI25:Rad:S}{38.9}
\setsymbol{LFI:knee:frequency:LFI26:Rad:S}{58.9}
\setsymbol{LFI:knee:frequency:LFI27:Rad:S}{104.4}
\setsymbol{LFI:knee:frequency:LFI28:Rad:S}{40.7}

\setsymbol{LFI:slope:70GHz:units}{$-1.03$\mHz}
\setsymbol{LFI:slope:44GHz:units}{$-0.89$\mHz}
\setsymbol{LFI:slope:30GHz:units}{$-0.87$\mHz}

\setsymbol{LFI:slope:70GHz}{$-1.03$}
\setsymbol{LFI:slope:44GHz}{$-0.89$}
\setsymbol{LFI:slope:30GHz}{$-0.87$}

\setsymbol{LFI:slope:LFI18:Rad:M:units}{$-1.04$\mHz}
\setsymbol{LFI:slope:LFI19:Rad:M:units}{$-1.09$\mHz}
\setsymbol{LFI:slope:LFI20:Rad:M:units}{$-0.69$\mHz}
\setsymbol{LFI:slope:LFI21:Rad:M:units}{$-1.56$\mHz}
\setsymbol{LFI:slope:LFI22:Rad:M:units}{$-1.01$\mHz}
\setsymbol{LFI:slope:LFI23:Rad:M:units}{$-0.96$\mHz}
\setsymbol{LFI:slope:LFI24:Rad:M:units}{$-0.83$\mHz}
\setsymbol{LFI:slope:LFI25:Rad:M:units}{$-0.91$\mHz}
\setsymbol{LFI:slope:LFI26:Rad:M:units}{$-0.95$\mHz}
\setsymbol{LFI:slope:LFI27:Rad:M:units}{$-0.87$\mHz}
\setsymbol{LFI:slope:LFI28:Rad:M:units}{$-0.88$\mHz}
\setsymbol{LFI:slope:LFI18:Rad:S:units}{$-1.15$\mHz}
\setsymbol{LFI:slope:LFI19:Rad:S:units}{$-1.00$\mHz}
\setsymbol{LFI:slope:LFI20:Rad:S:units}{$-0.95$\mHz}
\setsymbol{LFI:slope:LFI21:Rad:S:units}{$-0.92$\mHz}
\setsymbol{LFI:slope:LFI22:Rad:S:units}{$-1.01$\mHz}
\setsymbol{LFI:slope:LFI23:Rad:S:units}{$-0.95$\mHz}
\setsymbol{LFI:slope:LFI24:Rad:S:units}{$-0.73$\mHz}
\setsymbol{LFI:slope:LFI25:Rad:S:units}{$-1.16$\mHz}
\setsymbol{LFI:slope:LFI26:Rad:S:units}{$-0.79$\mHz}
\setsymbol{LFI:slope:LFI27:Rad:S:units}{$-0.82$\mHz}
\setsymbol{LFI:slope:LFI28:Rad:S:units}{$-0.91$\mHz}

\setsymbol{LFI:slope:LFI18:Rad:M}{$-1.04$}
\setsymbol{LFI:slope:LFI19:Rad:M}{$-1.09$}
\setsymbol{LFI:slope:LFI20:Rad:M}{$-0.69$}
\setsymbol{LFI:slope:LFI21:Rad:M}{$-1.56$}
\setsymbol{LFI:slope:LFI22:Rad:M}{$-1.01$}
\setsymbol{LFI:slope:LFI23:Rad:M}{$-0.96$}
\setsymbol{LFI:slope:LFI24:Rad:M}{$-0.83$}
\setsymbol{LFI:slope:LFI25:Rad:M}{$-0.91$}
\setsymbol{LFI:slope:LFI26:Rad:M}{$-0.95$}
\setsymbol{LFI:slope:LFI27:Rad:M}{$-0.87$}
\setsymbol{LFI:slope:LFI28:Rad:M}{$-0.88$}
\setsymbol{LFI:slope:LFI18:Rad:S}{$-1.15$}
\setsymbol{LFI:slope:LFI19:Rad:S}{$-1.00$}
\setsymbol{LFI:slope:LFI20:Rad:S}{$-0.95$}
\setsymbol{LFI:slope:LFI21:Rad:S}{$-0.92$}
\setsymbol{LFI:slope:LFI22:Rad:S}{$-1.01$}
\setsymbol{LFI:slope:LFI23:Rad:S}{$-0.95$}
\setsymbol{LFI:slope:LFI24:Rad:S}{$-0.73$}
\setsymbol{LFI:slope:LFI25:Rad:S}{$-1.16$}
\setsymbol{LFI:slope:LFI26:Rad:S}{$-0.79$}
\setsymbol{LFI:slope:LFI27:Rad:S}{$-0.82$}
\setsymbol{LFI:slope:LFI28:Rad:S}{$-0.91$}

\setsymbol{LFI:FWHM:70GHz:units}{13\parcm01}
\setsymbol{LFI:FWHM:44GHz:units}{27\parcm92}
\setsymbol{LFI:FWHM:30GHz:units}{32\parcm65}

\setsymbol{LFI:FWHM:70GHz}{13.01}
\setsymbol{LFI:FWHM:44GHz}{27.92}
\setsymbol{LFI:FWHM:30GHz}{32.65}

\setsymbol{LFI:FWHM:LFI18:units}{13\parcm39}
\setsymbol{LFI:FWHM:LFI19:units}{13\parcm01}
\setsymbol{LFI:FWHM:LFI20:units}{12\parcm75}
\setsymbol{LFI:FWHM:LFI21:units}{12\parcm74}
\setsymbol{LFI:FWHM:LFI22:units}{12\parcm87}
\setsymbol{LFI:FWHM:LFI23:units}{13\parcm27}
\setsymbol{LFI:FWHM:LFI24:units}{22\parcm98}
\setsymbol{LFI:FWHM:LFI25:units}{30\parcm46}
\setsymbol{LFI:FWHM:LFI26:units}{30\parcm31}
\setsymbol{LFI:FWHM:LFI27:units}{32\parcm65}
\setsymbol{LFI:FWHM:LFI28:units}{32\parcm66}

\setsymbol{LFI:FWHM:LFI18}{13.39}
\setsymbol{LFI:FWHM:LFI19}{13.01}
\setsymbol{LFI:FWHM:LFI20}{12.75}
\setsymbol{LFI:FWHM:LFI21}{12.74}
\setsymbol{LFI:FWHM:LFI22}{12.87}
\setsymbol{LFI:FWHM:LFI23}{13.27}
\setsymbol{LFI:FWHM:LFI24}{22.98}
\setsymbol{LFI:FWHM:LFI25}{30.46}
\setsymbol{LFI:FWHM:LFI26}{30.31}
\setsymbol{LFI:FWHM:LFI27}{32.65}
\setsymbol{LFI:FWHM:LFI28}{32.66}

\setsymbol{LFI:FWHM:uncertainty:LFI18:units}{0.170\arcm}
\setsymbol{LFI:FWHM:uncertainty:LFI19:units}{0.174\arcm}
\setsymbol{LFI:FWHM:uncertainty:LFI20:units}{0.170\arcm}
\setsymbol{LFI:FWHM:uncertainty:LFI21:units}{0.156\arcm}
\setsymbol{LFI:FWHM:uncertainty:LFI22:units}{0.164\arcm}
\setsymbol{LFI:FWHM:uncertainty:LFI23:units}{0.171\arcm}
\setsymbol{LFI:FWHM:uncertainty:LFI24:units}{0.652\arcm}
\setsymbol{LFI:FWHM:uncertainty:LFI25:units}{1.075\arcm}
\setsymbol{LFI:FWHM:uncertainty:LFI26:units}{1.131\arcm}
\setsymbol{LFI:FWHM:uncertainty:LFI27:units}{1.266\arcm}
\setsymbol{LFI:FWHM:uncertainty:LFI28:units}{1.287\arcm}

\setsymbol{LFI:FWHM:uncertainty:LFI18}{0.170}
\setsymbol{LFI:FWHM:uncertainty:LFI19}{0.174}
\setsymbol{LFI:FWHM:uncertainty:LFI20}{0.170}
\setsymbol{LFI:FWHM:uncertainty:LFI21}{0.156}
\setsymbol{LFI:FWHM:uncertainty:LFI22}{0.164}
\setsymbol{LFI:FWHM:uncertainty:LFI23}{0.171}
\setsymbol{LFI:FWHM:uncertainty:LFI24}{0.652}
\setsymbol{LFI:FWHM:uncertainty:LFI25}{1.075}
\setsymbol{LFI:FWHM:uncertainty:LFI26}{1.131}
\setsymbol{LFI:FWHM:uncertainty:LFI27}{1.266}
\setsymbol{LFI:FWHM:uncertainty:LFI28}{1.287}

\setsymbol{HFI:center:frequency:100GHz:units}{100\,GHz}
\setsymbol{HFI:center:frequency:143GHz:units}{143\,GHz}
\setsymbol{HFI:center:frequency:217GHz:units}{217\,GHz}
\setsymbol{HFI:center:frequency:353GHz:units}{353\,GHz}
\setsymbol{HFI:center:frequency:545GHz:units}{545\,GHz}
\setsymbol{HFI:center:frequency:857GHz:units}{857\,GHz}

\setsymbol{HFI:center:frequency:100GHz}{100}
\setsymbol{HFI:center:frequency:143GHz}{143}
\setsymbol{HFI:center:frequency:217GHz}{217}
\setsymbol{HFI:center:frequency:353GHz}{353}
\setsymbol{HFI:center:frequency:545GHz}{545}
\setsymbol{HFI:center:frequency:857GHz}{857}

\setsymbol{HFI:Ndetectors:100GHz}{8}
\setsymbol{HFI:Ndetectors:143GHz}{11}
\setsymbol{HFI:Ndetectors:217GHz}{12}
\setsymbol{HFI:Ndetectors:353GHz}{12}
\setsymbol{HFI:Ndetectors:545GHz}{3}
\setsymbol{HFI:Ndetectors:857GHz}{4}

\setsymbol{HFI:FWHM:Maps:100GHz:units}{9\parcm88}
\setsymbol{HFI:FWHM:Maps:143GHz:units}{7\parcm18}
\setsymbol{HFI:FWHM:Maps:217GHz:units}{4\parcm87}
\setsymbol{HFI:FWHM:Maps:353GHz:units}{4\parcm65}
\setsymbol{HFI:FWHM:Maps:545GHz:units}{4\parcm72}
\setsymbol{HFI:FWHM:Maps:857GHz:units}{4\parcm39}
\setsymbol{HFI:FWHM:Maps:100GHz}{9.88}
\setsymbol{HFI:FWHM:Maps:143GHz}{7.18}
\setsymbol{HFI:FWHM:Maps:217GHz}{4.87}
\setsymbol{HFI:FWHM:Maps:353GHz}{4.65}
\setsymbol{HFI:FWHM:Maps:545GHz}{4.72}
\setsymbol{HFI:FWHM:Maps:857GHz}{4.39}

\setsymbol{HFI:beam:ellipticity:Maps:100GHz}{1.15}
\setsymbol{HFI:beam:ellipticity:Maps:143GHz}{1.01}
\setsymbol{HFI:beam:ellipticity:Maps:217GHz}{1.06}
\setsymbol{HFI:beam:ellipticity:Maps:353GHz}{1.05}
\setsymbol{HFI:beam:ellipticity:Maps:545GHz}{1.14}
\setsymbol{HFI:beam:ellipticity:Maps:857GHz}{1.19}

\setsymbol{HFI:FWHM:Mars:100GHz:units}{9\parcm37}
\setsymbol{HFI:FWHM:Mars:143GHz:units}{7\parcm04}
\setsymbol{HFI:FWHM:Mars:217GHz:units}{4\parcm68}
\setsymbol{HFI:FWHM:Mars:353GHz:units}{4\parcm43}
\setsymbol{HFI:FWHM:Mars:545GHz:units}{3\parcm80}
\setsymbol{HFI:FWHM:Mars:857GHz:units}{3\parcm67}

\setsymbol{HFI:FWHM:Mars:100GHz}{9.37}
\setsymbol{HFI:FWHM:Mars:143GHz}{7.04}
\setsymbol{HFI:FWHM:Mars:217GHz}{4.68}
\setsymbol{HFI:FWHM:Mars:353GHz}{4.43}
\setsymbol{HFI:FWHM:Mars:545GHz}{3.80}
\setsymbol{HFI:FWHM:Mars:857GHz}{3.67}

\setsymbol{HFI:beam:ellipticity:Mars:100GHz}{1.18}
\setsymbol{HFI:beam:ellipticity:Mars:143GHz}{1.03}
\setsymbol{HFI:beam:ellipticity:Mars:217GHz}{1.14}
\setsymbol{HFI:beam:ellipticity:Mars:353GHz}{1.09}
\setsymbol{HFI:beam:ellipticity:Mars:545GHz}{1.25}
\setsymbol{HFI:beam:ellipticity:Mars:857GHz}{1.03}

\setsymbol{HFI:CMB:relative:calibration:100GHz}{$\lsim 1\%$}
\setsymbol{HFI:CMB:relative:calibration:143GHz}{$\lsim 1\%$}
\setsymbol{HFI:CMB:relative:calibration:217GHz}{$\lsim 1\%$}
\setsymbol{HFI:CMB:relative:calibration:353GHz}{$\lsim 1\%$}
\setsymbol{HFI:CMB:relative:calibration:545GHz}{}
\setsymbol{HFI:CMB:relative:calibration:857GHz}{}

\setsymbol{HFI:CMB:absolute:calibration:100GHz}{$\lsim 2\%$}
\setsymbol{HFI:CMB:absolute:calibration:143GHz}{$\lsim 2\%$}
\setsymbol{HFI:CMB:absolute:calibration:217GHz}{$\lsim 2\%$}
\setsymbol{HFI:CMB:absolute:calibration:353GHz}{$\lsim 2\%$}
\setsymbol{HFI:CMB:absolute:calibration:545GHz}{}
\setsymbol{HFI:CMB:absolute:calibration:857GHz}{}

\setsymbol{HFI:FIRAS:gain:calibration:accuracy:statistical:100GHz}{}
\setsymbol{HFI:FIRAS:gain:calibration:accuracy:statistical:143GHz}{}
\setsymbol{HFI:FIRAS:gain:calibration:accuracy:statistical:217GHz}{}
\setsymbol{HFI:FIRAS:gain:calibration:accuracy:statistical:353GHz}{2.5\%}
\setsymbol{HFI:FIRAS:gain:calibration:accuracy:statistical:545GHz}{1\%}
\setsymbol{HFI:FIRAS:gain:calibration:accuracy:statistical:857GHz}{0.5\%}

\setsymbol{HFI:FIRAS:gain:calibration:accuracy:systematic:100GHz}{}
\setsymbol{HFI:FIRAS:gain:calibration:accuracy:systematic:143GHz}{}
\setsymbol{HFI:FIRAS:gain:calibration:accuracy:systematic:217GHz}{}
\setsymbol{HFI:FIRAS:gain:calibration:accuracy:systematic:353GHz}{}
\setsymbol{HFI:FIRAS:gain:calibration:accuracy:systematic:545GHz}{7\%}
\setsymbol{HFI:FIRAS:gain:calibration:accuracy:systematic:857GHz}{7\%}

\setsymbol{HFI:FIRAS:zero:point:accuracy:100GHz:units}{0.8\MJysr}
\setsymbol{HFI:FIRAS:zero:point:accuracy:143GHz:units}{}
\setsymbol{HFI:FIRAS:zero:point:accuracy:217GHz:units}{}
\setsymbol{HFI:FIRAS:zero:point:accuracy:353GHz:units}{1.4\MJysr}
\setsymbol{HFI:FIRAS:zero:point:accuracy:545GHz:units}{2.2\MJysr}
\setsymbol{HFI:FIRAS:zero:point:accuracy:857GHz:units}{1.7\MJysr}

\setsymbol{HFI:FIRAS:zero:point:accuracy:100GHz}{0.8}
\setsymbol{HFI:FIRAS:zero:point:accuracy:143GHz}{}
\setsymbol{HFI:FIRAS:zero:point:accuracy:217GHz}{}
\setsymbol{HFI:FIRAS:zero:point:accuracy:353GHz}{1.4}
\setsymbol{HFI:FIRAS:zero:point:accuracy:545GHz}{2.2}
\setsymbol{HFI:FIRAS:zero:point:accuracy:857GHz}{1.7}

\setsymbol{HFI:unit:conversion:100GHz:units}{0.2415\MJysrmK}
\setsymbol{HFI:unit:conversion:143GHz:units}{0.3694\MJysrmK}
\setsymbol{HFI:unit:conversion:217GHz:units}{0.4811\MJysrmK}
\setsymbol{HFI:unit:conversion:353GHz:units}{0.2883\MJysrmK}
\setsymbol{HFI:unit:conversion:545GHz:units}{0.05826\MJysrmK}
\setsymbol{HFI:unit:conversion:857GHz:units}{0.002238\MJysrmK}

\setsymbol{HFI:unit:conversion:100GHz}{0.2415}
\setsymbol{HFI:unit:conversion:143GHz}{0.3694}
\setsymbol{HFI:unit:conversion:217GHz}{0.4811}
\setsymbol{HFI:unit:conversion:353GHz}{0.2883}
\setsymbol{HFI:unit:conversion:545GHz}{0.05826}
\setsymbol{HFI:unit:conversion:857GHz}{0.002238}

\setsymbol{HFI:colour:correction:alpha=-2:V1.01:100GHz}{0.9893}
\setsymbol{HFI:colour:correction:alpha=-2:V1.01:143GHz}{0.9759}
\setsymbol{HFI:colour:correction:alpha=-2:V1.01:217GHz}{1.0007}
\setsymbol{HFI:colour:correction:alpha=-2:V1.01:353GHz}{1.0028}
\setsymbol{HFI:colour:correction:alpha=-2:V1.01:545GHz}{1.0019}
\setsymbol{HFI:colour:correction:alpha=-2:V1.01:857GHz}{0.9889}

\setsymbol{HFI:colour:correction:alpha=0:V1.01:100GHz}{1.0008}
\setsymbol{HFI:colour:correction:alpha=0:V1.01:143GHz}{1.0148}
\setsymbol{HFI:colour:correction:alpha=0:V1.01:217GHz}{0.9909}
\setsymbol{HFI:colour:correction:alpha=0:V1.01:353GHz}{0.9888}
\setsymbol{HFI:colour:correction:alpha=0:V1.01:545GHz}{0.9878}
\setsymbol{HFI:colour:correction:alpha=0:V1.01:857GHz}{1.0014}

%% file: PIP_14_authors_and_institutes.tex
\author{\small
Planck Collaboration:
P.~A.~R.~Ade\inst{85}
\and
N.~Aghanim\inst{59}
\and
M.~Arnaud\inst{73}
\and
M.~Ashdown\inst{70, 6}
\and
F.~Atrio-Barandela\inst{19}
\and
J.~Aumont\inst{59}
\and
C.~Baccigalupi\inst{84}
\and
A.~Balbi\inst{37}
\and
A.~J.~Banday\inst{94, 9}
\and
R.~B.~Barreiro\inst{67}
% ...
}
\institute{\small
APC, AstroParticule et Cosmologie, Universit\'{e} Paris Diderot, CNRS/IN2P3, CEA/lrfu, Observatoire de Paris, Sorbonne Paris Cit\'{e}, 10, rue Alice Domon et L\'{e}onie Duquet, 75205 Paris Cedex 13, France\\
\and
Aalto University Mets\"{a}hovi Radio Observatory, Mets\"{a}hovintie 114, FIN-02540 Kylm\"{a}l\"{a}, Finland\\
\and
Academy of Sciences of Tatarstan, Bauman Str., 20, Kazan, 420111, Republic of Tatarstan, Russia\\
\and
Agenzia Spaziale Italiana Science Data Center, c/o ESRIN, via Galileo Galilei, Frascati, Italy\\
\and
Agenzia Spaziale Italiana, Viale Liegi 26, Roma, Italy\\
\and
Astrophysics Group, Cavendish Laboratory, University of Cambridge, J J Thomson Avenue, Cambridge CB3 0HE, U.K.\\
\and
Atacama Large Millimeter/submillimeter Array, ALMA Santiago Central Offices, Alonso de Cordova 3107, Vitacura, Casilla 763 0355, Santiago, Chile\\
\and
CITA, University of Toronto, 60 St. George St., Toronto, ON M5S 3H8, Canada\\
\and
CNRS, IRAP, 9 Av. colonel Roche, BP 44346, F-31028 Toulouse cedex 4, France\\
\and
California Institute of Technology, Pasadena, California, U.S.A.\\
\and
Centre of Mathematics for Applications, University of Oslo, Blindern, Oslo, Norway\\
\and
Centro de Astrof\'{\i}sica, Universidade do Porto, Rua das Estrelas, 4150-762 Porto, Portugal\\
\and
Centro de Estudios de F\'{i}sica del Cosmos de Arag\'{o}n (CEFCA), Plaza San Juan, 1, planta 2, E-44001, Teruel, Spain\\
\and
Computational Cosmology Center, Lawrence Berkeley National Laboratory, Berkeley, California, U.S.A.\\
\and
Consejo Superior de Investigaciones Cient\'{\i}ficas (CSIC), Madrid, Spain\\
\and
DSM/Irfu/SPP, CEA-Saclay, F-91191 Gif-sur-Yvette Cedex, France\\
\and
DTU Space, National Space Institute, Technical University of Denmark, Elektrovej 327, DK-2800 Kgs. Lyngby, Denmark\\
\and
D\'{e}partement de Physique Th\'{e}orique, Universit\'{e} de Gen\`{e}ve, 24, Quai E. Ansermet,1211 Gen\`{e}ve 4, Switzerland\\
\and
Departamento de F\'{\i}sica Fundamental, Facultad de Ciencias, Universidad de Salamanca, 37008 Salamanca, Spain\\
\and
Departamento de F\'{\i}sica, Universidad de Oviedo, Avda. Calvo Sotelo s/n, Oviedo, Spain\\
\and
Department of Astronomy and Geodesy, Kazan Federal University,  Kremlevskaya Str., 18, Kazan, 420008, Russia\\
\and
Department of Astrophysics, IMAPP, Radboud University, P.O. Box 9010, 6500 GL Nijmegen,  The Netherlands\\
\and
Department of Physics \& Astronomy, University of British Columbia, 6224 Agricultural Road, Vancouver, British Columbia, Canada\\
\and
Department of Physics and Astronomy, Dana and David Dornsife College of Letter, Arts and Sciences, University of Southern California, Los Angeles, CA 90089, U.S.A.\\
\and
Department of Physics, Gustaf H\"{a}llstr\"{o}min katu 2a, University of Helsinki, Helsinki, Finland\\
\and
Department of Physics, Princeton University, Princeton, New Jersey, U.S.A.\\
\and
Department of Physics, University of California, Berkeley, California, U.S.A.\\
\and
Department of Physics, University of California, One Shields Avenue, Davis, California, U.S.A.\\
\and
Department of Physics, University of California, Santa Barbara, California, U.S.A.\\
\and
Department of Physics, University of Illinois at Urbana-Champaign, 1110 West Green Street, Urbana, Illinois, U.S.A.\\
\and
Department of Statistics, Purdue University, 250 N. University Street, West Lafayette, Indiana, U.S.A.\\
\and
Dipartimento di Fisica e Astronomia G. Galilei, Universit\`{a} degli Studi di Padova, via Marzolo 8, 35131 Padova, Italy\\
\and
Dipartimento di Fisica, Universit\`{a} La Sapienza, P. le A. Moro 2, Roma, Italy\\
\and
Dipartimento di Fisica, Universit\`{a} degli Studi di Milano, Via Celoria, 16, Milano, Italy\\
\and
Dipartimento di Fisica, Universit\`{a} degli Studi di Trieste, via A. Valerio 2, Trieste, Italy\\
\and
Dipartimento di Fisica, Universit\`{a} di Ferrara, Via Saragat 1, 44122 Ferrara, Italy\\
\and
Dipartimento di Fisica, Universit\`{a} di Roma Tor Vergata, Via della Ricerca Scientifica, 1, Roma, Italy\\
\and
Dipartimento di Matematica, Universit\`{a} di Roma Tor Vergata, Via della Ricerca Scientifica, 1, Roma, Italy\\
\and
Discovery Center, Niels Bohr Institute, Blegdamsvej 17, Copenhagen, Denmark\\
\and
Dpto. Astrof\'{i}sica, Universidad de La Laguna (ULL), E-38206 La Laguna, Tenerife, Spain\\
\and
European Southern Observatory, ESO Vitacura, Alonso de Cordova 3107, Vitacura, Casilla 19001, Santiago, Chile\\
\and
European Space Agency, ESAC, Planck Science Office, Camino bajo del Castillo, s/n, Urbanizaci\'{o}n Villafranca del Castillo, Villanueva de la Ca\~{n}ada, Madrid, Spain\\
\and
European Space Agency, ESTEC, Keplerlaan 1, 2201 AZ Noordwijk, The Netherlands\\
\and
GEPI, Observatoire de Paris, Section de Meudon, 5 Place J. Janssen, 92195 Meudon Cedex, France\\
\and
Helsinki Institute of Physics, Gustaf H\"{a}llstr\"{o}min katu 2, University of Helsinki, Helsinki, Finland\\
\and
INAF - Osservatorio Astronomico di Padova, Vicolo dell'Osservatorio 5, Padova, Italy\\
\and
INAF - Osservatorio Astronomico di Roma, via di Frascati 33, Monte Porzio Catone, Italy\\
\and
INAF - Osservatorio Astronomico di Trieste, Via G.B. Tiepolo 11, Trieste, Italy\\
\and
INAF Istituto di Radioastronomia, Via P. Gobetti 101, 40129 Bologna, Italy\\
\and
INAF/IASF Bologna, Via Gobetti 101, Bologna, Italy\\
\and
INAF/IASF Milano, Via E. Bassini 15, Milano, Italy\\
\and
INFN, Sezione di Roma 1, Universit`{a} di Roma Sapienza, Piazzale Aldo Moro 2, 00185, Roma, Italy\\
\and
IPAG: Institut de Plan\'{e}tologie et d'Astrophysique de Grenoble, Universit\'{e} Joseph Fourier, Grenoble 1 / CNRS-INSU, UMR 5274, Grenoble, F-38041, France\\
\and
IUCAA, Post Bag 4, Ganeshkhind, Pune University Campus, Pune 411 007, India\\
\and
Imperial College London, Astrophysics group, Blackett Laboratory, Prince Consort Road, London, SW7 2AZ, U.K.\\
\and
Infrared Processing and Analysis Center, California Institute of Technology, Pasadena, CA 91125, U.S.A.\\
\and
Institut N\'{e}el, CNRS, Universit\'{e} Joseph Fourier Grenoble I, 25 rue des Martyrs, Grenoble, France\\
\and
Institut Universitaire de France, 103, bd Saint-Michel, 75005, Paris, France\\
\and
Institut d'Astrophysique Spatiale, CNRS (UMR8617) Universit\'{e} Paris-Sud 11, B\^{a}timent 121, Orsay, France\\
\and
Institut d'Astrophysique de Paris, CNRS (UMR7095), 98 bis Boulevard Arago, F-75014, Paris, France\\
\and
Institute for Space Sciences, Bucharest-Magurale, Romania\\
\and
Institute of Astro and Particle Physics, Technikerstrasse 25/8, University of Innsbruck, A-6020, Innsbruck, Austria\\
\and
Institute of Astronomy and Astrophysics, Academia Sinica, Taipei, Taiwan\\
\and
Institute of Astronomy, University of Cambridge, Madingley Road, Cambridge CB3 0HA, U.K.\\
\and
Institute of Theoretical Astrophysics, University of Oslo, Blindern, Oslo, Norway\\
\and
Instituto de Astrof\'{\i}sica de Canarias, C/V\'{\i}a L\'{a}ctea s/n, La Laguna, Tenerife, Spain\\
\and
Instituto de F\'{\i}sica de Cantabria (CSIC-Universidad de Cantabria), Avda. de los Castros s/n, Santander, Spain\\
\and
Jet Propulsion Laboratory, California Institute of Technology, 4800 Oak Grove Drive, Pasadena, California, U.S.A.\\
\and
Jodrell Bank Centre for Astrophysics, Alan Turing Building, School of Physics and Astronomy, The University of Manchester, Oxford Road, Manchester, M13 9PL, U.K.\\
\and
Kavli Institute for Cosmology Cambridge, Madingley Road, Cambridge, CB3 0HA, U.K.\\
\and
LAL, Universit\'{e} Paris-Sud, CNRS/IN2P3, Orsay, France\\
\and
LERMA, CNRS, Observatoire de Paris, 61 Avenue de l'Observatoire, Paris, France\\
\and
Laboratoire AIM, IRFU/Service d'Astrophysique - CEA/DSM - CNRS - Universit\'{e} Paris Diderot, B\^{a}t. 709, CEA-Saclay, F-91191 Gif-sur-Yvette Cedex, France\\
\and
Laboratoire Traitement et Communication de l'Information, CNRS (UMR 5141) and T\'{e}l\'{e}com ParisTech, 46 rue Barrault F-75634 Paris Cedex 13, France\\
\and
Laboratoire de Physique Subatomique et de Cosmologie, Universit\'{e} Joseph Fourier Grenoble I, CNRS/IN2P3, Institut National Polytechnique de Grenoble, 53 rue des Martyrs, 38026 Grenoble cedex, France\\
\and
Laboratoire de Physique Th\'{e}orique, Universit\'{e} Paris-Sud 11 \& CNRS, B\^{a}timent 210, 91405 Orsay, France\\
\and
Lawrence Berkeley National Laboratory, Berkeley, California, U.S.A.\\
\and
Max-Planck-Institut f\"{u}r Astrophysik, Karl-Schwarzschild-Str. 1, 85741 Garching, Germany\\
\and
Max-Planck-Institut f\"{u}r Extraterrestrische Physik, Giessenbachstra{\ss}e, 85748 Garching, Germany\\
\and
National University of Ireland, Department of Experimental Physics, Maynooth, Co. Kildare, Ireland\\
\and
Niels Bohr Institute, Blegdamsvej 17, Copenhagen, Denmark\\
\and
Observational Cosmology, Mail Stop 367-17, California Institute of Technology, Pasadena, CA, 91125, U.S.A.\\
\and
Optical Science Laboratory, University College London, Gower Street, London, U.K.\\
\and
SISSA, Astrophysics Sector, via Bonomea 265, 34136, Trieste, Italy\\
\and
School of Physics and Astronomy, Cardiff University, Queens Buildings, The Parade, Cardiff, CF24 3AA, U.K.\\
\and
Space Research Institute (IKI), Profsoyuznaya 84/32, Moscow, Russia\\
\and
Space Research Institute (IKI), Russian Academy of Sciences, Profsoyuznaya Str, 84/32, Moscow, 117997, Russia\\
\and
Space Sciences Laboratory, University of California, Berkeley, California, U.S.A.\\
\and
Stanford University, Dept of Physics, Varian Physics Bldg, 382 Via Pueblo Mall, Stanford, California, U.S.A.\\
\and
T\"{U}B\.{I}TAK National Observatory, Akdeniz University Campus, 07058, Antalya, Turkey\\
\and
UPMC Univ Paris 06, UMR7095, 98 bis Boulevard Arago, F-75014, Paris, France\\
\and
Universit\"{a}t Heidelberg, Institut f\"{u}r Theoretische Astrophysik, Albert-\"{U}berle-Str. 2, 69120, Heidelberg, Germany\\
\and
Universit\'{e} Denis Diderot (Paris 7), 75205 Paris Cedex 13, France\\
\and
Universit\'{e} de Toulouse, UPS-OMP, IRAP, F-31028 Toulouse cedex 4, France\\
% ...
}

%% file: WHIM_draftFinal.bbl
\begin{thebibliography}{66}
\expandafter\ifx\csname natexlab\endcsname\relax\def\natexlab#1{#1}\fi

\bibitem[{{Akahori} \& {Yoshikawa}(2008)}]{Akahori2008}
{Akahori}, T. \& {Yoshikawa}, K. 2008, PASJ, 60, 19

\bibitem[{{Andersson} \& {SPT Collaboration}(2010)}]{Andersson10}
{Andersson}, K. \& {SPT Collaboration}. 2010, in AAS/High Energy Astrophysics
  Division, Vol.~11, AAS/High Energy Astrophysics Division 11, 27.04

\bibitem[{{Arnaud}(1996)}]{Arnaud96}
{Arnaud}, K.~A. 1996, in Astronomical Society of the Pacific Conference Series,
  Vol. 101, Astronomical Data Analysis Software and Systems V, ed.
  {G.~H.~Jacoby \& J.~Barnes}, 17

\bibitem[{{Arnaud} {et~al.}(2010){Arnaud}, {Pratt}, {Piffaretti},
  {B{\"o}hringer}, {Croston}, \& {Pointecouteau}}]{Arnaud2010}
{Arnaud}, M., {Pratt}, G.~W., {Piffaretti}, R., {et~al.} 2010, \aap, 517, A92

\bibitem[{{Bersanelli} {et~al.}(2010){Bersanelli}, {Mandolesi}, {Butler},
  {Mennella}, {Villa}, {Aja}, {Artal}, {Artina}, {Baccigalupi}, {Balasini},
  {Baldan}, {Banday}, {Bastia}, {Battaglia}, {Bernardino}, {Blackhurst},
  {Boschini}, {Burigana}, {Cafagna}, {Cappellini}, {Cavaliere}, {Colombo},
  {Crone}, {Cuttaia}, {D'Arcangelo}, {Danese}, {Davies}, {Davis}, {de Angelis},
  {de Gasperis}, {de La Fuente}, {de Rosa}, {de Zotti}, {Falvella}, {Ferrari},
  {Ferretti}, {Figini}, {Fogliani}, {Franceschet}, {Franceschi}, {Gaier},
  {Garavaglia}, {Gomez}, {Gorski}, {Gregorio}, {Guzzi}, {Herreros},
  {Hildebrandt}, {Hoyland}, {Hughes}, {Janssen}, {Jukkala}, {Kettle},
  {Kilpi{\"a}}, {Laaninen}, {Lapolla}, {Lawrence}, {Lawson}, {Leahy},
  {Leonardi}, {Leutenegger}, {Levin}, {Lilje}, {Lowe}, {Lubin}, {Maino},
  {Malaspina}, {Maris}, {Marti-Canales}, {Martinez-Gonzalez}, {Mediavilla},
  {Meinhold}, {Miccolis}, {Morgante}, {Natoli}, {Nesti}, {Pagan}, {Paine},
  {Partridge}, {Pascual}, {Pasian}, {Pearson}, {Pecora}, {Perrotta},
  {Platania}, {Pospieszalski}, {Poutanen}, {Prina}, {Rebolo}, {Roddis},
  {Rubi{\~n}o-Martin}, {Salmon}, {Sandri}, {Seiffert}, {Silvestri},
  {Simonetto}, {Sjoman}, {Smoot}, {Sozzi}, {Stringhetti}, {Taddei}, {Tauber},
  {Terenzi}, {Tomasi}, {Tuovinen}, {Valenziano}, {Varis}, {Vittorio}, {Wade},
  {Wilkinson}, {Winder}, {Zacchei}, \& {Zonca}}]{Bersanelli2010}
{Bersanelli}, M., {Mandolesi}, N., {Butler}, R.~C., {et~al.} 2010, \aap, 520,
  A4+

\bibitem[{{Bertin} \& {Arnouts}(1996)}]{sextractor}
{Bertin}, E. \& {Arnouts}, S. 1996, \aaps, 117, 393

\bibitem[{{Bobin} {et~al.}(2008){Bobin}, {Moudden}, {Starck}, {Fadili}, \&
  {Aghanim}}]{gmca}
{Bobin}, J., {Moudden}, Y., {Starck}, J.-L., {Fadili}, J., \& {Aghanim}, N.
  2008, Statistical Methodology, 5, 307

\bibitem[{{Cen} \& {Ostriker}(1999)}]{Cen99}
{Cen}, R. \& {Ostriker}, J. 1999, ApJ, 514, 1

\bibitem[{{Dickey} \& {Lockman}(1990)}]{Dickey1990}
{Dickey}, J.~M. \& {Lockman}, F.~J. 1990, \araa, 28, 215

\bibitem[{{Diego} \& {Partridge}(2010)}]{diego2010}
{Diego}, J.~M. \& {Partridge}, B. 2010, \mnras, 402, 1179

\bibitem[{{Dolag} {et~al.}(2006){Dolag}, {Meneghetti}, {Moscardini}, {Rasia},
  \& {Bonaldi}}]{dolag2006}
{Dolag}, K., {Meneghetti}, M., {Moscardini}, L., {Rasia}, E., \& {Bonaldi}, A.
  2006, \mnras, 370, 656

\bibitem[{{Durret} {et~al.}(2011){Durret}, {Lagan{\'a}}, \& {Haider}}]{Durret}
{Durret}, F., {Lagan{\'a}}, T.~F., \& {Haider}, M. 2011, \aap, 529, A38

\bibitem[{{Eckert} {et~al.}(2011){Eckert}, {Vazza}, {Ettori}, {Molendi},
  {Nagai}, {Lau}, {Roncarelli}, {Rossetti}, {Snowden}, \&
  {Gastaldello}}]{eckert12}
{Eckert}, D., {Vazza}, F., {Ettori}, S., {et~al.} 2011, \aap in press

\bibitem[{{Fabian} {et~al.}(1997){Fabian}, {Peres}, \& {White}}]{Fabian1997}
{Fabian}, A., {Peres}, C., \& {White}, D. 1997, MNRAS, 285, 35

\bibitem[{{Flores-Cacho} {et~al.}(2009){Flores-Cacho},
  {Rubi{\~n}o-Mart{\'{\i}}n}, {Luzzi}, {Rebolo}, {de Petris}, {Yepes},
  {Lamagna}, {de Gregori}, {Battistelli}, {Coratella}, \&
  {Gottl{\"o}ber}}]{cacho2009}
{Flores-Cacho}, I., {Rubi{\~n}o-Mart{\'{\i}}n}, J.~A., {Luzzi}, G., {et~al.}
  2009, \mnras, 400, 1868

\bibitem[{{Fujita} {et~al.}(1996){Fujita}, {Koyama}, {Tsuru}, \&
  {Matsumoto}}]{Fujita1996}
{Fujita}, Y., {Koyama}, K., {Tsuru}, T., \& {Matsumoto}, H. 1996, PASJ, 48, 191

\bibitem[{{Fujita} {et~al.}(2008){Fujita}, {Tawa}, {Hayashida}, {Takizawa},
  {Matsumoto}, {Okabe}, \& {Reiprich}}]{Fujita2008}
{Fujita}, Y., {Tawa}, N., {Hayashida}, K., {et~al.} 2008, PASJ, 60, 343

\bibitem[{{G{\'o}rski} {et~al.}(2005){G{\'o}rski}, {Hivon}, {Banday},
  {Wandelt}, {Hansen}, {Reinecke}, \& {Bartelmann}}]{healpix}
{G{\'o}rski}, K.~M., {Hivon}, E., {Banday}, A.~J., {et~al.} 2005, \apj, 622,
  759

\bibitem[{{Hallman} {et~al.}(2007){Hallman}, {Burns}, {Motl}, \&
  {Norman}}]{Hallman07}
{Hallman}, E.~J., {Burns}, J.~O., {Motl}, P.~M., \& {Norman}, M.~L. 2007, \apj,
  665, 911

\bibitem[{{Harrison} \& {Coles}(2012)}]{HarrisonColes}
{Harrison}, I. \& {Coles}, P. 2012, \mnras, 421, L19

\bibitem[{{Hurier} {et~al.}(2010){Hurier}, {Hildebrandt}, \&
  {Macias-Perez}}]{milca}
{Hurier}, G., {Hildebrandt}, S.~R., \& {Macias-Perez}, J.~F. 2010, ArXiv
  e-prints

\bibitem[{{Kaastra}(1992)}]{kaas92}
{Kaastra}, J.~S. 1992, An X-ray spectral Code for Optically Thin Plasmas

\bibitem[{{Kalberla} {et~al.}(2005){Kalberla}, {Burton}, B., {Hartmann},
  {Arnal}, {Bajaja}, {Morras}, \& {P{\"o}ppel}}]{Kalberla2005}
{Kalberla}, P.~M.~W., {Burton}, B., W., {et~al.} 2005, \aap, 440, 775

\bibitem[{{Kravtsov} {et~al.}(2006){Kravtsov}, {Vikhlinin}, \&
  {Nagai}}]{Kravtsov06}
{Kravtsov}, A.~V., {Vikhlinin}, A., \& {Nagai}, D. 2006, The Astrophysical
  Journal, 650, 128

\bibitem[{{Kull} \& {B{\"o}hringer}(1999)}]{kull99}
{Kull}, A. \& {B{\"o}hringer}, H. 1999, \aap, 341, 23

\bibitem[{{Lamarre} {et~al.}(2010){Lamarre}, {Puget}, {Ade}, {Bouchet},
  {Guyot}, {Lange}, {Pajot}, {Arondel}, {Benabed}, {Beney}, {Beno{\^i}t},
  {Bernard}, {Bhatia}, {Blanc}, {Bock}, {Br{\'e}elle}, {Bradshaw}, {Camus},
  {Catalano}, {Charra}, {Charra}, {Church}, {Couchot}, {Coulais}, {Crill},
  {Crook}, {Dassas}, {de Bernardis}, {Delabrouille}, {de Marcillac}, {Delouis},
  {D{\'e}sert}, {Dumesnil}, {Dupac}, {Efstathiou}, {Eng}, {Evesque},
  {Fourmond}, {Ganga}, {Giard}, {Gispert}, {Guglielmi}, {Haissinski},
  {Henrot-Versill{\'e}}, {Hivon}, {Holmes}, {Jones}, {Koch}, {Lagard{\`e}re},
  {Lami}, {Land{\'e}}, {Leriche}, {Leroy}, {Longval},
  {Mac{\'{\i}}as-P{\'e}rez}, {Maciaszek}, {Maffei}, {Mansoux}, {Marty}, {Masi},
  {Mercier}, {Miville-Desch{\^e}nes}, {Moneti}, {Montier}, {Murphy},
  {Narbonne}, {Nexon}, {Paine}, {Pahn}, {Perdereau}, {Piacentini}, {Piat},
  {Plaszczynski}, {Pointecouteau}, {Pons}, {Ponthieu}, {Prunet}, {Rambaud},
  {Recouvreur}, {Renault}, {Ristorcelli}, {Rosset}, {Santos}, {Savini},
  {Serra}, {Stassi}, {Sudiwala}, {Sygnet}, {Tauber}, {Torre}, {Tristram},
  {Vibert}, {Woodcraft}, {Yurchenko}, \& {Yvon}}]{Lamarre2010}
{Lamarre}, J., {Puget}, J., {Ade}, P.~A.~R., {et~al.} 2010, \aap, 520, A9+

\bibitem[{{Leahy} {et~al.}(2010){Leahy}, {Bersanelli}, {D'Arcangelo}, {Ganga},
  {Leach}, {Moss}, {Keih{\"a}nen}, {Keskitalo}, {Kurki-Suonio}, {Poutanen},
  {Sandri}, {Scott}, {Tauber}, {Valenziano}, {Villa}, {Wilkinson}, {Zonca},
  {Baccigalupi}, {Borrill}, {Butler}, {Cuttaia}, {Davis}, {Frailis},
  {Francheschi}, {Galeotta}, {Gregorio}, {Leonardi}, {Mandolesi}, {Maris},
  {Meinhold}, {Mendes}, {Mennella}, {Morgante}, {Prezeau}, {Rocha},
  {Stringhetti}, {Terenzi}, \& {Tomasi}}]{Leahy2010}
{Leahy}, J.~P., {Bersanelli}, M., {D'Arcangelo}, O., {et~al.} 2010, \aap, 520,
  A8+

\bibitem[{{Liedahl} {et~al.}(1995){Liedahl}, {Osterheld}, \&
  {Goldstein}}]{lied95}
{Liedahl}, D.~A., {Osterheld}, A.~L., \& {Goldstein}, W.~H. 1995, \apj, 438,
  L115

\bibitem[{{Mandolesi} {et~al.}(2010){Mandolesi}, {Bersanelli}, {Butler},
  {Artal}, {Baccigalupi}, {Balbi}, {Banday}, {Barreiro}, {Bartelmann},
  {Bennett}, {Bhandari}, {Bonaldi}, {Borrill}, {Bremer}, {Burigana}, {Bowman},
  {Cabella}, {Cantalupo}, {Cappellini}, {Courvoisier}, {Crone}, {Cuttaia},
  {Danese}, {D'Arcangelo}, {Davies}, {Davis}, {de Angelis}, {de Gasperis}, {de
  Rosa}, {de Troia}, {de Zotti}, {Dick}, {Dickinson}, {Diego}, {Donzelli},
  {D{\"o}rl}, {Dupac}, {En{\ss}lin}, {Eriksen}, {Falvella}, {Finelli},
  {Frailis}, {Franceschi}, {Gaier}, {Galeotta}, {Gasparo}, {Giardino}, {Gomez},
  {Gonzalez-Nuevo}, {G{\'o}rski}, {Gregorio}, {Gruppuso}, {Hansen}, {Hell},
  {Herranz}, {Herreros}, {Hildebrandt}, {Hovest}, {Hoyland}, {Huffenberger},
  {Janssen}, {Jaffe}, {Keih{\"a}nen}, {Keskitalo}, {Kisner}, {Kurki-Suonio},
  {L{\"a}hteenm{\"a}ki}, {Lawrence}, {Leach}, {Leahy}, {Leonardi}, {Levin},
  {Lilje}, {L{\'o}pez-Caniego}, {Lowe}, {Lubin}, {Maino}, {Malaspina}, {Maris},
  {Marti-Canales}, {Martinez-Gonzalez}, {Massardi}, {Matarrese}, {Matthai},
  {Meinhold}, {Melchiorri}, {Mendes}, {Mennella}, {Morgante}, {Morigi},
  {Morisset}, {Moss}, {Nash}, {Natoli}, {Nesti}, {Paine}, {Partridge},
  {Pasian}, {Passvogel}, {Pearson}, {P{\'e}rez-Cuevas}, {Perrotta}, {Polenta},
  {Popa}, {Poutanen}, {Prezeau}, {Prina}, {Rachen}, {Rebolo}, {Reinecke},
  {Ricciardi}, {Riller}, {Rocha}, {Roddis}, {Rohlfs}, {Rubi{\~n}o-Martin},
  {Salerno}, {Sandri}, {Scott}, {Seiffert}, {Silk}, {Simonetto}, {Smoot},
  {Sozzi}, {Sternberg}, {Stivoli}, {Stringhetti}, {Tauber}, {Terenzi},
  {Tomasi}, {Tuovinen}, {T{\"u}rler}, {Valenziano}, {Varis}, {Vielva}, {Villa},
  {Vittorio}, {Wade}, {White}, {White}, {Wilkinson}, {Zacchei}, \&
  {Zonca}}]{Mandolesi2010}
{Mandolesi}, N., {Bersanelli}, M., {Butler}, R.~C., {et~al.} 2010, \aap, 520,
  A3+

\bibitem[{{Mathiesen} {et~al.}(1999){Mathiesen}, {Evrard}, \&
  {Mohr}}]{1999ApJ...520L..21M}
{Mathiesen}, B., {Evrard}, A.~E., \& {Mohr}, J.~J. 1999, \apjl, 520, L21

\bibitem[{{Melin} {et~al.}(2012){Melin}, {Aghanim}, {Bartelmann}, {Bartlett},
  {Betoule}, {Bobin}, {Carvalho}, {Chon}, {Delabrouille}, {Diego}, {Harrison},
  {Herranz}, {Hobson}, {Kneissl}, {Lasenby}, {Le Jeune}, {Lopez-Caniego},
  {Mazzotta}, {Rocha}, {Schaefer}, {Starck}, {Waizmann}, \& {Yvon}}]{Melin2012}
{Melin}, J.-B., {Aghanim}, N., {Bartelmann}, M., {et~al.} 2012, ArXiv e-prints

\bibitem[{{Mennella et al.}(2011)}]{Planck2011-1.4}
{Mennella et al.} 2011, \aap, 536, A3

\bibitem[{{Mewe} {et~al.}(1985){Mewe}, {Gronenschild}, \& {van den
  Oord}}]{mewe85}
{Mewe}, R., {Gronenschild}, E. H. B.~M., \& {van den Oord}, G. H.~J. 1985,
  \aaps, 62, 197

\bibitem[{{Mewe} {et~al.}(1986){Mewe}, {Lemen}, R., \& {van den Oord}}]{mewe86}
{Mewe}, R., {Lemen}, R., J., \& {van den Oord}, G. H.~J. 1986, \aaps, 65, 511

\bibitem[{{Morrison} \& {McCammon}(1983)}]{morr83}
{Morrison}, R. \& {McCammon}, D. 1983, \apj, 270, 119

\bibitem[{{Murgia} {et~al.}(2010){Murgia}, {Govoni}, {Feretti}, \&
  {Giovannini}}]{Murgia2010}
{Murgia}, M., {Govoni}, F., {Feretti}, L., \& {Giovannini}, G. 2010, A\&A, 509,
  86

\bibitem[{{Nagai} {et~al.}(2007){Nagai}, {Kravtsov}, \&
  {Vikhlinin}}]{Nagai2007}
{Nagai}, D., {Kravtsov}, A.~V., \& {Vikhlinin}, A. 2007, \apj, 668, 1

\bibitem[{{Piffaretti} {et~al.}(2011){Piffaretti}, {Arnaud}, {Pratt},
  {Pointecouteau}, \& {Melin}}]{Piffaretti2011}
{Piffaretti}, R., {Arnaud}, M., {Pratt}, G.~W., {Pointecouteau}, E., \&
  {Melin}, J.-B. 2011, \aap, 534, A109

\bibitem[{{Planck Collaboration} {et~al.}(2012){Planck Collaboration}, {Ade},
  {Aghanim}, {Arnaud}, {Ashdown}, {Atrio-Barandela}, {Aumont}, {Baccigalupi},
  {Balbi}, {Banday}, \& et~al.}]{ppp_paper}
{Planck Collaboration}, {Ade}, P.~A.~R., {Aghanim}, N., {et~al.} 2012,
  2012arXiv1207.4061P

\bibitem[{{Planck Collaboration} {et~al.}(2011{\natexlab{a}}){Planck
  Collaboration}, {Ade}, {Aghanim}, {Arnaud}, {Ashdown}, {Aumont},
  {Baccigalupi}, {Baker}, {Balbi}, {Banday}, \& et~al.}]{planckmission}
{Planck Collaboration}, {Ade}, P.~A.~R., {Aghanim}, N., {et~al.}
  2011{\natexlab{a}}, \aap, 536, A1

\bibitem[{{Planck Collaboration} {et~al.}(2011{\natexlab{b}}){Planck
  Collaboration}, {Ade}, {Aghanim}, {Arnaud}, {Ashdown}, {Aumont},
  {Baccigalupi}, {Balbi}, {Banday}, {Barreiro}, \&
  et~al.}]{2011A&A...536A...8P}
{Planck Collaboration}, {Ade}, P.~A.~R., {Aghanim}, N., {et~al.}
  2011{\natexlab{b}}, \aap, 536, A8

\bibitem[{{Planck Collaboration} {et~al.}(2011{\natexlab{c}}){Planck
  Collaboration}, {Aghanim}, {Arnaud}, {Ashdown}, {Aumont}, {Baccigalupi},
  {Balbi}, {Banday}, {Barreiro}, {Bartelmann}, {Bartlett}, {Battaner},
  {Benabed}, {Beno{\^i}t}, {Bernard}, {Bersanelli}, {Bhatia}, {Bock},
  {Bonaldi}, {Bond}, {Borrill}, {Bouchet}, {Brown}, {Bucher}, {Burigana},
  {Cabella}, {Cardoso}, {Catalano}, {Cay{\'o}n}, {Challinor}, {Chamballu},
  {Chary}, {Chiang}, {Chiang}, {Chon}, {Christensen}, {Churazov}, {Clements},
  {Colafrancesco}, {Colombi}, {Couchot}, {Coulais}, {Crill}, {Cuttaia}, {da
  Silva}, {Dahle}, {Danese}, {de Bernardis}, {de Gasperis}, {de Rosa}, {de
  Zotti}, {Delabrouille}, {Delouis}, {D{\'e}sert}, {Diego}, {Dolag},
  {Donzelli}, {Dor{\'e}}, {D{\"o}rl}, {Douspis}, {Dupac}, {Efstathiou},
  {En{\ss}lin}, {Finelli}, {Flores-Cacho}, {Forni}, {Frailis}, {Franceschi},
  {Fromenteau}, {Galeotta}, {Ganga}, {G{\'e}nova-Santos}, {Giard}, {Giardino},
  {Giraud-H{\'e}raud}, {Gonz{\'a}lez-Nuevo}, {G{\'o}rski}, {Gratton},
  {Gregorio}, {Gruppuso}, {Harrison}, {Henrot-Versill{\'e}},
  {Hern{\'a}ndez-Monteagudo}, {Herranz}, {Hildebrandt}, {Hivon}, {Hobson},
  {Holmes}, {Hovest}, {Hoyland}, {Huffenberger}, {Jaffe}, {Jones}, {Juvela},
  {Keih{\"a}nen}, {Keskitalo}, {Kisner}, {Kneissl}, {Knox}, {Kurki-Suonio},
  {Lagache}, {Lamarre}, {Lasenby}, {Laureijs}, {Lawrence}, {Leach}, {Leonardi},
  {Linden-V{\o}rnle}, {L{\'o}pez-Caniego}, {Lubin}, {Mac{\'{\i}}as-P{\'e}rez},
  {MacTavish}, {Maffei}, {Maino}, {Mandolesi}, {Mann}, {Maris}, {Marleau},
  {Mart{\'{\i}}nez-Gonz{\'a}lez}, {Masi}, {Matarrese}, {Matthai}, {Mazzotta},
  {Melchiorri}, {Melin}, {Mendes}, {Mennella}, {Mitra},
  {Miville-Desch{\^e}nes}, {Moneti}, {Montier}, {Morgante}, {Mortlock},
  {Munshi}, {Murphy}, {Naselsky}, {Natoli}, {Netterfield},
  {N{\o}rgaard-Nielsen}, {Noviello}, {Novikov}, {Novikov}, {Osborne}, {Pajot},
  {Pasian}, {Patanchon}, {Perdereau}, {Perotto}, {Perrotta}, {Piacentini},
  {Piat}, {Pierpaoli}, {Piffaretti}, {Plaszczynski}, {Pointecouteau},
  {Polenta}, {Ponthieu}, {Poutanen}, {Pratt}, {Pr{\'e}zeau}, {Prunet}, {Puget},
  {Rebolo}, {Reinecke}, {Renault}, {Ricciardi}, {Riller}, {Ristorcelli},
  {Rocha}, {Rosset}, {Rubi{\~n}o-Mart{\'{\i}}n}, {Rusholme}, {Sandri},
  {Santos}, {Schaefer}, {Scott}, {Seiffert}, {Smoot}, {Starck}, {Stivoli},
  {Stolyarov}, {Sunyaev}, {Sygnet}, {Tauber}, {Terenzi}, {Toffolatti},
  {Tomasi}, {Tristram}, {Tuovinen}, {Valenziano}, {Vibert}, {Vielva}, {Villa},
  {Vittorio}, {Wandelt}, {White}, {White}, {Yvon}, {Zacchei}, \&
  {Zonca}}]{2011A&A...536A..10P}
{Planck Collaboration}, {Aghanim}, N., {Arnaud}, M., {et~al.}
  2011{\natexlab{c}}, \aap, 536, A10

\bibitem[{{Planck Collaboration} {et~al.}(2011{\natexlab{d}}){Planck
  Collaboration}, {Aghanim}, {Arnaud}, {Ashdown}, {Aumont}, {Baccigalupi},
  {Balbi}, {Banday}, {Barreiro}, {Bartelmann}, \& et~al.}]{2011A&A...536A...9P}
{Planck Collaboration}, {Aghanim}, N., {Arnaud}, M., {et~al.}
  2011{\natexlab{d}}, \aap, 536, A9

\bibitem[{{Planck Collaboration I}(2011)}]{Planck2011-1.1}
{Planck Collaboration I}. 2011, \aap, 536, A1

\bibitem[{{Planck Collaboration II}(2011)}]{Planck2011-1.3}
{Planck Collaboration II}. 2011, \aap, 536, A2

\bibitem[{{Planck HFI Core Team}(2011{\natexlab{a}})}]{Planck2011-1.5}
{Planck HFI Core Team}. 2011{\natexlab{a}}, \aap, 536, A4

\bibitem[{{Planck HFI Core Team}(2011{\natexlab{b}})}]{Planck2011-1.7}
{Planck HFI Core Team}. 2011{\natexlab{b}}, \aap, 536, A6

\bibitem[{{Planck HFI Core Team} {et~al.}(2011){Planck HFI Core Team}, {Ade},
  {Aghanim}, {Ansari}, {Arnaud}, {Ashdown}, {Aumont}, {Banday}, {Bartelmann},
  {Bartlett}, {Battaner}, {Benabed}, {Beno{\^i}t}, {Bernard}, {Bersanelli},
  {Bock}, {Bond}, {Borrill}, {Bouchet}, {Boulanger}, {Bradshaw}, {Bucher},
  {Cardoso}, {Castex}, {Catalano}, {Challinor}, {Chamballu}, {Chary}, {Chen},
  {Chiang}, {Church}, {Clements}, {Colley}, {Colombi}, {Couchot}, {Coulais},
  {Cressiot}, {Crill}, {Crook}, {de Bernardis}, {Delabrouille}, {Delouis},
  {D{\'e}sert}, {Dolag}, {Dole}, {Dor{\'e}}, {Douspis}, {Dunkley},
  {Efstathiou}, {Filliard}, {Forni}, {Fosalba}, {Ganga}, {Giard}, {Girard},
  {Giraud-H{\'e}raud}, {Gispert}, {G{\'o}rski}, {Gratton}, {Griffin}, {Guyot},
  {Haissinski}, {Harrison}, {Helou}, {Henrot-Versill{\'e}},
  {Hern{\'a}ndez-Monteagudo}, {Hildebrandt}, {Hills}, {Hivon}, {Hobson},
  {Holmes}, {Huffenberger}, {Jaffe}, {Jones}, {Kaplan}, {Kneissl}, {Knox},
  {Kunz}, {Lagache}, {Lamarre}, {Lange}, {Lasenby}, {Lavabre}, {Lawrence}, {Le
  Jeune}, {Leroy}, {Lesgourgues}, {Mac{\'{\i}}as-P{\'e}rez}, {MacTavish},
  {Maffei}, {Mandolesi}, {Mann}, {Marleau}, {Marshall}, {Masi}, {Matsumura},
  {McAuley}, {McGehee}, {Melin}, {Mercier}, {Mitra}, {Miville-Desch{\^e}nes},
  {Moneti}, {Montier}, {Mortlock}, {Murphy}, {Nati}, {Netterfield},
  {N{\o}rgaard-Nielsen}, {North}, {Noviello}, {Novikov}, {Osborne}, {Pajot},
  {Patanchon}, {Peacocke}, {Pearson}, {Perdereau}, {Perotto}, {Piacentini},
  {Piat}, {Plaszczynski}, {Pointecouteau}, {Ponthieu}, {Pr{\'e}zeau}, {Prunet},
  {Puget}, {Reach}, {Remazeilles}, {Renault}, {Riazuelo}, {Ristorcelli},
  {Rocha}, {Rosset}, {Roudier}, {Rowan-Robinson}, {Rusholme}, {Saha}, {Santos},
  {Savini}, {Schaefer}, {Shellard}, {Spencer}, {Starck}, {Stolyarov},
  {Stompor}, {Sudiwala}, {Sunyaev}, {Sutton}, {Sygnet}, {Tauber}, {Thum},
  {Torre}, {Touze}, {Tristram}, {van Leeuwen}, {Vibert}, {Vibert}, {Wade},
  {Wandelt}, {White}, {Wiesemeyer}, {Woodcraft}, {Yurchenko}, {Yvon}, \&
  {Zacchei}}]{2011A&A...536A...6P}
{Planck HFI Core Team}, {Ade}, P.~A.~R., {Aghanim}, N., {et~al.} 2011, \aap,
  536, A6

\bibitem[{{Remazeilles} {et~al.}(2011){Remazeilles}, {Delabrouille}, \&
  {Cardoso}}]{nilc}
{Remazeilles}, M., {Delabrouille}, J., \& {Cardoso}, J.-F. 2011, \mnras, 410,
  2481

\bibitem[{{Richter} {et~al.}(2008){Richter}, {Paerels}, \&
  {Kaastra}}]{Richter08}
{Richter}, P., {Paerels}, F., \& {Kaastra}, J. 2008, Space Science Reviews,
  134, 25

\bibitem[{{Rosset} {et~al.}(2010){Rosset}, {Tristram}, {Ponthieu}, {Ade},
  {Aumont}, {Catalano}, {Conversi}, {Couchot}, {Crill}, {D{\'e}sert}, {Ganga},
  {Giard}, {Giraud-H{\'e}raud}, {Ha{\"i}ssinski}, {Henrot-Versill{\'e}},
  {Holmes}, {Jones}, {Lamarre}, {Lange}, {Leroy}, {Mac{\'{\i}}as-P{\'e}rez},
  {Maffei}, {de Marcillac}, {Miville-Desch{\^e}nes}, {Montier}, {Noviello},
  {Pajot}, {Perdereau}, {Piacentini}, {Piat}, {Plaszczynski}, {Pointecouteau},
  {Puget}, {Ristorcelli}, {Savini}, {Sudiwala}, {Veneziani}, \&
  {Yvon}}]{Rosset2010}
{Rosset}, C., {Tristram}, M., {Ponthieu}, N., {et~al.} 2010, \aap, 520, A13+

\bibitem[{{Rozo} {et~al.}(2012){Rozo}, {Vikhlinin}, \& {More}}]{Rozo12}
{Rozo}, E., {Vikhlinin}, A., \& {More}, S. 2012, ArXiv e-prints

\bibitem[{{Sakelliou} \& {Ponman}(2004)}]{Sakelliou2004}
{Sakelliou}, I. \& {Ponman}, T. 2004, MNRAS, 351, 1439

\bibitem[{{Salpeter}(1955)}]{1955ApJ...121..161S}
{Salpeter}, E.~E. 1955, \apj, 121, 161

\bibitem[{{Sereno} {et~al.}(2006){Sereno}, {De Filippis}, {Longo}, \&
  {Bautz}}]{sereno2006}
{Sereno}, M., {De Filippis}, E., {Longo}, G., \& {Bautz}, M.~W. 2006, \apj,
  645, 170

\bibitem[{{Snowden} {et~al.}(1994){Snowden}, {McCammon}, {Burrows}, \&
  {Mendenhall}}]{snowden94}
{Snowden}, S.~L., {McCammon}, D., {Burrows}, D.~N., \& {Mendenhall}, J.~A.
  1994, \apj, 424, 714

\bibitem[{{Springel} \& {Hernquist}(2003)}]{Springel2003}
{Springel}, V. \& {Hernquist}, L. 2003, \mnras, 339, 289

\bibitem[{{Sunyaev} \& {Zeldovich}(1972)}]{SZ}
{Sunyaev}, R.~A. \& {Zeldovich}, Y.~B. 1972, Comments on Astrophysics and Space
  Physics, 4, 173

\bibitem[{{Tauber} {et~al.}(2010){Tauber}, {Mandolesi}, {Puget}, {Banos},
  {Bersanelli}, {Bouchet}, {Butler}, {Charra}, {Crone}, {Dodsworth}, \&
  et~al.}]{tauber2010a}
{Tauber}, J.~A., {Mandolesi}, N., {Puget}, J., {et~al.} 2010, \aap, 520, A1+

\bibitem[{{Tittley} \& {Henriksen}(2001)}]{Tittley2001}
{Tittley}, E.~R. \& {Henriksen}, M. 2001, \apj, 563, 673

\bibitem[{{Tornatore} {et~al.}(2007){Tornatore}, {Borgani}, {Dolag}, \&
  {Matteucci}}]{2007MNRAS.382.1050T}
{Tornatore}, L., {Borgani}, S., {Dolag}, K., \& {Matteucci}, F. 2007, \mnras,
  382, 1050

\bibitem[{{Tornatore} {et~al.}(2004){Tornatore}, {Borgani}, {Matteucci},
  {Recchi}, \& {Tozzi}}]{2004MNRAS.349L..19T}
{Tornatore}, L., {Borgani}, S., {Matteucci}, F., {Recchi}, S., \& {Tozzi}, P.
  2004, \mnras, 349, L19

\bibitem[{{Vikhlinin} {et~al.}(1999){Vikhlinin}, {Forman}, \&
  {Jones}}]{vikhlinin99}
{Vikhlinin}, A., {Forman}, W., \& {Jones}, C. 1999, \apj, 525, 47

\bibitem[{{Werner} {et~al.}(2008){Werner}, {Finoguenov}, {Kaastra},
  {Simionescu}, {Dietrich}, {Vink}, \& {B{\"o}hringer}}]{Werner2008}
{Werner}, N., {Finoguenov}, A., {Kaastra}, J.~S., {et~al.} 2008, \aap, 482, L29

\bibitem[{{Zacchei} {et~al.}(2011){Zacchei}, {Maino}, {Baccigalupi},
  {Bersanelli}, {Bonaldi}, {Bonavera}, {Burigana}, {Butler}, {Cuttaia}, {de
  Zotti}, {Dick}, {Frailis}, {Galeotta}, {Gonz{\'a}lez-Nuevo}, {G{\'o}rski},
  {Gregorio}, {Keih{\"a}nen}, {Keskitalo}, {Knoche}, {Kurki-Suonio},
  {Lawrence}, {Leach}, {Leahy}, {L{\'o}pez-Caniego}, {Mandolesi}, {Maris},
  {Matthai}, {Meinhold}, {Mennella}, {Morgante}, {Morisset}, {Natoli},
  {Pasian}, {Perrotta}, {Polenta}, {Poutanen}, {Reinecke}, {Ricciardi},
  {Rohlfs}, {Sandri}, {Suur-Uski}, {Tauber}, {Tavagnacco}, {Terenzi}, {Tomasi},
  {Valiviita}, {Villa}, {Zonca}, {Banday}, {Barreiro}, {Bartlett}, {Bartolo},
  {Bedini}, {Bennett}, {Binko}, {Borrill}, {Bouchet}, {Bremer}, {Cabella},
  {Cappellini}, {Chen}, {Colombo}, {Cruz}, {Curto}, {Danese}, {Davies},
  {Davis}, {de Gasperis}, {de Rosa}, {de Troia}, {Dickinson}, {Diego},
  {Donzelli}, {D{\"o}rl}, {Efstathiou}, {En{\ss}lin}, {Eriksen}, {Falvella},
  {Finelli}, {Franceschi}, {Gaier}, {Gasparo}, {G{\'e}nova-Santos}, {Giardino},
  {G{\'o}mez}, {Gruppuso}, {Hansen}, {Hell}, {Herranz}, {Hovest}, {Huynh},
  {Jewell}, {Juvela}, {Kisner}, {Knox}, {L{\"a}hteenm{\"a}ki}, {Lamarre},
  {Leonardi}, {Le{\'o}n-Tavares}, {Lilje}, {Lubin}, {Maggio}, {Marinucci},
  {Mart{\'{\i}}nez-Gonz{\'a}lez}, {Massardi}, {Matarrese}, {Meharga},
  {Melchiorri}, {Migliaccio}, {Mitra}, {Moss}, {N{\o}rgaard-Nielsen}, {Pagano},
  {Paladini}, {Paoletti}, {Partridge}, {Pearson}, {Pettorino}, {Pietrobon},
  {Pr{\'e}zeau}, {Procopio}, {Puget}, {Quercellini}, {Rachen}, {Rebolo},
  {Robbers}, {Rocha}, {Rubi{\~n}o-Mart{\'{\i}}n}, {Salerno}, {Savelainen},
  {Scott}, {Seiffert}, {Silk}, {Smoot}, {Sternberg}, {Stivoli}, {Stompor},
  {Tofani}, {Toffolatti}, {Tuovinen}, {T{\"u}rler}, {Umana}, {Vielva},
  {Vittorio}, {Vuerli}, {Wade}, {Watson}, {White}, \&
  {Wilkinson}}]{2011A&A...536A...5Z}
{Zacchei}, A., {Maino}, D., {Baccigalupi}, C., {et~al.} 2011, \aap, 536, A5

\bibitem[{{Zacchei et al.}(2011)}]{Planck2011-1.6}
{Zacchei et al.} 2011, \aap, 536, A5

\end{thebibliography}
